\documentclass[twocolumn]{aastex63}
\pdfoutput=1 
\usepackage[T1]{fontenc}
\usepackage{amsmath,amstext}
\usepackage{apjfonts} 
\usepackage[figure,figure*]{hypcap}
\usepackage{microtype}
\usepackage{tablefootnote}
\usepackage{booktabs}  
\usepackage{longtable}

\begin{document}

\shortauthors{Eftekhari et al.}
\shorttitle{Millimeter Transients in the Era of CMB Surveys}

\uppercase{\title{\normalfont Extragalactic Millimeter Transients in the Era of Next Generation CMB Surveys}}

\author[0000-0003-0307-9984]{T.~Eftekhari}
\affiliation{Center for Interdisciplinary Exploration and Research in Astrophysics (CIERA) and Department of Physics and Astronomy,\\ Northwestern University, Evanston, IL 60208, USA}
\affiliation{Center for Astrophysics | Harvard \& Smithsonian, Cambridge, MA 02138, USA}

\author[0000-0002-9392-9681]{E.~Berger}
\affiliation{Center for Astrophysics | Harvard \& Smithsonian, Cambridge, MA 02138, USA}

\author[0000-0002-4670-7509]{B.~D.~Metzger}
\affiliation{Department of Physics and Columbia Astrophysics Laboratory, Columbia University, New York, NY 10027, USA}
\affil{Center for Computational Astrophysics, Flatiron Institute, 162 5th Ave, New York, NY 10010, USA} 

\author[0000-0003-1792-2338]{T.~Laskar}
\affiliation{Department of Physics, University of Bath, Claverton Down, Bath BA2 7AY, UK}
\affiliation{Department of Astrophysics, Radboud University, 6525 AJ Nijmegen, The Netherlands}

\author[0000-0002-5814-4061]{V.~A.~Villar}
\affiliation{Department of Astronomy \& Astrophysics, The Pennsylvania State University, University Park, PA 16802, USA}
\affiliation{Institute for Computational \& Data Sciences, The Pennsylvania State University, University Park, PA 16802, USA}
\affiliation{Institute for Gravitation and the Cosmos, The Pennsylvania State University, University Park, PA 16802, USA}

\author[0000-0002-8297-2473]{K.~D.~Alexander}
\affiliation{Center for Interdisciplinary Exploration and Research in Astrophysics (CIERA) and Department of Physics and Astronomy,\\ Northwestern University, Evanston, IL 60208, USA}

\author[0000-0002-0463-6394]{G.~P.~Holder}
\affiliation{Department of Physics, University of Illinois Urbana-Champaign, 1110 West Green Street, Urbana, IL, 61801, USA}

\author[0000-0001-7192-3871]{J.~D.~Vieira}
\affiliation{Department of Physics, University of Illinois Urbana-Champaign, 1110 West Green Street, Urbana, IL, 61801, USA}
\affiliation{Astronomy Department, University of Illinois at Urbana-Champaign, 1002 W. Green Street, Urbana, IL 61801, USA}
\affiliation{Center for AstroPhysical Surveys, National Center for Supercomputing Applications, Urbana, IL, 61801, USA}

\author[0000-0002-3157-0407]{N.~Whitehorn}
\affiliation{Department of Physics and Astronomy, Michigan State University, 567 Wilson Road, East Lansing, MI 48824}
\affiliation{Department of Physics and Astronomy, University of California, Los Angeles, CA 90095, USA}

\author[0000-0003-3734-3587]{P.~K.~G.~Williams}
\affiliation{Center for Astrophysics | Harvard \& Smithsonian, Cambridge, MA 02138, USA}
\affiliation{American Astronomical Society, 1667 K Street NW, Suite 800 Washington, DC 20006 USA}

\begin{abstract}
The next generation of wide-field cosmic microwave background (CMB) surveys are uniquely poised to open a new window for time-domain astronomy in the millimeter band. Here we explore the discovery phase space for extragalactic transients with near-term and future CMB experiments to characterize the expected population. We use existing millimeter-band light curves of known transients (gamma-ray bursts, tidal disruption events, fast blue optical transients, neutron star mergers) and theoretical models, in conjunction with known and estimated volumetric rates. Using Monte Carlo simulations of various CMB survey designs (area, cadence, depth, duration) we estimate the detection rates and the resulting light curve characteristics.  We find that existing and near-term surveys will find tens to hundreds of long-duration gamma-ray bursts (LGRBs), driven primarily by detections of the reverse shock emission, and including off-axis LGRBs.  Next-generation experiments (CMB-S4, CMB-HD) will find tens of fast blue optical transients (FBOTs) in the nearby universe and will detect a few tidal disruption events. CMB-HD will additionally detect a small number of short gamma-ray bursts, where these will be discovered within the detection volume of next generation gravitational wave experiments like Cosmic Explorer.
\end{abstract}


\section{Introduction}
\label{sec:intro}

Our view of the transient sky has been revolutionized in the past few decades by the advent of untargeted wide-field optical surveys and all-sky X-ray/$\gamma$-ray satellites. More recently, wide-field centimeter-band radio searches have been successful at finding millisecond-duration transients (fast radio bursts; \citealt{Shannon2018,CHIME2019,CHIME2021}), as well as a few long-duration ($\sim$years) transients \citep{Law2019,Anderson2020,Dong2021,Ravi2021,Stroh2021}. Outside of the electromagnetic (EM) band, gravitational wave (GW) observatories have detected various types of compact object mergers \citep{LIGO2017,LIGO2019,Abbott2021}, and neutrino detectors have found some evidence for flaring neutrino emission from active galactic nuclei (AGN) \citep{IceCube2018}. These various approaches have significantly increased the discovery rate of known events, enabled the discovery of new phenomena, and have led to multi-messenger signals from some events.  Upcoming powerful facilities such as the Vera Rubin Observatory (VRO;  \citealt{LSST2009}) and the Roman Space Telescope \citep{Foley2019} will continue to reshape the landscape of time-domain astronomy. 

At centimeter wavelengths, wide-field searches are in their infancy, with surveys such as the Australian Square Kilometer Array Pathfinder (ASKAP) Survey for Variables and Slow Transients (VAST; \citealt{Murphy2013}) and the Very Large Array Sky Survey (VLASS; \citealt{Lacy2020}) only just beginning to provide the depth and sky coverage required for the discovery of extragalactic transients. In the coming decade, the improved sensitivity and wide areal coverage of facilities such as the Square Kilometer Array (SKA) will enable a broader range of time-domain discoveries \citep{Metzger2015,Dobie2021,Leung2021}.

At millimeter wavelengths, the study of transients has largely been limited to targeted follow-up of events discovered at other wavelengths. These include relativistic sources like gamma-ray bursts (GRBs; e.g., \citealt{Berger2003}) and some tidal disruption events (TDEs; e.g., \citealt{Berger2012,Cendes2021b}), as well as a few supernovae in the nearby universe (e.g., \citealt{Soderberg2008,Horesh2013}). The direct {\it discovery} of transients in the millimeter band may be enabled by the next generation of cosmic microwave background (CMB) experiments, as these surveys will cover a large fraction of the sky at high cadence and with unprecedented sensitivity, complementing centimeter-band and optical surveys \citep{Metzger2015}. Indeed, the discovery of millimeter transients at early times in CMB surveys will uniquely facilitate the follow-up of sources across the electromagnetic spectrum, lending to improved localizations and robust source characterizations. 

There are already possible hints of an interesting discovery space from on-going CMB surveys. The first search for transients in a CMB survey led to the detection of a single event with the South Pole Telescope polarization-sensitive SPTpol survey \citep{Whitehorn2016}; the lack of a clear host galaxy and the low detection significance, however, prevented a conclusive determination of the nature of the event. A similar search using the third generation camera on the South Pole Telescope, SPT-3G, revealed two extragalactic transient events with timescales of a few weeks, some evidence of light curve variability, and potential host galaxies in the localization regions \citep{Guns2021}. The nature of these sources remains unclear, although the observed variability may be indicative of an AGN origin. Finally, using archival Planck data, \citet{Yuan2016} discovered a bright millimeter source coincident with the TDE IGR J12580+0134, suggesting the event may represent one of two known TDEs with off-axis jetted emission (see also \citealt{Mattila2018}). An increasing number of Galactic stellar flares have also been serendipitously detected in CMB surveys, but with much shorter durations and obvious stellar counterparts \citep{Guns2021,Naess2021}.

Given the potential of CMB surveys to discover transients, it is essential to explore their actual discovery space and the survey strategies that would maximize the discovery rate. This has been previously discussed in \citet{Metzger2015}, although with a focus on centimeter-band transients and surveys. Here, we follow the approach of \citet{Metzger2015} and present for the first time a comprehensive exploration of the discovery of millimeter-band transients in CMB surveys. We consider a wide range of phenomena that span a broad range of energies and ambient densities (and hence timescales and luminosities). We use known and estimated volumetric rates and their cosmological evolution, and define a robust detection criterion to explore how detection rates vary as a function of survey depth, cadence, and sky coverage. The wide range of known transients considered here, along with the basic physical constraints of synchrotron emission, provide a robust characterization of the parameter space. 

We note that we do not consider Galactic transients such as stellar flares, and we further distinguish between bona-fide transients and variable sources like AGN. Although in principle wide-field experiments may lead to the serendipitous discovery of new populations of previously unknown events, as we show, the physical constraints of synchrotron emission limit the allowed phase space for such sources. The broad range of energies and environment densities probed by the models we consider here thus provide an adequate characterization of the allowed parameter space. 

The paper is structured as follows. In \S\ref{sec:transients} we review the landscape of known  transients with detected millimeter emission, and present basic arguments that govern the timescales and luminosities of synchrotron sources (\S\ref{sec:sync}). In \S\ref{sec:models}, we discuss the properties and rates of various classes of transients considered in our analysis. We provide a brief overview of relevant CMB surveys in \S\ref{sec:surveys}, and describe our simulations in \S\ref{sec:sims}. Our results are presented in \S\ref{sec:results}, and we summarize our main findings in \S\ref{sec:conclusions}. Throughout the paper, we use the following Planck cosmological parameters:
$H_0 = 67.8 \ \rm km \ s^{-1} \  Mpc^{-1}$, $\Omega_m = 0.308$, and $\Omega_\Lambda = 0.692$ \citep{Planck2016}.

\begin{figure}
\includegraphics[width=\columnwidth]{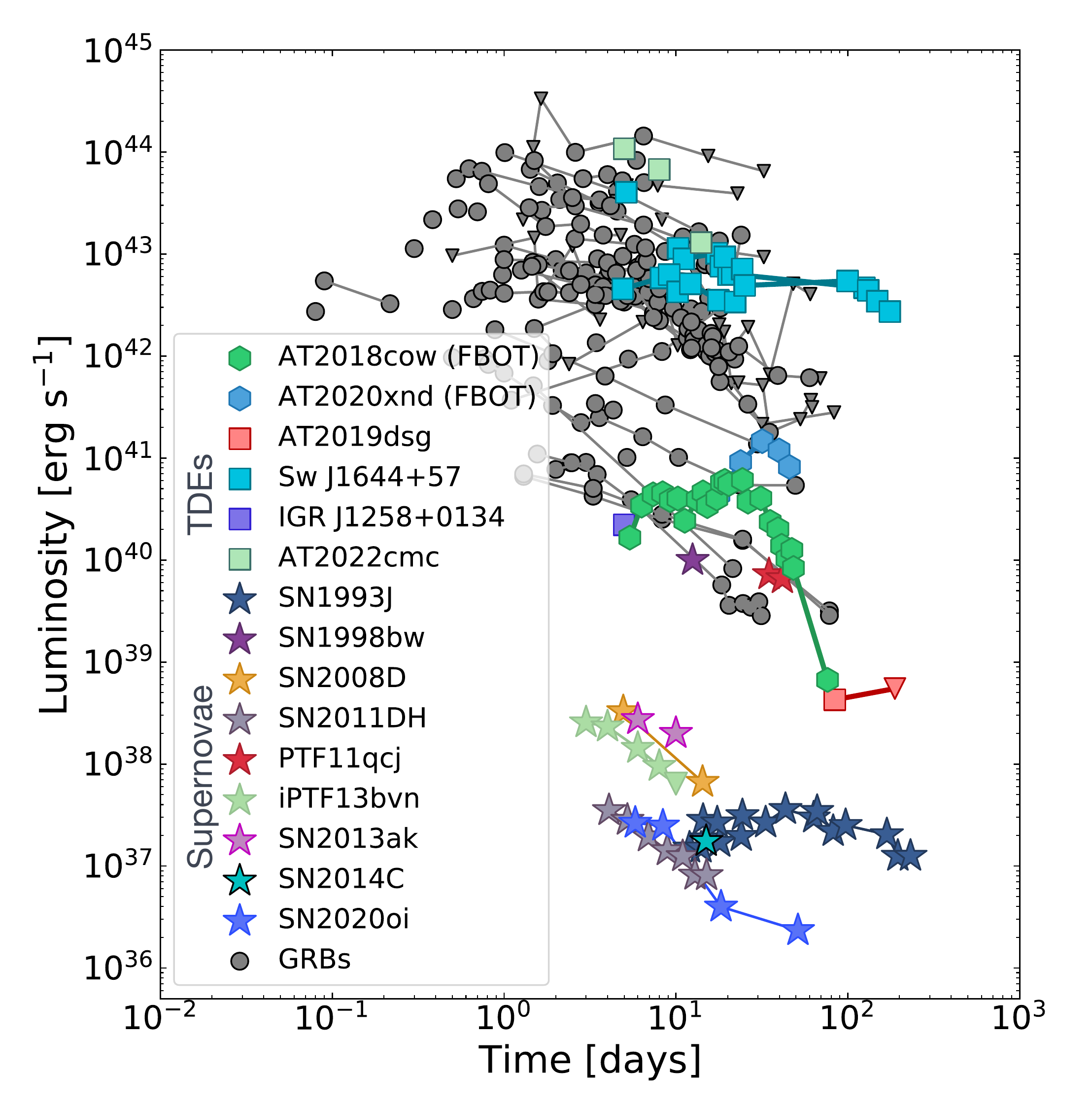}
\caption{Millimeter light curves of extragalactic transients, including tidal disruption events (AT2019dsg: \citealt{Cendes2021a}, Sw J1644+57: \citealt{Berger2012}, IGR J1258+0134: \citealt{Yuan2016}, AT2022cmc: \citealt{Alexander2022,Perley2022,Smith2022}), the fast blue optical transients AT2018cow \citep{Ho2019} and AT20202xnd \citep{Ho2021xnd}, core-collapse supernovae (stars), and long GRBs (gray circles; see Table~\ref{tab:lit}). Limits correspond to $3\sigma$.}
\label{fig:lcs}
\end{figure}

\section{Extragalactic Transients}
\label{sec:transients}

We consider a wide variety of extragalactic synchrotron transients that span a range of timescales and luminosities in the millimeter band, corresponding to a fiducial range of $\sim 30 - 300$ GHz. These include detected sources, which span over seven orders of magnitude in luminosity ($L\sim 10^{37} - 10^{44} \ \rm erg \ s^{-1}$) and three orders of magnitude in time ($t\sim 10^{-1} - 10^2$ d); see Figure~\ref{fig:lcs}.  We also consider events that are theoretically predicted despite no unambiguous direct detections to date (e.g., off-axis ``orphan'' GRB afterglows), as well as more speculative transients (neutron star mergers leaving stable magnetar remnants).

In Table~\ref{tab:lit} and Figure~\ref{fig:lcs} we provide a complete list of extragalactic transients with millimeter detections from the literature through March 2022 using the NASA Astrophysics Data System (ADS).  Long GRBs, which dominate the sample, span luminosities of $L\sim 10^{41} - 10^{44} \ \rm erg \ s^{-1}$ (see \citet{deUgarte2012} for a review). These LGRB detections have also revealed reverse shock emission at early times with ALMA \citep{Laskar2013,Laskar2018,Laskar2019}.

A small, but growing, sample of TDEs have also been detected in the millimeter. The most luminous examples are the relativistic TDEs Swift J164449.3+573451 (hereafter, Sw J1644+57; \citealt{Bloom2011,Burrows2011,Levan2011,Zauderer2011}) and AT2022cmc \citep{Andreoni2022} with a luminosity comparable to typical LGRBs detected over a timescale $\sim 200$ days (\citealt{Berger2012}; \citealt{Alexander2022,Perley2022,Smith2022}). More recently, \citet{Cendes2021b} used ALMA to reveal the first millimeter detection of a non-relativistic TDE, AT2019dsg, with a luminosity of $\sim {\rm few}\times 10^{38}$ erg s$^{-1}$.

Finally, in the nearby universe, a small number of supernovae have also been detected in the millimeter (e.g., \citealt{Weiler2007,Soderberg2008,Horesh2013}), with luminosities of $\sim 10^{37}-10^{38}$ erg s$^{-1}$. At higher luminosities ($\sim 10^{41}$ erg s$^{-1}$), a growing number of luminous, rapidly evolving blue transients (see \S\ref{sec:fbots}) have also been discovered (e.g., \citealt{Ho2019,Margutti2019,Coppejans2020,Ho2021xnd}).


Despite the wide diversity of physical phenomena, the production of synchrotron emission in these various transients is a generic feature of shock mediated outflows. Synchrotron emission is generated at the interface between the outflowing ejecta and a dense ambient environment. The wide diversity in luminosities and timescales reflects the range of outflow velocities, kinetic energies, and geometries, as well as a range of ambient environments. For example, energetic relativistic outflows viewed on-axis (e.g., LGRBs) produce luminous millimeter emission peaking on $\sim$day timescales, while non-relativistic outflows with modest kinetic energies (e.g., supernovae) result in low luminosities and longer durations.

\subsection{Synchrotron Emission: Timescales and Luminosities}
\label{sec:sync}

We begin with a general investigation of the luminosities and timescales expected for millimeter-band synchrotron emitting transients.  The discussion below is meant to provide a broad overview of the landscape of transients, and not a detailed derivation for any particular class of events. We consider an outflow with kinetic energy $E_K$ and initial Lorentz factor $\Gamma\equiv (1-\beta^2)^{-1/2}$ where $\beta = v/c$ for an ejecta velocity $v$ expanding into an ambient medium with constant density $n$. The outflow sweeps up a mass comparable to its own, and transfers its energy to the surrounding medium, at the deceleration radius $R_{\rm dec}$ given by \citep{Nakar2011}:
\begin{equation}
\begin{split}
R_{\rm dec} = & \bigg(\frac{3 E_{K}}{4 \pi n m_p c^2 \beta^2 \Gamma^2}\bigg)^{1/3} \\
 &\approx 10^{18} \ {\rm cm} \ E_{K,52}^{1/3} n^{-1/3} \beta^{-2/3} \Gamma^{-2/3}, 
\end{split}
\end{equation}
where we have scaled the kinetic energy to $E_K=10^{52}$ erg and density to $n = 1 \rm \ cm^{-3}$. The deceleration time is given by:
\begin{equation}
\label{eq:tdec}
t_{\rm dec} = \frac{R_{\rm dec}}{2 c\beta^2 \Gamma^2} \approx 225\ {\rm days} \ E_{K,52}^{1/3} n^{-1/3} \beta^{-5/3} \Gamma^{-8/3}.
\end{equation}
Before $t_{\rm dec}$ the light curve will rapidly brighten, before subsequently declining following the Sedov-Taylor self-similar solution. This timescale therefore defines a minimum peak time for millimeter transients; we note that relativistic events will exhibit much shorter timescales (i.e., $t_{\rm dec}\propto \Gamma^{-8/3}$). 
For an observing frequency (here, $\nu_{\rm obs} = 100$ GHz) above the characteristic synchrotron frequency, $\nu_m$, and the self-absorption frequency, $\nu_a$, the peak flux at $t_{\rm dec}$ is given by \citep{Nakar2011}:
\begin{equation}
\label{eq:fdec}
F_{\rm \nu, dec} \approx 3\ {\rm mJy} \ E_{K,52} n^{0.83} \epsilon_{\rm e,-1}^{1.3} \epsilon_{\rm B,-2}^{0.83} \beta^{2.3} d_{\rm L,27}^{-2} \nu_{\rm 100\ GHz}^{-0.65},
\end{equation}
where $\epsilon_e$ and $\epsilon_B$ are the fractions of post-shock energy in the radiating electrons and magnetic field, respectively, $d_L = 10^{27} \ d_{\rm L,27}$ is the luminosity distance, and we adopt a power-law index $p=2.3$ for the electron energy distribution given by $N(\gamma) \propto\gamma^{-p}$ for $\gamma$ above a minimal Lorentz factor $\gamma_m = \Gamma\epsilon_{\rm e}m_p(p-2)/m_e(p-1) \approx 40 \epsilon_{e,-1} \Gamma$ \citep{Sari1998}, where $m_p/m_e \approx 1800$ is the ratio of proton and electron masses. 

For non-relativistic transients, or initially off-axis relativistic events that reach sub-relativistic velocities at late times, the flux density therefore scales linearly with energy, while the timescale scales as $E_{K}^{1/3}$. High density environments similarly lead to brighter events ($F_{\rm \nu,dec}\propto n^{0.83}$) and overall shorter timescales ($t_{\rm dec} \propto n^{-1/3}$). 

In the case where the density of the ambient medium is shaped by the progenitor's stellar wind (e.g., $\rho = Ar^{-2}$, where $A \equiv \dot{M}/4 \pi v_w = 5 \times 10^{11} A_{\star}$ for a mass-loss rate $\dot{M} = 10^{-5} \ M_{\odot} \ \rm yr^{-1}$ and a wind velocity $v_w = 10^3 \ \rm km \ s^{-1}$), the deceleration time increases and is given by
\begin{equation}
t_{\rm dec} \approx 1025 \ {\rm days} \ A_{\star} E_{K,52} \beta^{-3} \Gamma^{-4}.
\end{equation}

For on-axis relativistic sources, the relevant timescale to reach peak brightness is not $t_{\rm dec}$ but the timescale when $\nu_{\rm obs} = \nu_m$, where the value of $\nu_m$ is determined by the minimum Lorentz factor $\gamma_m$, and is given by \citep{Granot2002}:
\begin{equation}\label{eq:vm}
\nu_m \approx 100 \ {\rm GHz} \ E_{\rm K,iso,54}^{1/2} \epsilon_{e,\rm -1}^{2} \epsilon_{B,\rm -2}^{1/2} t_{1}^{-3/2},
\end{equation}
where we have scaled to $t=10$ days, and re-scaled to an isotropic-equivalent kinetic energy $E_{\rm K,iso}=10^{54}$ erg assuming a beaming correction of $0.01$.  Equation~\ref{eq:vm} makes clear that relativistic sources reach peak brightness in the millimeter-band on a timescale of $\sim 10$ d, as indeed is observed in LGRBs (Figure~\ref{fig:lcs}). This is in contrast to the centimeter band where events evolve on much longer ($\sim 100$ days) timescales. While the longer timescales at GHz frequencies present a challenge for transient detection in radio surveys \citep{Metzger2015}, the short durations at $\sim 100$ GHz suggest that relativistic transients can be more readily identified in CMB surveys.

\begin{figure}
\includegraphics[width=\columnwidth]{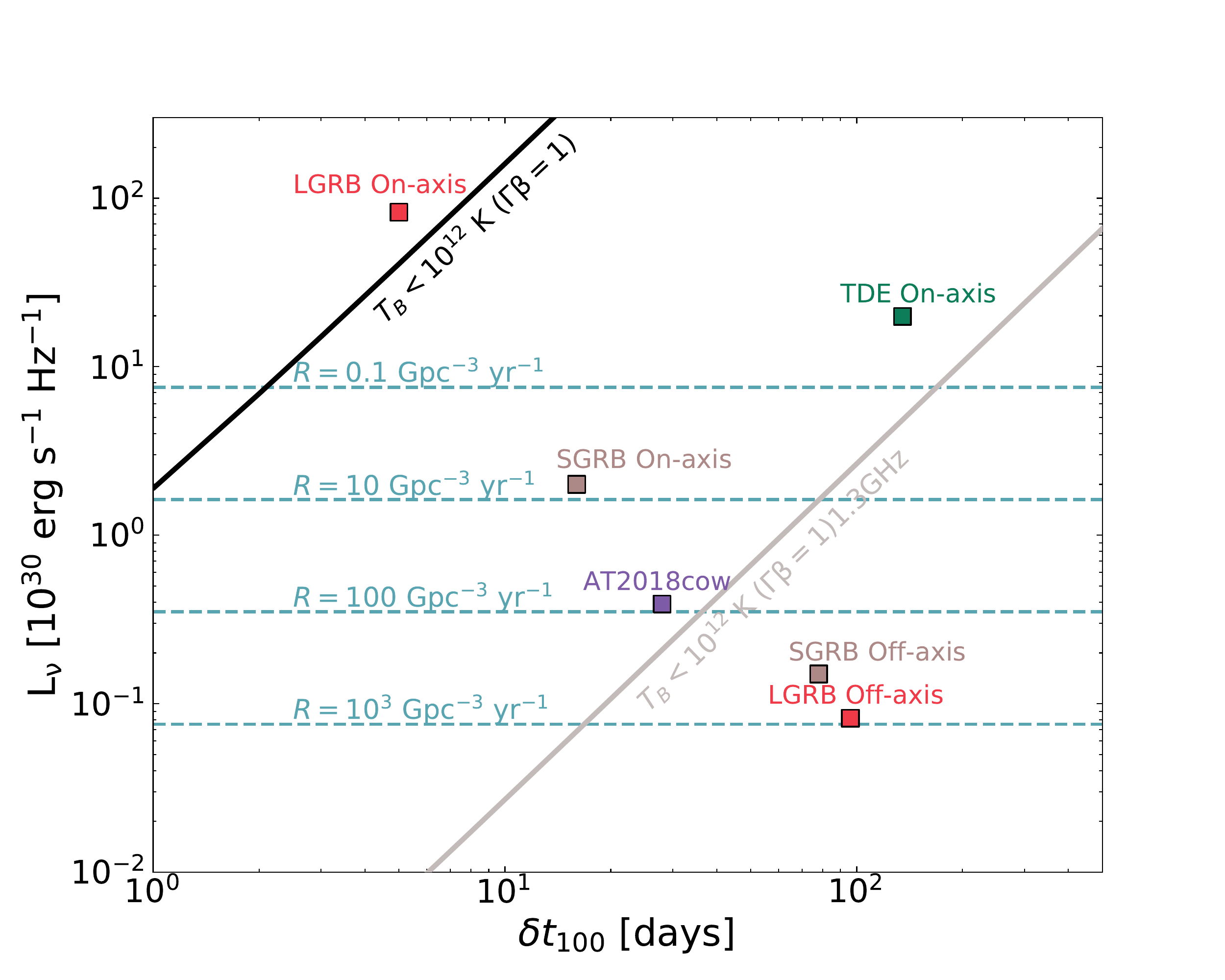}
\caption{Duration-luminosity phase space for extragalactic synchrotron transients at 100 GHz. Timescales for individual sources are defined as the time interval over which the flux is within a factor of 2 of the peak flux (Table~\ref{tab:rates}). Off-axis events correspond to $\theta_{\rm obs} = 0.8$. The black line indicates the brightness temperature constraint $T_{B} < 10^{12}$ K, above which inverse Compton scattering dominates the luminosity. With the exception of relativistic events (e.g., on-axis LGRBs), synchrotron sources obey $T_B < 10^{12}$ K. We also show this limit at 1.3 GHz (gray line) which restricts sources to longer timescales in the centimeter band. Horizontal dotted lines correspond to the minimum volumetric event rate leading to one transient detection across the entire sky for a seven year survey and a limiting sensitivity of 60 mJy ($10 \sigma$). Sources with lower luminosities (and hence shorter timescales) require large intrinsic rates to compensate for the decreased detection volume.}
\label{fig:phase}
\end{figure}

In addition to the above arguments, the brightness temperature $T_B$ of a synchrotron source sets a limit on the minimum observed variability timescale. The inverse Compton catastrophe requires that synchrotron sources obey $T_B < 10^{12}$ K; above this temperature, electrons rapidly lose energy as inverse Compton scattering of synchrotron photons dominates the energy output \citep{Kellermann1969}. From \citet{Kulkarni1998}, the brightness temperature of an expanding synchrotron source in the plasma frame is given by:
\begin{equation}
T_B' = \frac{L_\nu}{8 \pi^2 k} \frac{\Gamma^2}{\mathcal{D} \beta^2 \nu^2 t^2} \lesssim 10^{12} {\rm K}
\end{equation}
where $\mathcal{D} = [ \Gamma (1- \beta \ {\rm cos}\ \theta)]^{-1}$ is the Doppler factor and is equal to $\Gamma$ for the fastest moving material. This constraint implies that a source with a given emission radius cannot vary on a timescale shorter than the light travel time, where $\delta t_{\rm min} \approx t\beta/2\Gamma^2$ for relativistic expansion, corresponding to a minimum timescale for a source of luminosity $L_\nu$ given by \citep{Metzger2015}:
\begin{equation}
\delta t_{\rm min} \approx \bigg[\frac{L_\nu}{ 16 \pi^2 k T_{B}' \nu^2 \Gamma^3 } \bigg]^{1/2} \approx 0.8 \ {\rm days} \ \Gamma^{-3/2} L_{\nu,\rm 30}^{1/2} \nu_{\rm 100 \ GHz}^{-1}.
\end{equation}
The minimum variability timescale for sources above $L_{\nu} \gtrsim 10^{30} \ \rm erg \ s^{-1} \ Hz^{-1}$ at 100 GHz is therefore of order days, while the characteristic timescales at frequencies of a few GHz are $\sim 100$ days \citep{Metzger2015}.  Figure~\ref{fig:phase} shows the boundary corresponding to $T_B < 10^{12}$ K for $\Gamma \beta = 1$ compared to the timescales and luminosities for a range of millimeter transients. The shorter variability timescales at millimeter wavelengths further emphasizes the need for faster cadence in CMB surveys. 

Finally, we calculate the limiting luminosity ($L_\nu$) leading to a single detection over the course of a given survey by estimating the instantaneous number of events across the sky above a flux density threshold ($F_{\rm \nu, lim}$): 
\begin{equation}
N = \frac{4\pi}{3} \bigg(\frac{L_\nu}{4\pi F_{\rm \nu,lim}} \bigg)^{3/2} \mathcal{R} t_{\rm dur} 
\end{equation}
where $\mathcal{R}$ is the volumetric event rate, and we adopt $t_{\rm dur} = 7$ yr, motivated by the typical survey duration of planned and upcoming CMB surveys. In Figure~\ref{fig:phase}, we plot the contours corresponding to $N=1$ and unique values of $\mathcal{R}$ assuming $F_{\nu,\rm lim} = 60$ mJy (i.e., a $10\sigma$ detection in CMB-S4; see \S\ref{sec:cmbs4}). The results suggest that transients varying on timescales of $\lesssim 1$ day require large volumetric rates of $\mathcal{R} \gtrsim 10 \ \rm Gpc^{-3} \ yr^{-1}$ in order to result in a single detection for nominal CMB survey designs. Conversely, for a volumetric rate of $\mathcal{R} = 0.1 \ \rm Gpc^{-3} \ yr^{-1}$, comparable to the rate of relativistic transients (Table~\ref{tab:rates}), events with timescales as short as $\sim 2$ days are detectable. Thus while the combination of event rates and minimum variability timescales precludes the detection of short-duration ($\lesssim 100$ d) centimeter wavelength transients \citep{Metzger2015}, millimeter surveys will be sensitive to sources varying on timescales as short as a few days. 

\begin{figure*}
\center
\includegraphics[width=0.8\textwidth]{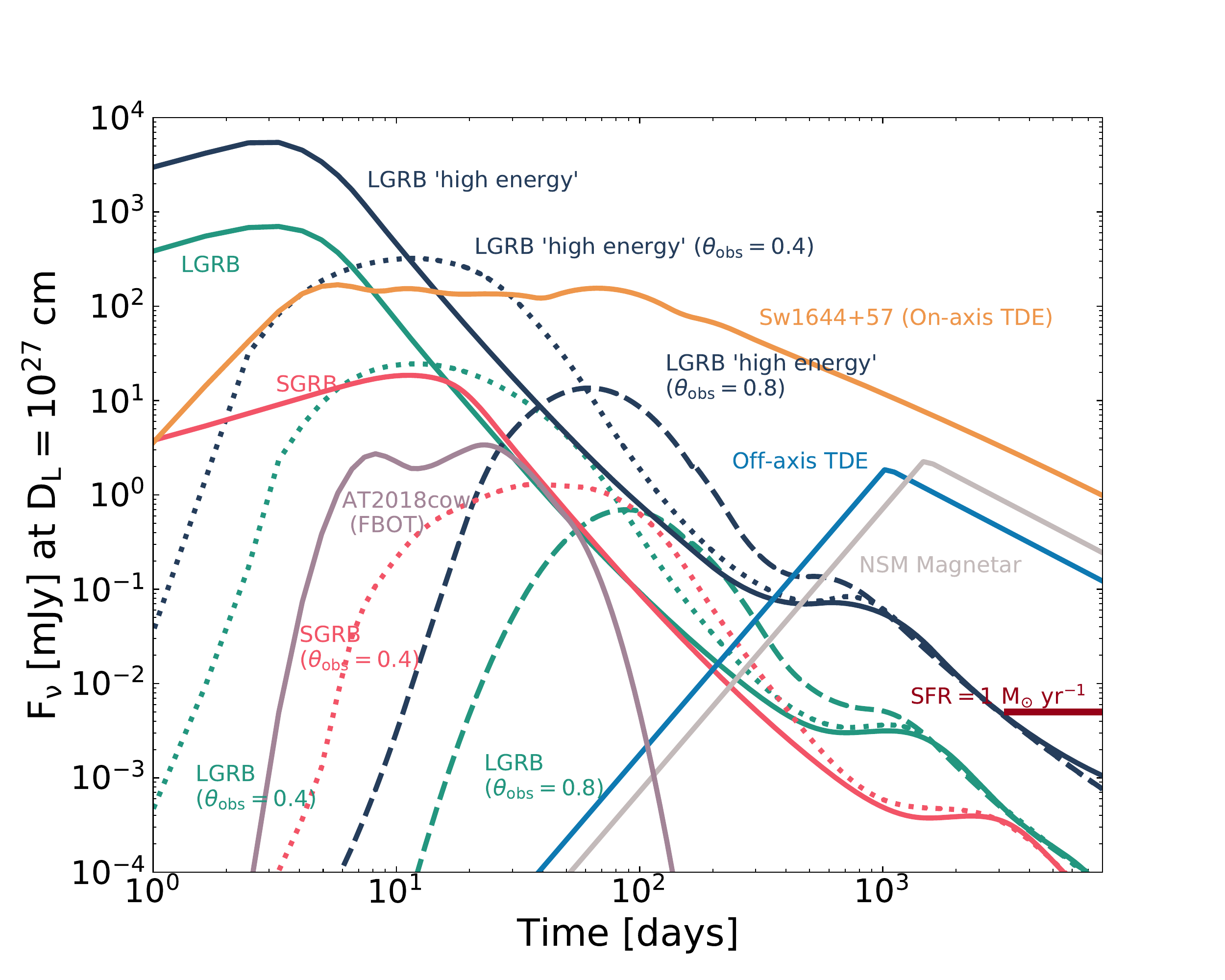}
\caption{Light curves of the extragalactic synchrotron transients considered in this work (Table~\ref{tab:rates}) at an observer frequency of 100 GHz. Sources include on- and off-axis long and short GRBs, on-axis relativistic TDEs (Sw1644+57), off-axis TDEs, FBOTs (AT2018cow), and neutron star binary mergers leaving behind a stable magnetar remnant. We also show the 100 GHz flux density corresponding to a typical star forming galaxy with a star formation rate of $1 \ \rm M_{\odot} \ yr^{-1}$ (horizontal red bar). Flux densities are normalized to $D_L = 10^{27}$ cm ($z \approx 0.07$).} 
\label{fig:100ghzlcs}
\end{figure*}

\subsection{Models and Volumetric Rates}
\label{sec:models}

In Figure~\ref{fig:100ghzlcs} we show the model 100 GHz light curves for the extragalactic transients we consider in this work, scaled to a common distance of $10^{27}$ cm ($z \approx 0.07$). This includes both on- and off-axis long and short GRBs, on- and off-axis TDEs, binary neutron star mergers with stable magnetar remnants, and fast blue optical transients (FBOTs). We also show lightcurves for the reverse shock models which peak at much earlier times (\S\ref{sec:lgrbs}) in Figure~\ref{fig:rs}. While a handful of SNe have been detected in the millimeter, the low luminosities of these events ($L \sim 10^{37} - 10^{38} \rm erg \ s^{-1}$; Figure~\ref{fig:lcs}) precludes their detection in CMB surveys. Indeed, given the sensitivity of CMB experiments (Table~\ref{tab:surveys}), except in the most extreme cases, SNe would only be detected to a distance of a few Mpc, thus we do not consider them further here \citep{Yadlapalli2022}. 

To determine the detection rates we also consider the volumetric rates and their redshift evolution for each class of transients; see Table~\ref{tab:rates}.  For transients involving massive stars or compact objects, we assume that the volumetric rates track the cosmic star formation rate density, given by \citep{Madau2014}:
\begin{equation}\label{eq:sfh}
{\rm SFH}(z) = 0.015 \dfrac{(1+z)^{2.7}}{1 + [(1+z)/2.9]^{5.6}}\ \rm M_{\odot}\ Mpc^{-3} \ yr^{-1}.
\end{equation}
For these systems, the volumetric rate for a given transient class is normalized to the cosmic star formation history and assumed to evolve with redshift in the same manner. For TDEs, where the event rate depends on the number density of supermassive black holes (SMBH), we use the black hole accretion rate density from \citet{Sijacki2015} (their Figure 2) for SMBHs in the mass range $10^6 - 10^7 \ \rm M_{\odot}$. 

Below we provide a brief overview of the different classes and models considered in our simulations.

\subsubsection{Long Gamma-ray Bursts}
\label{sec:lgrbs}
LGRBs mark the energetic explosions following the core-collapse of stripped massive stars \citep{Woosley2006}. These events produce highly collimated outflows or relativistic jets that power afterglow emission across the electromagnetic spectrum as the jets decelerate and expand into the surrounding circumstellar medium (CSM) \citep{Meszaros1997}. 

LGRBs have beaming-corrected kinetic energies of $\sim 10^{51}-10^{52}$ erg with jet opening angles of $\theta_j\sim 10^\circ$ \citep{Frail2001,Berger2003}. The environments around LGRBs span densities of $\sim 0.1-100$ cm$^{-3}$ (e.g., \citealt{Panaitescu2002}). Here we consider both the forward shock (FS) and reverse shock (RS) components of the afterglow emission. The RS propagates into the ejecta as the jet decelerates, providing a unique probe of the baryon content and magnetization of the ejecta \citep{Nakar2004,Granot2005}. Unlike the FS component which peaks on timescales of days to months depending on the observer viewing angle, the RS component is expected to peak in the millimeter band on timescales of $\lesssim$ 1 day \citep{Laskar2013,Laskar2018,Laskar2019}.

For the FS emission, we generate LGRB millimeter-band light curves using the two-dimensional relativistic hydrodynamical code \texttt{Boxfit v2} \citep{vanEerten2012}. We consider two cases: $E_K = 10^{51}$ erg and $n = 1 \ \rm cm^{-3}$ (``typical'') and $E_K = 10^{52}$ erg and $n = 10 \ \rm cm^{-3}$ (``high energy''). In both cases, we assume typical parameters for the jet opening angle $\theta_j = 0.2$, the synchrotron power law index $p=2.5$, and the microphysical parameters describing the fraction of energy in electric and magnetic fields $\epsilon_e = 0.1$ and $\epsilon_B = 0.01$, respectively.  We further consider events viewed both on- and off-axis, noting that off-axis events comprise the vast majority of the intrinsic LGRB population, despite a paucity of detections to date (owing to their lack of prompt $\gamma$-ray emission) \citep{Ghirlanda2014}. For the off-axis case, we generate models with observer viewing angles of $\theta_{\rm obs} = 0.4$ and $0.8$.

We compute RS light curves at 100~GHz for on-axis top-hat jets in an ISM environment for both typical and high energy models, each with two sets of jet Lorentz factors at the shell crossing time, $\Gamma(t_{\rm dec}) = 200$ and 50. We assume a ``thin shell'' model with a non-relativistic RS, such that $t_{\rm dec}$ corresponds to the RS shell crossing time \citep{kob00}. In this model, the shell expands adiabatically at the sound speed after $t_{\rm dec}$ and the Lorentz factor of the shocked shell evolves as $\Gamma\propto R^{-g}$. We set $g=3.5$, corresponding to the minimum value expected in an ISM environment as the shocked shell must lag behind the FS \citep{ks00}. We scale the RS cooling frequency, and characteristic synchrotron frequency, and RS peak flux to that of the FS at $t_{\rm dec}$ using the relations in \cite{kz03}, using $p=2.5$ and fiducial values for the microphysical parameters, $\epsilon_{\rm e}=0.1$ and $\epsilon_{\rm B}=0.01$, as for the FS. For simplicity, we assume equal electron acceleration efficiency in both RS and FS regions and no additional fireball magnetization ($\epsilon_{\rm e, RS}/\epsilon_{\rm e,FS}=1$ and $\epsilon_{\rm B, RS}/\epsilon_{\rm B,FS}=1$) or magnetic field decay (which may suppress the RS; e.g. \citealt{sry+21}). RS synchrotron self-absorption is not included as it does not have a strong effect on the millimeter-band light curves for these parameters.

For the local volumetric rate of on-axis LGRBs, we use $\mathcal{R_{\rm LGRB}} = 0.2^{+0.03}_{-0.02} \ \rm Gpc^{-3} \ yr^{-1}$ for events with $L_{\rm iso}\gtrsim 5\times 10^{51} \ \rm erg \ s^{-1}$ (corresponding to the ``typical'' model) and $\mathcal{R_{\rm LGRB}} = 0.1^{+0.01}_{-0.01} \ \rm Gpc^{-3} \ yr^{-1}$ for our ``high energy'' events with $L_{\rm iso} \gtrsim 5 \times 10^{52} \ \rm erg \ s^{-1}$ \citep{Sun2015}. For off-axis events, we quantify the fraction of events, $f_{\rm obs} = \int_{\theta_{1}}^{\theta_{2}} sin\theta \,d\theta$, contributing to the total event rate as a function of viewing angle $\theta_{\rm obs}$. Here the integration is carried out from $\theta_1 = \theta_{\rm obs}/2$ to $\theta_2 = \theta_{\rm obs}$, corresponding to the range of solid angles subtended for an observer with a viewing angle $\theta \approx \theta_{\rm obs}$. The off-axis rate is then given by $\mathcal{R_{\rm LGRB}}(\theta_{\rm obs}) = \mathcal{R_{\rm LGRB}} f_{\rm obs}^{-1}$. Finally, we scale the rates to the cosmic star formation rate density using Equation~\ref{eq:sfh}.

\subsubsection{Short Gamma-ray Bursts}

SGRBs are produced by binary neutron star mergers \citep{Berger2014}, but otherwise their afterglow emission physics is similar to that of LGRBs \citep{Sari1995}, albeit with lower energy jets ($E_K \approx 10^{49} - 10^{50}$ erg) and generally lower density environments ($n\approx 10^{-3} - 10^{-2}\ \rm cm^{-3}$; \citealt{Berger2014,Fong2015}). Such sources are expected to peak in the millimeter at early times when the synchrotron peak frequency is located in the millimeter band. Moreover, for environments with $n \gtrsim 10^{-2} \ \rm cm^{-3}$, self-absorption will initially suppress the centimeter band emission. 

We generate the afterglow models using \texttt{Boxfit} with a beaming-corrected energy $E_K = 10^{50}$ erg, $\theta_j = 0.2$, and $n = 10^{-2} \ \rm cm^{-3}$ (and microphysical parameters identical to the LGRB case). We also consider off-axis events with $\theta_{\rm obs} = 0.4$.

We adopt a local volumetric rate of on-axis SGRBs of $\mathcal{R_{\rm SGRB}} = 1.3^{+0.4}_{-0.3} \ \rm Gpc^{-3} \ yr^{-1}$ for events above $L_{\rm iso} \gtrsim 10^{50} \ \rm erg \ s^{-1}$ \citep{Sun2015}, and scale it with redshift using Equation~\ref{eq:sfh}.  We note that the event rate for SGRBs is depdendent on the assumed merger delay model. Here we use the rate as derived for a lognormal delay model, consistent with existing data \citep{Wanderman2015}. 
For the off-axis rates, we scale the on-axis rate as in \S\ref{sec:lgrbs}.

\begin{figure}
\includegraphics[width=\columnwidth]{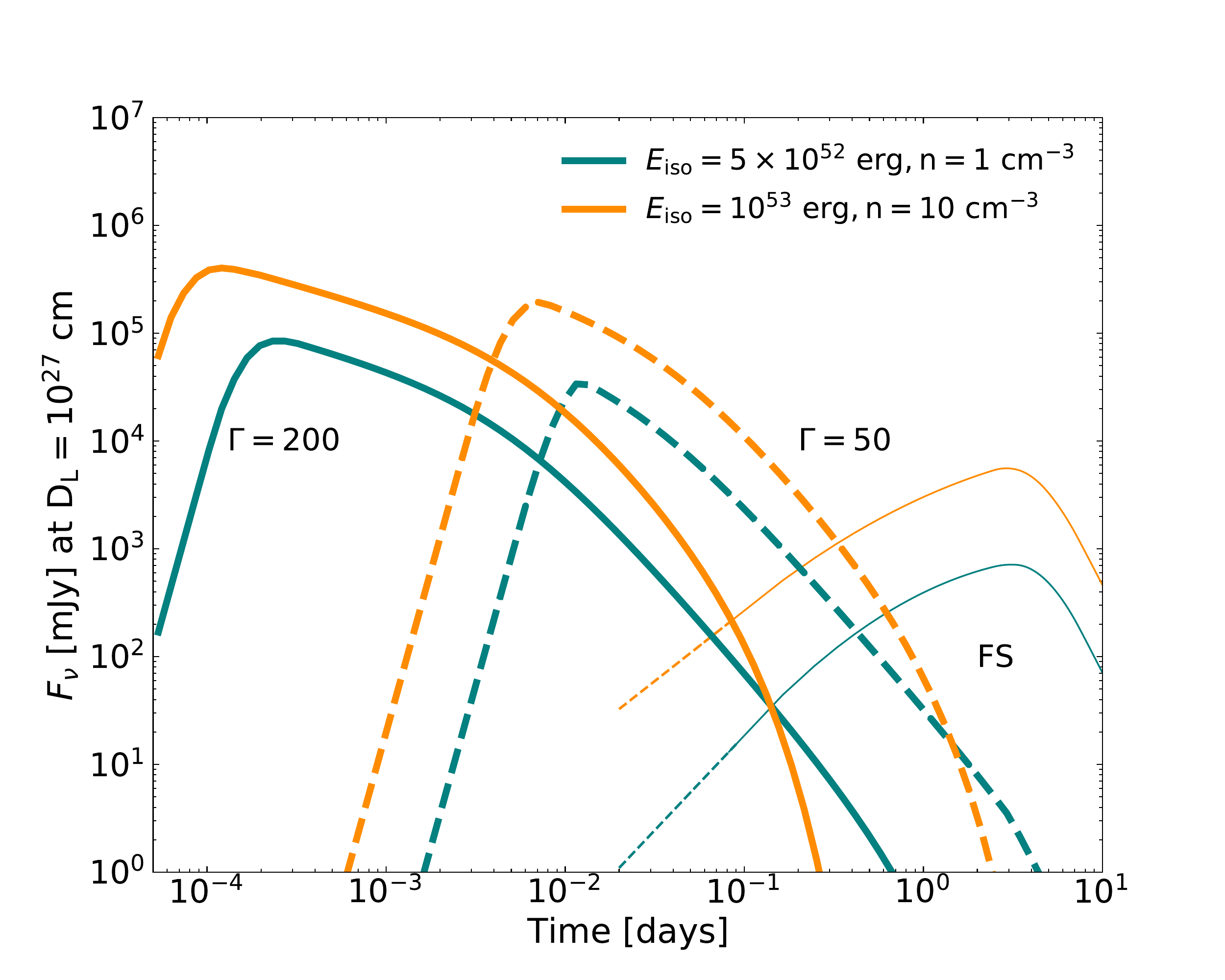}
\caption{Reverse shock models at 100 GHz for the LGRB jet energies and densities considered in this work. Solid and dashed lines correspond to initial Lorentz factors of $\Gamma = 200$ and $50$, respectively. Thin lines correspond to the forward shock component which peaks at later times.}
\label{fig:rs}
\end{figure}

\subsubsection{Neutron Star Mergers with a Stable Magnetar Remnant}
\label{sec:nsmmagnetar}

The merger of two neutron stars may lead to the formation of a long-lived, rapidly spinning magnetar (e.g., \citealt{Metzger2008}). For a strong dipole magnetic field $B\gtrsim 10^{13}$ G and millisecond spin period, the neutron star will transfer its rotational energy, $E_{\rm rot}\sim 10^{52}-10^{53}$ erg, to the merger ejecta via magnetic spin-down, accelerating it to mildly relativistic velocities and leading to bright radio and millimeter emission (e.g., \citealt{Metzger2014}).

We model the millimeter-band emission using the spherical blastwave model of \citet{Nakar2011}, with $E_K = 3\times 10^{52}$ erg, $n=0.1 \ \rm cm^{-3}$, $p=2.3$, $\epsilon_e = 0.1$, $\epsilon_B = 0.01$, and $\beta_i=1$, but note that such systems may instead produce less energetic, sub-relativistic outflows which would produce much fainter millimeter emission. Furthermore, losses due to gravitational waves may limit the total available kinetic energy for synchrotron radiation \citep{Corsi2009}.

For the volumetric rate of neutron star mergers leaving behind a stable remnant, we assume that the extragalactic population of binary neutron star mergers tracks the rate of Galactic double neutron star systems, and consider that recent constraints on the neutron star equation of state imply that at most a few percent of mergers leave stable remnants \citep{Margalit2019}, consistent with late-time radio upper limits of SGRBs \citep{Metzger2014,Horesh+16,Schroeder+20}.  We therefore adopt a total volumetric rate of $R_{\rm NSM} = 2.5 \ \rm Gpc^{-3} \ yr^{-1}$, corresponding to $\sim 0.5\%$ the total neutron star merger (NSM) rate, as in \citet{Metzger2015}. The actual rate of NSMs leaving stable remnants depends on the highly uncertain binary neutron star mass distribution and neutron star equation of state.  We scale the event rate with redshift using Equation~\ref{eq:sfh}.

\subsubsection{On-axis TDEs}
\label{sec:onaxistdes}

The discovery of the $\gamma$-ray/X-ray transient Sw J1644+57 provided the first observational evidence for relativistic jet formation from a TDE \citep{Bloom2011,Burrows2011,Levan2011,Zauderer2011}. Radio and millimeter observations were initiated within $0.3$ d of the \textit{Swift} trigger, revealing a peak in the millimeter band and a steep spectral slope at lower frequencies \citep{Zauderer2011}. The millimeter emission remained detectable for about a year, while the centimeter-band emission is still detectable a decade later \citep{Berger2012,Zauderer2013,Eftekhari2018,Cendes2021b}. Since the discovery of Sw J1644+57, three additional relativistic TDEs have been discovered \citep{Cenko2012,Pasham2015,Brown2015,Brown2017,Andreoni2022}.

Here we use the observed light curve of Sw J1644+57 at 87 GHz from \citet{Berger2012} as a template. 
We assume an on-axis rate of TDEs based on the rate of \textit{Swift} detected events, with $\mathcal{R_{\rm Sw,TDE}}=0.03^{+0.04}_{-0.02} \ \rm Gpc^{-3} \ yr^{-1}$ for events above $L_{\rm iso}\gtrsim 10^{48} \ \rm erg \ s^{-1}$ \citep{Sun2015} and note that the rate of radio non-detections of thermally-detected TDEs sets an approximate upper limit of $\mathcal{R_{\rm on-axis \  TDE}}=1 \ \rm Gpc^{-3} \ yr^{-1}$ (assuming a beaming fraction of $100$ and that $\lesssim 10\%$ of TDEs produce powerful jets; \citealt{Bower2013}). We scale the TDE rate with redshift using the black hole accretion rate density of \citet{Sijacki2015} for black holes in the mass range $10^6-10^7 \rm M_{\odot}$.

\subsubsection{Off-axis TDEs}
\label{sec:offaxistdes}

The vast majority of TDEs are radio-quiet, with luminosities $\nu L_{\nu} \lesssim 10^{40} \ \rm erg \ s^{-1}$ \citep{Alexander2020}. In such cases, the radio emission may arise from a non-relativistic quasi-spherical outflow or an initially off-axis relativistic jet that will decelerate and expand into the observer's line of sight at late times \citep{Giannios2011}. Indeed, to date, there have been two examples of radio TDEs interpreted as harboring off-axis jets \citep{Lei2016,Mattila2018}. Nevertheless, deep radio limits of TDEs suggest that off-axis jets are not ubiquitous \citep{Bower2013,vanVelzen2013,Generozov2017}.

We model the emission from off-axis TDEs using the formalism of \citet{Nakar2011} for mildly relativistic, quasi-spherical emission, which provides a good approximation for the late-time emission from jetted TDEs viewed off-axis. Following \citet{Metzger2015}, we select model parameters $E_K = 10^{52}$ erg, $p=2.3$, $\epsilon_e = 0.2$, $\epsilon_B = 0.01$, and $n = 0.1 \ \rm cm^{-3}$, based on the late-time emission from Sw J1644+57 \citep{Metzger2012}.

For the local volumetric rate, we adopt $\mathcal{R_{\rm TDE, off-axis}} = 3^{+4}_{-2} \ \rm Gpc^{-3} \ yr^{-1}$, a factor $f_b^{-1} \sim 100$ times the on-axis rate. We scale the event rate according to the SMBH accretion rate density \citep{Sijacki2015}.

\subsubsection{Fast Blue Optical Transients} \label{sec:fbots}

Rapid cadence surveys have led to the discovery of a subset of luminous transients that evolve on short timescales and exhibit strong blue continuum \citep{Drout2014}. Commonly referred to as ``fast blue optical transients'' (FBOTs), these events reach typical luminosities of $-15 > M > -20$ on timescales of $\lesssim 10$ days \citep{Pursiainen2018}. The overall sample of FBOTs is largely heterogeneous, and exhibits wide diversity in the observed spectral features (e.g., \citealt{Arcavi2016,Whitesides2017}). A number of theories have been put forward to reconcile the range of luminosities, observed spectral diversity, and short timescales, including shock breakout in a dense stellar envelope \citep{Drout2014}, CSM interaction \citep{Ofek2010}, white dwarf detonation \citep{Arcavi2016}, and energy deposition by a central engine \citep{Margutti2019}.

 A small but growing sample of FBOTs have been detected at radio and millimeter wavelengths \citep{Coppejans2020,Ho2020,Perley2021}. The discovery of the nearby ($\sim 60$ Mpc) fast-rising transient AT2018cow \citep{Prentice2018,Smartt2018} provided a unique opportunity to study one of these events in detail \citep{Ho2019,Margutti2019,Perley2019}. A follow-up  campaign initiated across the electromagnetic spectrum revealed long-lasting ($\sim 100$ d) millimeter emission with an optically thick spectrum below $\nu \lesssim 100$ GHz, pointing to the presence of a sub-relativistic outflow in a dense medium \citep{Ho2019}. Coupled with highly variable X-ray emission, the properties of the event implicate a central engine, in the form of a compact object or embedded internal shock \citep{Margutti2019}. While the majority of known FBOTs have not been detected in the radio, this may in part be due to the fact that most FBOTs have been discovered archivally in optical surveys and thus have not been observed in the millimeter bands at early times when the emission is expected to peak \citep{Ho2021}.

To model the millimeter emission from FBOTs, we use the empirical light curve of AT2018cow \citep{Ho2019}. We scale the light curve at 215 - 243 GHz to 100 GHz assuming $F_\nu \propto \nu^{-0.7}$ and a center frequency of 230 GHz, designed to match the 100 GHz data at peak ($\delta t \sim 22$ d). The volumetric rate for FBOTs is constrained to $4-7\%$ the core-collapse supernovae rate \citep{Drout2014}, while the rate of AT2018cow-like events is shown to be $0.01 - 0.1\%$ the core-collapse supernovae rate \citep{Ho2021} (see also \citealt{Coppejans2020}). We therefore adopt $\mathcal{R}_{\rm FBOT} = 70 \ \rm Gpc^{-3} \ yr^{-1}$ ($0.1\%$ the core-collapse rate). Finally, we scale the event rate to the cosmic star formation history. 

\section{CMB Surveys}
\label{sec:surveys}

Here we give a brief overview of upcoming and long-term CMB surveys and their technical capabilities as they pertain to transient searches:  point source sensitivity, cadence, survey area, survey duration, and angular resolution. The key survey parameters for various CMB experiments are summarized in Table~\ref{tab:surveys}. While most of these surveys operate at multiple frequencies, we use the parameters corresponding to the 90 or 100 GHz bands for our simulations. At lower frequencies, the sensitivity of CMB surveys typically degrades,  while frequencies above 100 GHz will provide largely comparable sensitivity and higher spatial resolution. Thus while we focus here on the 100 GHz band for uniformity, we note that CMB transient searches will benefit from the use of higher frequencies for cross-checking any detections at 100 GHz. 

\subsection{Atacama Cosmology Telescope}\label{sec:act}
The Atacama Cosmology Telescope (ACT) in Chile first began operations in 2007 \citep{Fowler2007}. Since 2016, the instrument has been surveying roughly $40\%$ of the sky between 98 and 225 GHz with a variable cadence of $2-8$ days. The point source sensitivity at 90 GHz is approximately $17$ mJy (assuming 2 arrays and a 4 minute exposure; Sigurd Naess, private communication),\footnote{This is comparable to the per-visit noise levels reported in \citet{Naess2021}.} with an average beam FWHM of $\theta_{\rm res}\approx 2'$. For the purpose of our simulations, we adopt a 5 day cadence and a 7 year survey duration. 

\subsection{South Pole Telescope}\label{sec:spt}
The South Pole Telescope (SPT) in Antarctica is designed to operate in three frequency bands at 95, 150 and 220 GHz, with an angular resolution of about $1.5'$ and $1'$ at 95 and 220 GHz, respectively \citep{Bender2018}. The third generation SPT-3G survey covers an area of $\sim 1500 \ \rm deg^2$, with an average cadence of 12 hours; for simplicity we assume here a 0.5 day cadence. The integration time for a given point on the sky is 20 minutes, achieving an rms noise of about 6 mJy per visit at 95 GHz \citep{Guns2021}. The total survey duration is 5.5 years.  

\subsection{Simons Observatory}\label{sec:simons}
The Simons Observatory (SO), located on Cerro Toco in Chile, is under construction and expected to achieve first light in 2023 \citep{Ade2019}. The experiment will consist of both small and large aperture telescopes (LAT) to optimally probe the CMB across a wide range of angular scales. The fractional sky coverage for the LAT is  projected to be $40\%$ and designed to maximally overlap with LSST and the Dark Energy Spectroscopic Instrument (DESI) survey. The observatory will operate in 6 frequency bands between 27 and 280 GHz, and an angular resolution of about $2.2'$. As a rough approximation for the per-visit sensitivity, we scale the ACT rms by the inverse square root of the ratio of the number of arrays for ACT and SO (Sigurd Naess, private communication), leading to an rms of about 10 mJy. The total survey duration is 5 years.

\subsection{CMB-S4}\label{sec:cmbs4}
The Stage-4 CMB experiment (CMB-S4) represents the next generation of CMB surveys \citep{Abazajian2019}. With planning underway, construction is slated for completion in 2029. CMB-S4 observatories will be located in both Chile and the South Pole, and will consist of an ultra deep survey, covering $3\%$ of the sky, and a deep and wide, high-resolution legacy survey covering over half the sky with a daily cadence; we use the latter here. The frequency range will be $30 - 280$ GHz, with a projected resolution of $7.4'$ and $0.9'$ at 30 and 280 GHz, respectively. For our simulations, we adopt a resolution of 2.2' at 93 GHz and a daily rms of $6$ mJy, corresponding to an anticipated total noise depth of 1.97 $\mu$K-arcmin in this frequency band. The total survey duration is expected to be 7 years.

\subsection{CMB-HD}\label{sec:cmbhd}
CMB-HD is a proposed CMB experiment submitted to the Astro2020 Decadal Survey \citep{Sehgal2019}. The instrument resolution will improve markedly on previous experiments with $0.4'$ at 90 GHz. The project is forecast to cover over half the sky with a daily cadence and a per-visit rms of $2$ mJy at 90 GHz for a total survey duration of 7.5 years. 

\section{Simulated CMB Surveys}
\label{sec:sims}

Here we describe our methodology for determining the rate of transient detections in CMB surveys. We define a detection criterion and combine the theoretical and empirical light curves from \S\ref{sec:models} with the various survey parameters in \S\ref{sec:surveys} to determine the number of transients detected in each survey, as well as their basic light curve properties.

\subsection{Detection Criteria}
\label{sec:detections}

Our sole criterion for detecting and identifying a source as a transient event is at least one $\ge 10\sigma$ detection in a single visit. The large detection significance is motivated by the large number of independent measurements (beams, or resolution elements) in wide-field CMB surveys, and the likelihood for spurious detections as evidenced by past radio surveys \citep{Bower2007,Ofek2010a,Frail2012}. In Table~\ref{tab:surveys}, we estimate the total number of independent beams searched over the course of each survey, $\rm N_{\rm beam}$, and use this to calculate the expected number of false detections assuming a $5\sigma$ detection threshold, $N_{5\sigma}$; we find that for most CMB surveys $N_{5\sigma}\gtrsim 10^4$. We also calculate the minimum source detection threshold leading to one false detection per year, $\sigma_{\rm 1\,f.d.}$, which we find to be $\approx 6.4-7.1$.  In addition to Gausssian noise, data errors and mis-calibration may further increase the number of false detections. We therefore adopt $10\sigma$ as a conservative threshold to mitigate false detections, especially in the presence of systematic uncertainties.

\begin{figure*}
\center
\hspace*{-2cm}
\includegraphics[width=0.9\textwidth]{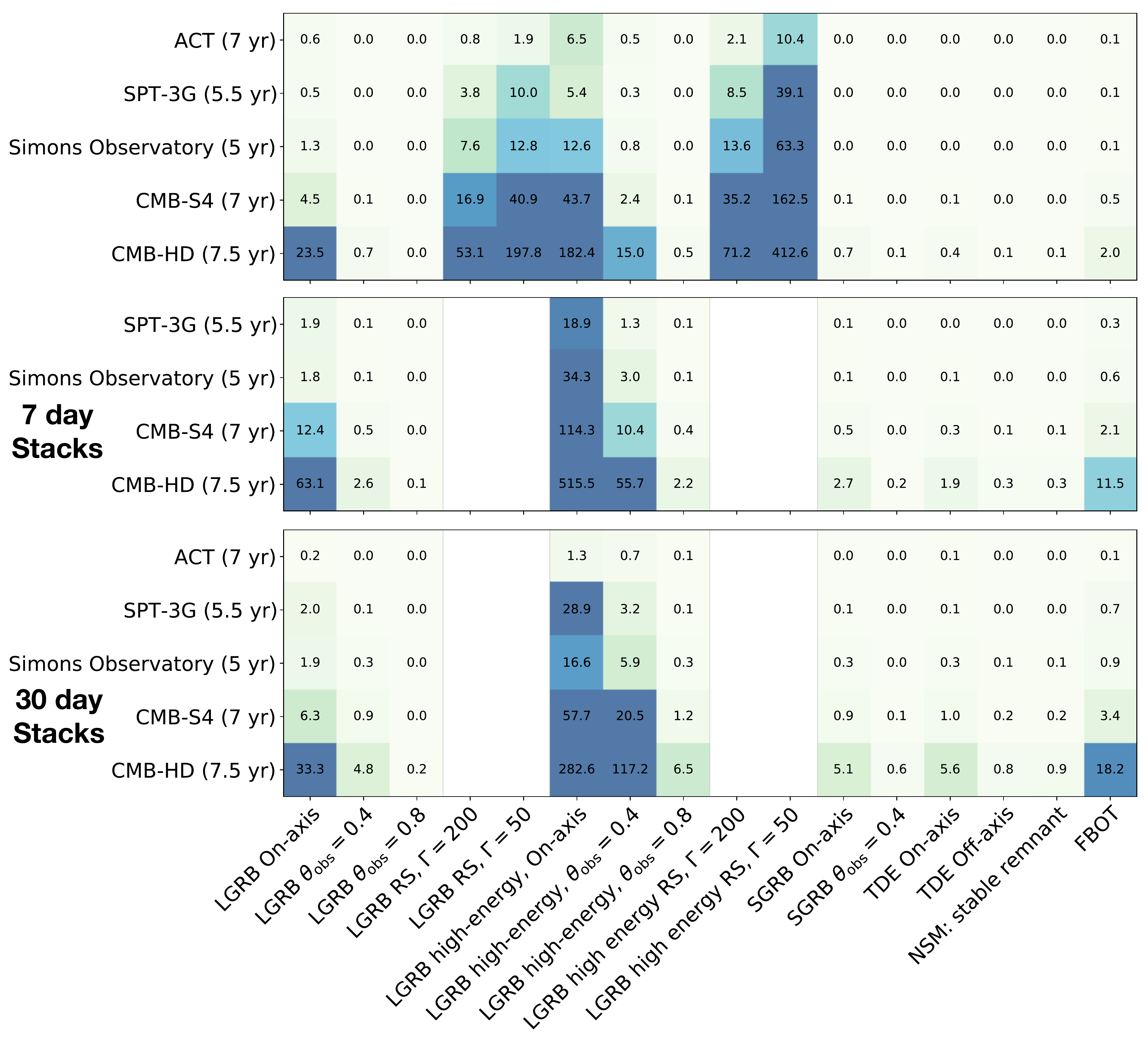}
\caption{Total number of predicted event detections (column $N$ in Tables~\ref{tab:act}-\ref{tab:spt}) for each survey and event class, including results for the stacked surveys. We note that the stacked analyses do not include the reverse shocks as these are generally too fast to be detectable.}
\label{fig:heatmap}
\end{figure*}

Previous work focused on the centimeter band additionally required a factor of two change in brightness for identifying events as being transient \citep{Metzger2015}. Given that the majority of transients evolve more rapidly in the millimeter band (days to weeks) relative to the CMB survey durations, we do not include this condition here. Furthermore, we note that once a source is detected at high significance, it may be recovered in multiple epochs at lower significance, leading to source classifications and inferences of their physical properties. 

Given that the cadence of most CMB surveys is about a day, shorter than the duration of most transients (Figure~\ref{fig:lcs}, \ref{fig:100ghzlcs}), we also consider searches in ``stacked'' images.  Specifically, we explore the impact of averaging the data to 7 day and 30 day cadences, thus increasing the overall depth of the searches by about a factor of 2.5 and 5.5, respectively. This effectively accounts for the fact that a source may be detected at $<10\sigma$ significance in multiple epochs and thus be recovered. Indeed, such an approach has been adopted by the SPT-3G collaboration \citep{Guns2021}. In principle, the limiting timescale for producing stacked surveys is set by the transient duration and the timescale for which there is at least one detection above a given detection threshold. Thus we note that a number of transients that we consider in this work may benefit from generating stacked images on even longer timescales. 

Finally, we consider the effect of contamination due to emission from host galaxies. This effect is less prevalent at 100 GHz given that the non-thermal power-law spectrum of star-forming galaxies, which peaks at a few GHz, follows $F_\nu\propto \nu^{-0.7}$.  At 100 GHz thermal dust emission is also minimal, although it may become problematic at a few hundred GHz \citep{Carilli1999}. For a star formation rate of $1 \ \rm M_{\odot} \ yr^{-1}$, we estimate a typical 100 GHz luminosity of $L_\nu \approx 10^{38} \rm \ erg \ s^{-1}$. This is roughly two orders of magnitude lower than the luminosity of the faintest transients considered here, and thus we conclude that host galaxy contamination is not likely to be an issue for the majority of transients at 100 GHz. We note that while present day CMB surveys like ACT and SPT-3G have demonstrated that stellar flares will dominate the total transient detection rates (when considering both Galactic and extragalactic events), these sources will have obvious stellar associations, and therefore can be easily identified and removed in extragalactic transient detection pipelines \citep{Guns2021,Naess2021}. We discuss the issue of AGN contamination, and how this might be alleviated, in \S\ref{sec:multi}.

\subsection{Monte Carlo Simulations}
\label{sec:mc}

We run a series of Monte Carlo simulations to estimate the detection rates and measured properties for various transients as a function of CMB survey strategy. For each transient and survey, we inject $\gtrsim 10^{4}$ mock light curves using the models described above with a random start time relative to the survey duration and a random distance weighted by the comoving cosmological volume. We scale the volumetric event rate by the cosmic star formation history or SMBH density (see Section~\ref{sec:models}). The light curves are calculated at each redshift, taking into account time-dilation and a ($1+z$) bandpass correction. We ignore the effects of $K$-corrections, which are largely negligible for the moderate range of redshifts accessible to CMB surveys. An absolute detection rate is calculated for each transient class and CMB survey by weighting the mock detections by the volumetric event rate and survey duration.

\section{Results}
\label{sec:results}

\subsection{Detection Rates}

The results of the simulations are summarized in Tables~\ref{tab:act}-\ref{tab:cmbhd} and in Figures~\ref{fig:heatmap} and \ref{fig:redshifts}. For each survey and transient event class, we report the total number of detected events over the course of the survey ($N$), the mean redshift of detected events ($\bar{z}$), and the mean number of $3\sigma$ and $10\sigma$ detections ($\bar{n}_{3\sigma}$ and $\bar{n}_{10\sigma}$, respectively). We additionally quantify the fraction of events with a measured rise ($f_{\rm rise}$) and decline ($f_{\rm fall}$) times, defined as those with at least a factor of $2$ change in brightness. Finally, we quantify the fraction of events for which the peak of the light curve is captured during the survey duration ($f_{\rm peak}$).

\subsection{Detection Rates and Redshift Distributions}\label{sec:detrates}

 We find that the overall detection rates are dominated by LGRBs. In daily cadence searches, LGRBs are detected most readily by the RS emission, while the FS emission dominates the detection rates for the weekly and monthly stacked cadences. Indeed, the high luminosity associated with the RS (a factor of $\sim 10^2$ times brighter than the FS; Figure~\ref{fig:rs}) lends to a large number of detections in next generation experiments like CMB-S4 and CMB-HD, where the detection rates reach values of $\sim 160 - 400$ for the most luminous ``high-energy'' RS events ($\Gamma = 50$). Even for the fiducial LGRB energy RS models, current and upcoming surveys like ACT, SPT-3G, and SO are expected to detect a few events, and as high as $\sim 15 - 60$ events in the case of the ``high-energy'' models. Given that the timescale for RS emission is $\lesssim 1$ day, after which point emission from the FS dominates, we do not explore RS detections in the stacked survey cadences, but we note that a few such RS detections are still possible with the ACT nominal $5$-day cadence.

We caution that our analysis assumes that all on-axis LGRBs produce a detectable RS. Although there are not yet robust estimates constraining the fraction of LGRBs with detectable RS emission, empirical estimates currently indicate a lower limit of $\gtrsim 30\%$ based on targeted observations (e.g., \citealt{Laskar2013,Perley2014,Laskar2016,Laskar2018_140304,Laskar2018,Laskar2019,Laskar2019_grb190114c}). If we scale our RS detection rates by $\sim 1/3$, we instead find that the number of LGRB detections based on FS emission and RS detections are comparable. 

\begin{figure}
\center
\includegraphics[width=\columnwidth]{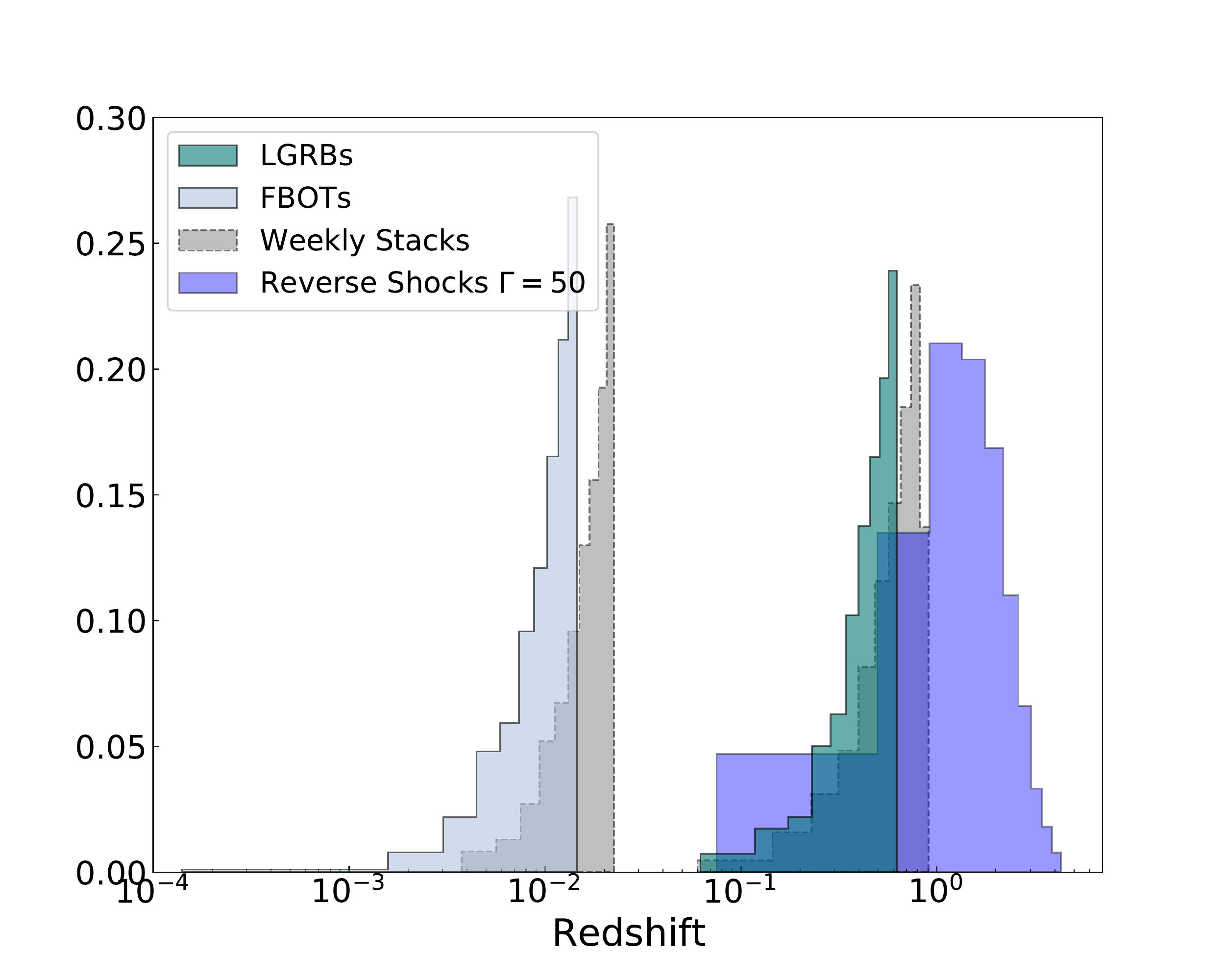}\caption{Normalized redshift distributions for "high-energy" LGRBs (including the RS emission) and FBOTs for CMB-S4. Also shown in grey are the distributions for the stacked weekly cadences which push the total detection volume out to higher redshifts.}
\label{fig:redshifts}
\end{figure}

The average redshifts of detected on-axis LGRBs (based on their FS emission) span $z\sim 0.1 - 0.4$, while surveys like CMB-S4 and CMB-HD will detect the most energetic LGRBs to $z\sim 1$; see Figure~\ref{fig:redshifts}. For events that are detected based on their bright RS emission, the  detection range increases to $z\sim 4$. In Figure~\ref{fig:redshifts}, we plot the normalized redshift distributions for detected on-axis ``high-energy'' LGRBs for both the FS and RS emission for CMB-S4. Although the redshift distribution for RS detections extend to higher redshifts, it eventually drops off following the rate of cosmic star formation at high redshift. We note that in the case of the FS component, searches in stacked data increase the average redshift of detected events across all surveys, and thus the total detection volume (Figure~\ref{fig:redshifts}).

The light curves of detected on-axis LGRBs are well-characterized, with $\bar{n}_{3\sigma}\sim 10$ epochs, given the rapid daily cadence of most CMB surveys (Table~\ref{tab:surveys}). Although ACT may detect a small number of LGRBs, the irregular cadence suggests that such events will not have rise times that are well-characterized. Conversely, the $0.5$ day cadence of SPT-3G implies that the light curves of all detected LGRBs will be well-sampled. While the stacked surveys result in significantly fewer detections per event, in practice, once a source is identified in the stacked images, additional epochs before and after the stacked detection may be recovered by relaxing the $10\sigma$ detection threshold. In the case of RS detections, the short timescales of these events relative to the typical CMB survey cadence implies only a single detection, demonstrating the importance of rapid ($\sim$ hour timescale) alerts to facilitate multi-wavelength follow-up observations. 

For less energetic LGRBs, CMB-S4 and CMB-HD will detect $5-24$ events with average redshifts of $\bar{z}\sim 0.2-0.3$, while the sensitivities of ACT and SPT-3G and the combination of reduced sensitivity and smaller areal coverage for SO relative to CMB-S4, largely preclude such detections. On the other hand, weekly and monthly stacks of SPT-3G and SO data may recover a few such sources during the entire survey duration. 

Of particular interest are the detection rates for off-axis LGRBs, which will not be recovered in $\gamma$-ray searches. We find that CMB-S4 and CMB-HD will detect $\sim 2 - 15$ such events in the daily cadence data, or $\sim 10 - 56$ events using weekly stacks and as high as $\sim 20 - 100$ in the monthly stacks. The lower luminosities of off-axis events lowers their detection ranges to $z\lesssim 0.2$ and $z \lesssim 0.4$ for the daily and stacked surveys, respectively. The longer timescales associated with off-axis events on the other hand lends to a larger number of detection epochs ($\bar{n_{\rm 3\sigma}}\approx 60 - 200$).

We further find that CMB surveys will detect a small number of FBOTs similar to AT2018cow, and that such detections will typically be made in the stacked survey data given the low luminosity of these events relative to the survey sensitivities. We find that surveys like CMB-S4 and CMB-HD will detect a few to $\approx 20$ events in the weekly and monthly stacked cadences. 

We show the distance distributions for FBOTs in the daily and weekly stacked searches in Figure~\ref{fig:redshifts}. Unlike LGRBs, these events will be exclusively detected at nearby distances of $\lesssim 50$ Mpc, comparable to the distance of AT2018cow (60 Mpc). 

While we find that such sources will not be detected in surveys like ACT, SPT-3G, and SO, we emphasize that the rate of FBOTs with millimeter emission is highly uncertain. If millimeter emission is more ubiquitous in the overall population of FBOTs, then the enhanced volumetric event rate will lead to detections across most CMB surveys.

Beyond LGRBs and FBOTs, we do not expect other transients to be detected in CMB surveys in appreciable numbers.  The exception is CMB-HD, which may detect a few TDEs, SGRBs, and NSMs at redshifts $z\sim 0.04 - 0.3$, while CMB-S4 may similarly recover $\sim 1$ TDE and $\sim 1$ SGRB during the survey duration by searching the stacked images (30 day cadence). We note that we consider the number of TDE detections across all surveys a lower limit based on the uncertain event rate; if the rate of on-axis TDEs is closer to the upper limit of $\mathcal{R_{\rm on-axis \  TDE}}=1 \ \rm Gpc^{-3} \ yr^{-1}$, then we expect roughly $30 \times$ the number of detections. We note that the extended timescales ($\gtrsim 100$ days) associated with on-axis TDEs (and both SGRBs and NSMs) in particular suggests that CMB surveys may benefit from stacking the data on longer ($\gtrsim 30$ days) timescales. Indeed, as demonstrated by Tables~\ref{tab:cmbs4} and \ref{tab:cmbhd}, on-axis TDEs that are detected will be detected in $\sim 20$ epochs even in the 30 day stacks. 

Finally, we note that we can quantify the typical post-explosion time of detection by considering both the transient duration ($\delta t_{100}$) and the survey cadence ($\Delta T$). Namely, if the transient duration is comparable to (e.g., on-axis LGRBs) or shorter than (e.g., LGRB RS emission) the survey cadence, then the typical post-explosion time is set by the survey cadence. Conversely, if the transient duration is much longer than the survey cadence, the typical post-explosion time is set by the transient duration. We discuss optimal survey strategies and implications for multi-wavelength follow-up in light of these results.

\subsection{Optimal Survey Strategies}\label{sec:multi}

Based on our simulations and in light of the brightness temperature constraints discussed in \S\ref{sec:transients}, we find that the optimal survey design for millimeter transient detection prioritizes a rapid ($\sim 1$ day timescale) cadence. Such a rapid cadence lends to the detection of LGRBs (which dominate the detection rates due to their intrinsic brightness), as well as less luminous, longer duration events (such as TDEs and off-axis LGRBs) that can be recovered in the stacked data streams. We emphasize that this is in contrast to the centimeter-band where characteristic timescales of $\gtrsim 100$ days \citep{Metzger2015} generally preclude the detection of extragalactic transients at early times, further highlighting the utility of rapid-timescale millimeter surveys to detect these sources early in their evolution.

Furthermore, we note that observations on an even faster cadence (e.g., 2$\times$ per night) would allow for multiple detections of the RS emission from LGRBs. The typical short duration ($\lesssim 1$ day) associated with the RS implies that follow-up millimeter observations (e.g., with ALMA) on the relevant timescales will be challenging. Thus, rapid cadence CMB surveys provide the potential to characterize the early time millimeter light curves from LGRBs, while also reaching the requisite sensitivity to detect less luminous, longer-lived transients using stacked images.

\subsection{Multi-wavelength Follow-up and Detections}\label{sec:multi}

Beyond the question of detectability we address here the question of localization and source classification. Obtaining host galaxy identifications and hence a distance will be key for constraining the energy scale of detected millimeter transients and elucidating their origin (Figure~\ref{fig:lcs}). The next generation of CMB experiments will drastically improve in angular resolution, leading to more precise localizations that will enable host identifications. Given our $10\sigma$ threshold for source detection and a fiducial beam size of $\sim 120''$ in CMB surveys, typical localizations will be $\sim 10''$ ($\sim \theta_{\rm res}/$SNR). Within such a region, there are nominally a few galaxies brighter than $m_r \approx 25$ mag \citep{Driver2016}. Absent any additional observations, identifying host galaxies purely via chance coincidence arguments may therefore be feasible, allowing for a measurement of the redshift and thus source luminosity \citep{Eftekhari2017}. In general, given a detection threshold in CMB surveys, sources will be readily detected with ALMA follow-up using short integration times, leading to sub-arcsecond localizations and definitive host galaxy associations.  

We further consider the latency requirements for multi-wavelength follow-up to facilitate source classifications, taking into account the typical post-explosion time of detection for each source class (\S\ref{sec:detrates}). With early-time millimeter detections of LGRBs ($\sim$ day) and FBOTs ($\sim$ weeks), these sources will be ideally suited for multi-wavelength follow-up and transient classification. Given the small distances of any detected FBOTs and their typical post-explosion time of detection ($\sim$ weeks), these events are likely to be detected (and potentially classified) first in optical surveys. Nevertheless, CMB survey detections will lead to an unbiased sample of millimeter FBOTs for the first time, as we discuss below, and it will also draw attention to specific events based on the presence of quasi-relativistic outflows given that most optical transients are likely to remain unclassified.

While both on- and off-axis LGRBs are expected to pass the detection threshold for transient alerts in wide-field optical surveys like LSST, given an LSST survey cadence of $\sim 3-5$ days, many of these events will be detected initially in CMB surveys. In general, the rapid daily cadence of CMB surveys will constrain the explosion time to within $\lesssim 1$ day, which will be sufficient for triggering both optical, cm, and mm-band observations to confirm an afterglow origin. While rapid response ALMA observations to detect the LGRB RS emission may be challenging, the sensitivity of ALMA will enable detections of the afterglow forward shock associated with any RS detections in CMB surveys. We note that the short timescales associated with the LGRB RS underlines the need for rapid alerts on $\sim$hour timescales. Indeed, among the sources we consider here, LGRB reverse shocks impose the strictest latency requirement for follow-up, requiring immediate ALMA observations on $\sim$ day  timescales \citep{Laskar2013,Laskar2016,Laskar2019,Laskar2019_grb190114c}. Finally, accounting for the sky coverage of \textit{Swift}, we expect  $\sim 15\%$ of on-axis LGRBs detected with CMB surveys to be detected initially with gamma-ray satellites. Millimeter detections of on-axis GRBs within the \textit{Swift} field of view that are unaccompanied by high energy gamma-ray emission will constrain the presence of ``dirty fireballs'' for the first time. 

For SGRBs, the small number of events that may be detected with CMB-HD will be well within the detection volume of next generation GW experiments like Cosmic Explorer and Einstein Telescope and are therefore likely to have associated GW emission \citep{Hall2020} and thus can be readily classified. In the absence of dedicated target-of-opportunity (TOO) programs at optical wavelengths, CMB surveys may provide a useful tool for identifying and localizing EM counterparts to future GW events. Even with dedicated optical follow-up programs, the reduced number of contaminants in millimeter surveys may provide an edge over optical searches. In \citet{Chen2021}, the authors estimate that an LSST TOO program using 30 s integration times will lead to $\sim 40\%$ of GW sources with detected EM counterparts within $700$ Mpc (roughly the detection volume for CMB-HD). Thus, a small fraction of GW events that go undetected in optical searches may be uniquely localized with CMB experiments.

Although on-axis TDEs will be detected with typical post-explosion times of $\sim$ months, the rarity of on-axis events suggests that they may be unambiguously associated with gamma-ray triggers in archival searches. High-cadence wide-field optical surveys may also identify the jet synchrotron afterglow as a rapidly fading, red transient \citep{Andreoni2022}. Absent such gamma-ray or optical counterparts, such sources will be difficult to distinguish from AGN with limited multi-wavelength data, as we discuss below. This further emphasizes the utility of community alerts to enable a wide range of multi-wavelength follow-up, noting that in the case of TDEs, the prolonged timescales relaxes the latency requirement for alerts. 

Finally, we highlight a few key multi-wavelength diagnostics that will be particularly useful for distinguishing between transients and variable AGN. Initial results from SPT-3G suggest that AGN flares may dominate over extragalactic transients \citep{Guns2021}. To first order, any sources detected with CMB surveys should be cross-matched against all-sky radio, millimeter, and submillimeter catalogs to constrain the presence of pre-existing AGN. Similarly, cross-matching against existing WISE data in CMB transient detection pipelines can further help to distinguish between transients and AGN using color-based selections. Following these initial checks, a number of follow-up observations can be leveraged to definitively constrain or rule out an AGN origin. Given the low redshifts of detected events, high resolution ALMA observations can provide source positions within the host galaxy and determine whether they are offset from the galaxy nucleus. This will be feasible for most events below $z\lesssim 0.4$ but will become challenging at higher redshifts and for smaller galaxy sizes. Similarly, ground-based optical spectroscopy can be used to probe high-ionization emission lines from AGN.

\subsection{Science Questions Addressed by Discoveries}

The detection of transient sources in CMB surveys will facilitate a wide range of scientific discovery, providing answers to key questions governing the nature and phenomenology of transient sources. Here we highlight a number of scientific questions that can be uniquely addressed with millimeter detections in CMB surveys. 

\subsubsection{How common is reverse shock emission in LGRBs?}

RS detections in LGRBs by CMB surveys will lead to an unbiased probe of such emission for the first time, providing a constraint on the ubiquity of RS emission among the population of LGRBs.  Targeted observations to search for RS emission from known LGRBs are limited in nature, relying on initial detections by gamma-ray satellites, which cover a small fraction of the sky, and are subject to limited availability and response times with millimeter telescopes such as ALMA. Thus, while a growing sample of LGRBs with detectable RS emission are now being detected in targeted millimeter surveys \citep{Laskar2013,Laskar2019_grb190114c,Laskar2019}, the simulations presented here demonstrate the utility of CMB surveys to drastically improve the existing landscape of RS detections. 

Obtaining a large sample of RS detections will facilitate detailed studies of the mechanisms that drive jet launching and collimation \citep{Zhang2005}. While the afterglow FS emission from LGRBs is insensitive to the properties of GRB jets, RS detections provide critical insight into the jet magnetization and initial Lorentz factor. 

Finally, we note that in addition to blind transient searches afforded by CMB surveys, targeted searches for millimeter emission from GRBs that are initially detected with gamma-ray satellites will provide an alternate means of enabling high impact science. By searching the CMB data streams for a millimeter source each time there is a GRB alert, the coincidence criterion allows for a lower detection threshold. This will not only help to facilitate the identification of strong candidates for continued multi-wavelength follow-up, but also provide automatic constraints for the jet energies and ambient densities of gamma-ray detected GRBs.

\subsubsection{What fraction of TDEs produce relativistic jets?}

Radio follow-up observations have been conducted for several tens of TDEs, revealing a diversity in terms of luminosity and outflow properties (e.g., \citealt{Alexander2020}). These observations indicate that TDEs span over six orders of magnitude in radio luminosity, and generally bifurcate into radio-loud (relativistic) and radio-quiet (non-relativistic) events. 

Key open questions include the fraction of TDEs that produce relativistic jets, as well as the fraction that drive non-relativistic outflows. While on-axis events may produce bright gamma-ray emission at early times (e.g., Sw J1644+57), in the case of off-axis TDEs gamma-ray emission will not be detectable. In contrast, the millimeter emission is expected to persist on timescales of months, highlighting the utility of wide-field millimeter surveys for TDE detections. In particular, CMB surveys will facilitate the first non-optical or X-ray selected sample of TDEs. Searches for radio and millimeter emission from TDEs routinely rely on initial detections at optical and/or gamma-ray surveys. However, the discovery of a potential TDE in the Caltech-NRAO Stripe 82 Survey (CNSS) suggests that wide-field radio surveys can uniquely discover TDEs that are otherwise missed at other wavelengths \citep{Anderson2020}.

\subsubsection{The Radio Diversity of FBOTs}
\label{sec:fbotsdisc}
Targeted radio and millimeter observations of the most luminous FBOTs indicate that at least a subset of these sources produce associated radio and millimeter emission \citep{Ho2019,Coppejans2020,Ho2020}. Indeed, the present day sample of four radio-detected FBOTs trace out a region of parameter space corresponding to the shortest duration and highest luminosity events \citep{Ho2021}. To date, however, the vast majority of FBOTs have been detected archivally in optical survey data, and thus radio observations on the relevant timescales and with adequate sensitivity have been limited \citep{Ho2021}. This in turn has hindered progress in our understanding of what drives the observed radio diversity in the FBOT population. 

The wide areal coverage of CMB surveys will provide an unprecedented opportunity to characterize the millimeter emission from FBOTs in an unbiased sample for the first time. While our work here demonstrates that a few such detections are expected in the stacked data streams from CMB surveys, in principle, these surveys may detect additional events if radio and millimeter emission is prevalent in less optically luminous FBOTs which dominate the observed event rates \citep{Drout2014}. Even if radio/millimeter loud FBOTs are strictly correlated with the most luminous events, the daily cadence, improved sensitivity, and large fractional sky coverage of CMB surveys will ensure coverage for nearly all detected events, allowing us to determine whether all FBOTs generate trans-relativistic outflows.

\section{Conclusions}
\label{sec:conclusions}

We characterized the expected extragalactic transient detection rates in wide-field current and future CMB surveys for a diverse set of transient classes spanning a wide range of timescales and luminosities. Our key results can be summarized as follows: 

\begin{itemize}

    \item The basic physical arguments for synchrotron emission imply that relativistic millimeter transients will peak on short characteristic timescales of $\lesssim {\rm few}$ days. This is in contrast to the centimeter band, where events will reach peak brightness on much longer timescales of $\sim 100$ d. Events with initial off-axis orientations will peak at later times and evolve more slowly, while high densities will lead to more luminous and short-lived events. Conversely, non-relativistic events exhibit much longer rise times of hundreds of days. 

    \item CMB experiments will detect a large number of on-axis LGRBs, ranging from $\sim$ a few to tens of events in surveys like SPT-3G, ACT, and Simons Observatory, to $\sim 20 - 400$ in upcoming next generation experiments like CMB-S4 and CMB-HD. The detection rates on daily cadences will be driven primarily by the reverse shock emission, whereas the forward shock component dominates the detection rates for the time-averaged stacked data streams. We note however that our RS detection rates assume that all on-axis LGRBs produce detectable RS emission. Events detected via their forward shock emission will be limited to redshifts $z \lesssim 1$, but with well-sampled lightcurves given the daily cadence afforded by most surveys. Conversely, events detected on the basis of their reverse shock emission will be detected out to $z \sim 4$. CMB-S4 and CMB-HD will further detect a large number of off-axis LGRBs. The stacked survey data will lead to $N\sim 8 - 44$ events at $z \lesssim 0.3$.
    
    \item In addition to LGRBs, CMB surveys will detect a small number of FBOTs, assuming a fiducial FBOT rate of $\sim 0.1\%$ the core-collapse supernova rate. Indeed, CMB-S4 and CMB-HD will detect between $\sim 1-2$ FBOTs in the daily cadence data, while these numbers increase to a few to $\sim 20$ in the time-averaged stacked data streams. We consider these numbers lower limits given the highly uncertain rate of FBOTs with associated millimeter counterparts and that less luminous FBOTs may also produce millimeter emission. 
    
    \item The redshift distributions of detected LGRBs and FBOTs correspond to unique detection volumes. While LGRBs will be detected out to $z \sim 1$ with CMB-HD (and $z \sim 4$ in the case of RS detections), FBOTs will be detected exclusively in the local universe ($\lesssim 50$ Mpc). Thus FBOTs are likely to be detected initially in wide-field optical surveys, whereas for LGRBs, while the limited $3-5$ day cadence of wide-field optical surveys like LSST implies that they may be detected in larger numbers in CMB surveys.
        
    \item On-axis TDEs may be detected in sizeable numbers in CMB surveys depending on the highly uncertain event rate. We find at most a few TDE detections with surveys like CMB-S4 and CMB-HD, but note that these numbers may be $\approx 30$ times larger depending on whether the event rate more closely tracks the lower ($0.03 \ \rm Gpc^{-3} \ yr^{-1}$) or upper ($1 \ \rm Gpc^{-3} \ yr^{-1}$) limit based on existing observations. Detections in CMB surveys will allow for constraints on an unbiased rate of jetted TDEs.
    
    \item CMB-HD may detect a few SGRBs out to $\sim 500$ Mpc, well within the detection volume of next generation gravitational wave experiments. Such sources are therefore likely to have gravitational wave counterparts. While this is also within the detection volume of proposed targeted programs to search for kilonovae with the VRO, the efficiency for optical detections is $\sim 0.4$. Coupled with the fact that there are fewer contaminants in millimeter surveys, CMB experiments may uniquely localize a handful of GW events.
    
    \item Typical transient localizations will be of order $\sim 10''$, which may be sufficient for identifying a host galaxy and hence redshift. However, multi-wavelength follow-up will play a key role in classifying events and distinguishing variable AGN from true transients. ALMA observations will be ideally suited for resolving source locations, given the nearby distances, while ground-based spectroscopy, coupled with information from existing radio, millimeter, and submillimeter catalogs, will help to identify AGN. 
    
    \item Inherently unique to millimeter surveys is the potential to answer key questions relating to transient phenomena, including how common is reverse shock emission in LGRBs, what fraction of TDEs produce relativistic jets, and what is the radio diversity in an unbiased sample of FBOTs?
    
\end{itemize}

While our results broadly characterize the detection landscape for millimeter transients in CMB surveys, it is worth noting that there are a number of parameters that imply large uncertainties on the absolute detection rates. For example, in our models we make the simplifying assumption that a given transient class can be characterized by a single energy and density. Moreover, the volumetric event rates for certain classes of events, particularly those with few (on-axis TDEs) or no (neutron star mergers) known examples, are ill-constrained, and subject to large uncertainties. Nevertheless, our results indicate the utility of CMB surveys for transient science in the millimeter band.

\acknowledgments 
We thank the referee for their helpful comments. We thank Sam Guns and Sigurd Naess for helpful discussions regarding the point source sensitivity of SPT-3G, ACT, and SO. K.D.A. acknowledges support from NASA through NASA Hubble Fellowship grant \#HST-HF2-51403.001-A awarded by the Space Telescope Science Institute, which is operated by the Association of Universities for Research in Astronomy, Inc., for NASA, under contract NAS5-26555.

\software{Astropy \citep{astropy2018}, Boxfit \citep{vanEerten2012}}

\startlongtable
\begin{deluxetable*}{lcccc}\label{tab:lit}
\tablecaption{Millimeter detections of extragalactic transients from the literature}
\tablecolumns{5}
\tablehead{
\colhead{Source} &
\colhead{Type} &
\colhead{$z$} & 
\colhead{Frequencies [GHz]} &
\colhead{Reference} 
}  
\startdata
AT2018cow & FBOT & 0.013 & 90 - 360 & \citealt{Ho2019}\\
AT2020xnd & FBOT & 0.2433 & 100 - 200 & \citealt{Ho2021xnd}\\
SN2008D & SN Ib & 0.007 & 95 & \citealt{Soderberg2008,Gorosabel2010}\\
PTF11qcj & SN Ic-BL &  0.029& 93 & \citealt{Corsi2014}\\
SN1993J & SN IIb & 0.0008 & 87, 99 & \citealt{Weiler2007} \\
SN1998bw & SN Ic-BL & 0.0085 & 150 & \citealt{Kulkarni1998}\\
SN2011DH & SN IIb & 0.002 & 93, 107 & \citealt{Horesh2013} \\
iPTF13bvn & SN Ib/c& 0.0059& 95&\citealt{Horesh2013_iptf}\\
SN2013ak & SN IIb & 0.0037 & 85, 270 & \citealt{sn2013ak,sn2013ak_2}\\
SN2014C & SN Ib & 0.0027 & 85 & \citealt{Zauderer2014} \\
SN2020oi & SN Ic & 0.0052 & 100 & \citealt{Maeda2021}\\
GRB 970508 & GRB & 0.835 & 86 & \citealt{Bremer1998}\\
GRB 980329 & GRB & $\lesssim 5$ & 350 & \citealt{Smith1999}\\
GRB 991208 & GRB & 0.707 & 100, 250 & \citealt{Galama2000}\\
GRB 000301C & GRB & 2.329& 100, 250 & \citealt{Berger2000}\\
GRB 021004 & GRB & 2.329 & 90, 230 & \citealt{deUgarte2005} \\
GRB 030329 & GRB & 0.169 & 86, 100, 235, 350 & \citealt{Sheth2003,Kuno2004,Kohno2005,Resmi2005}\\
GRB 041219A & GRB & 0.31 & 99 &  \citealt{deUgarte2012}\\
GRB 050904 & GRB & 6.29 & 89 & \citealt{deUgarte2012} \\
GRB 051022 & GRB & 0.809 & 90 & \citealt{Castro2007}\\
GRB 060904B & GRB & 0.70 & 92 & \citealt{deUgarte2012}\\
GRB 070125 & GRB & 1.55 & 95, 250 & \citealt{Chandra2008,deUgarte2012} \\
GRB 070306 & GRB & 1.5 & 86 & \citealt{deUgarte2012} \\
GRB 071003 & GRB & 1.6 & 86 & \citealt{deUgarte2012} \\
GRB 080129 & GRB & 4.35 & 250 & \citealt{Greiner2009}\\
GRB 080319B & GRB & 0.937 & 97 & \citealt{Pandey2009}\\
GRB 090313 & GRB & 3.38 & 87, 92, 105 & \citealt{Bock2009,Mealandri2010}\\
GRB 090404 & GRB & -- & 87, 108 & \citealt{Castro2009a}\\
GRB 090423 & GRB & 8.2 & 97 & \citealt{Castro2009b}\\
GRB 100418A & GRB & 0.62 & 100, 345 & \citealt{Martin2010,deUgarte2018}\\
GRB 100621A & GRB & 0.54 & 345 & \citealt{Greiner2013}\\
GRB 110715A & GRB & 0.825 & 345 & \citealt{Sanchez2017}\\
GRB 120326A & GRB & 1.798 & 92.5, 230 & \citealt{Urata2012,Urata2014,Laskar2015_eneinjection}\\
GRB 130215A & GRB & 0.597 & 93 &  \citealt{Perley2013_grb130215a}\\
GRB 130418A & GRB & 1.218 & 93 & \citealt{Perley2013_grb130418a}\\
GRB 130427A & GRB & 0.34 & 90 & \citealt{Perley2014}\\
GRB 130702A & GRB & 0.145 & 93 & \citealt{Singer2013}\\
GRB 131108A & GRB & 2.4 & 93 & \citealt{Perley2013_grb131108a}\\
GRB 131231A & GRB & 0.642 & 93 & \citealt{Perley2014_grb131231A}\\
GRB 140304A & GRB & 5.283 & 86 & \citealt{Laskar2018_140304}\\
GRB 140311A & GRB & 4.954 & 86 & \citealt{Laskar2018_140311} \\
GRB 160623A & GRB & 0.367 & 230 & \citealt{Chen2020}\\
GRB 161023A & GRB & 2.71 & 100, 345 & \citealt{grb161023a}\\
GRB 161219B & GRB & 0.148 & 91, 103 & \citealt{Laskar2018}\\
GRB 171205A & GRB & 0.037 & 100, 345 & \citealt{Perley2017,Smith2017,Laskar2020} \\
GRB 181201A & GRB & 0.450 & 97 & \citealt{Laskar2019}\\
GRB 190114C & GRB & 0.435 & 97.5 & \citealt{Laskar2019_grb190114c} \\
GRB 190829A & GRB & 0.0785 & 100, 343.5 & \citealt{Dichiara2022}\\
GRB 191221B & GRB & 1.148 & 90.5 & \citealt{Laskar2019_grb191221b}\\
GRB 210610B & GRB & 1.135 & 90.5 & \citealt{Laskar2021}\\
GRB 210619B & GRB & 1.937 & 97.5 & \citealt{grb210619b}\\
GRB 210702A & GRB & 1.1757 & 97.5 & \citealt{grb210702a_a,grb210702a_b}\\
GRB 210905A & GRB & 6.318 & 97.5 & \citealt{grb210905a}\\
GRB 220101A & GRB & 4.618 & 97.5 & \citealt{grb220101a}\\
AT2019dsg & TDE & 0.051 & 97.5 & \citealt{Cendes2021a}\\
Sw J1644+57 & TDE & 0.354 & 87, 200, 230, 345 & \citealt{Berger2012}\\
IGR J1258+0134 & TDE & 0.004 & 100 & \citealt{Yuan2016} \\
AT2022cmc & TDE & 1.193 & 90, 230, 350 & \citealt{Alexander2022,Perley2022,Smith2022}\\
\enddata
\end{deluxetable*}

\begin{deluxetable*}{lccccccc}
\small
\tablecolumns{7}
\caption{Extragalactic Transient Models}
\tablehead{
\colhead{Transient} &
\colhead{$\mathcal{R} \ (z=0)$ } &
\colhead{$E_K$} & 
\colhead{$n$} &
\colhead{$\beta_i$} & 
\colhead{$\delta t_{100}^{\rm c}$} & 
\colhead{Model Type} & 
\colhead{Reference}\\ 
\colhead{} &
\colhead{(Gpc$^{-3}$ yr$^{-1}$)} &
\colhead{(erg)} &
\colhead{(cm$^{-3}$)} & 
\colhead{} &
\colhead{(days)} &
\colhead{} & 
\colhead{} 
}  
\startdata
LGRB, On-Axis & $0.2^{+0.02}_{-0.03}$$^{\rm a}$ & $10^{51}$ & 1 & 1 & 5 & Theoretical & \citealt{vanEerten2012}\\
LGRB, $\theta_{\rm obs} = 0.4$ & $0.8^{+0.1}_{-0.1}$$^{\rm a}$ & $10^{51}$ & 1 & 1 & 23 &  Theoretical & \citealt{vanEerten2012}\\
LGRB, $\theta_{\rm obs} = 0.8$ & $3.0^{+0.5}_{-0.3}$$^{\rm a}$ & $10^{51}$ & 1 & 1 & 96 &  Theoretical &\citealt{vanEerten2012}\\
LGRB high energy, On-axis & $0.1^{+0.01}_{-0.01}$$^{\rm a}$ & $3 \times 10^{51}$ & 10 & 1 & 4 &  Theoretical &\citealt{vanEerten2012} \\
LGRB high energy, $\theta_{\rm obs} = 0.4$ & $0.4^{+0.04}_{-0.04}$$^{\rm a}$ & $3 \times 10^{51}$ & 10 & 1 & 21  & Theoretical & \citealt{vanEerten2012}\\
LGRB high energy, $\theta_{\rm obs} = 0.8$ & $1.5^{+0.2}_{-0.2}$$^{\rm a}$ & $3 \times 10^{51}$ & 10 & 1 & 72 & Theoretical & \citealt{vanEerten2012}\\
SGRB, On-Axis & 1.3$^{+0.4}_{-0.3}$$^{\rm a}$ & $10^{50}$ & 0.01 & 1 & 16 & Theoretical & \citealt{vanEerten2011}\\
SGRB, $\theta_{\rm obs} = 0.4$ & $5.0^{+1.7}_{-1.0}$$^{\rm a}$ & $10^{50}$ & 0.01 & 1 & 78 &  Theoretical & \citealt{vanEerten2011}\\
TDE, On-Axis & 0.03$^{+0.04}_{-0.02}$$^{\rm b}$ & $10^{52}$ & 0.1 & 1 & 135 & Empirical &\citealt{Berger2012}\\
TDE, Off-Axis (spherical) & $3^{+4}_{-2}$$^{\rm b}$ & $10^{52}$ & 0.1 &  0.8 & 900 & Theoretical & \citealt{Giannios2011} \\
NSM: stable remnant & 2.5$^{\rm a}$ & $3 \times 10^{52}$ & 0.1 & 1 & 1300 & Theoretical & \citealt{Metzger2014}\\
FBOTs & 70& $4\times 10^{48}$ & $3\times 10^{5}$&0.13  & 28 & Empirical &\citealt{Ho2019}\\
\enddata
\tablecomments{\\
$^{\rm a}$ Scaled with redshift according to the cosmic star formation history \citep{Madau2014}.\\
$^{\rm b}$ Scaled with redshift according to the SMBH accretion rate density \citep{Sijacki2015}.\\
$^{\rm c}$ Light curve duration at observer frequency $\nu = 100$ GHz.}
\label{tab:rates}
\end{deluxetable*}

\begin{deluxetable*}{lcccccccccc}
\small
\tablecolumns{11}
\caption{CMB Survey Parameters}
\tablehead{
\colhead{Name} &
\colhead{$\theta_{\rm res}$ } &
\colhead{$\nu$} &
\colhead{$\sigma^{\rm a}$} &
\colhead{Area} & 
\colhead{$T_{\rm dur}^{\rm b}$} & 
\colhead{$\Delta T^{\rm c}$} &
\colhead{$\rm N_{obs}^{\rm d}$} & 
\colhead{$\rm N_{beam}^{\rm e}$} & 
\colhead{$\sigma_{\rm 1 f.d.}^{\rm f}$} & 
\colhead{$\rm N_{5\sigma}^{\rm g}$}\\
\colhead{} &
\colhead{(arcsec)} &
\colhead{(GHz)} &
\colhead{(mJy)} & 
\colhead{(deg$^2$)} &
\colhead{(yr)} &
\colhead{(days)} &
\colhead{} & 
\colhead{} & 
\colhead{($\sigma$)} 
}  
\startdata
ACT & 123 &  98, 150, 225 & 17 & $1.7 \times 10^{4}$ & 7 & 5 & 511  & $6.6 \times 10^9$ & 6.4 & 3785 \\
SPT-3G & 92.4 & 95, 150, 220 & 6 & 1500 & 5.5 &  0.5 & 4020 & $8.1 \times 10^{9}$& 6.4 & 4645\\ 
Simons Observatory & 132 & 27, 39, 93, 145, 225, 280 & 10 & $1.7\times 10^4$& 5 & 1 & 1825 & $2 \times 10^{10}$ & 6.6 & 11500\\ 
CMB-S4 & 132 & 27, 39, 93, 145, 225, 278 & 6 & $2\times 10^4$ & 7 & 1 & 2555 & $3.4 \times 10^{10}$ & 6.4 & 19500\\
CMB-HD & 25 & 30, 40, 90, 150, 220, 280 & 2 & $2\times 10^4$ & 7.5 & 1& 2740 & $1.0 \times 10^{12}$ & 7.13 & $5.7\times 10^5$  \\
\enddata
\tablecomments{We list all frequency bands for each experiment for completeness but note that the angular resolution and sensitivity are given at the frequency band nearest 100 GHz.\\
$^{\rm a}$ 1$\sigma$ rms sensitivity.\\
$^{\rm b}$ Survey duration.\\
$^{\rm c}$ Survey cadence.\\
$^{\rm d}$ Number of independent observations.\\
$^{\rm e}$ Number of independent synthesized beams searched over the course of the survey.\\
$^{\rm f}$ Source detection threshold leading to one false detection per year due to thermal fluctuations.\\
$^{\rm g}$ Number of false detections over the course of the survey if the detection threshold is $5\sigma$.
}
\label{tab:surveys}
\end{deluxetable*}

\begin{deluxetable*}{lccccccccccccccc}
\tablecolumns{8}
\caption{ACT}
\tablehead{
\colhead{Transient} &
\colhead{N$^{\rm a}$} &
\colhead{$\bar{z}^{\rm b}$} &
\colhead{$\bar{n}_{3\sigma}^{\rm c}$} & 
\colhead{$\bar{n}_{10\sigma}^{\rm d}$} &  
\colhead{$f_{\rm peak}^{\rm e}$} &
\colhead{$f_{\rm rise}^{\rm f}$} &
\colhead{$f_{\rm fall}^{\rm g}$} &
\colhead{N}&
\colhead{$\bar{z}$} &
\colhead{$\bar{n}_{3\sigma}$} & 
\colhead{$\bar{n}_{10\sigma}$} &  
\colhead{$f_{\rm peak}$} &
\colhead{$f_{\rm rise}$} &
\colhead{$f_{\rm fall}$} 
}  
\startdata
& & \multicolumn{7}{c}{5 Day Cadence} & \multicolumn{7}{c}{30 Day Stack} \\
\multicolumn8c{--------------------------------------------------------------------------------------------------------} &  \multicolumn8c{--------------------------------------------------------} \\
LGRBs (on-axis) & 0.61 & 0.10 & 2.2 & 1.4 & 0.12 & 0.0 & 0.66 & 0.17 & 0.09 & 1.0 & 1.0 & 0.0 & 0.0 & 0.0\\
LGRBs ($\theta_{\rm obs} = 0.4$) & 0.02 & 0.02 & 11.9 & 5.5 & 0.19 & 0.34 & 0.99 & 0.03 & 0.03 & 1.0 & 1.0 & 0.0 & 0.0 & 0.01\\
LGRBs ($\theta_{\rm obs} = 0.8$) & 0.0 & 0.004 & 45.3 & 19.4 & 0.22 & 0.97 & 0.91 & 0.0 & 0.01 & 2.1 & 1.4 & 0.0 & 0.0 & 0.27\\
LGRBs RS, $\Gamma = 200$ &0.8& 0.75& 1.0 & 1.0& \nodata & \nodata & \nodata & \nodata & \nodata & \nodata & \nodata & \nodata & \nodata & \nodata\\
LGRBs RS, $\Gamma = 50$ &1.9 &0.58 &1.0  &1.0& \nodata & \nodata & \nodata & \nodata & \nodata & \nodata & \nodata & \nodata & \nodata & \nodata\\
LGRBs high energy, (on-axis) &6.5 & 0.3 & 2.0 & 1.3 & 0.12 & 0.0 & 0.62 & 1.3 & 0.2 & 1.0 & 1.0 & 0.01 & 0.0 & 0.0\\
LGRBs high energy, $\theta_{\rm obs} = 0.4$ & 0.46 & 0.07 & 9.7 & 5.1 & 0.21 & 0.33 & 0.99 & 0.65& 0.08 & 1.0 & 1.0 & 0.0 & 0.0 & 0.0\\
LGRBs high energy, $\theta_{\rm obs} = 0.8$ & 0.02 & 0.02 & 33.6 & 15.6 & 0.26 & 0.99 & 0.98 & 0.10 & 0.03 & 1.6 & 1.1 & 0.0 & 0.0 & 0.61\\
LGRBs high energy RS, $\Gamma = 200$ & 2.1 & 1.2&1.0  & 1.0& \nodata & \nodata & \nodata & \nodata & \nodata & \nodata & \nodata & \nodata & \nodata & \nodata\\
LGRBs high energy RS, $\Gamma = 50$ & 10.4& 1.1& 1.0 &1.0 & \nodata & \nodata & \nodata & \nodata & \nodata & \nodata & \nodata & \nodata & \nodata & \nodata\\
SGRB, On-axis & 0.02 & 0.02 & 7.6 & 4.0 & 0.21 & 0.40 & 0.99 & 0.02 & 0.02 & 1.0 & 1.0 & 0.0 & 0.0 & 0.0 \\
SGRB ($\theta_{\rm obs} = 0.4$) & 0.0 & 0.004  & 33.8 & 18.3 & 0.23 & 0.98 & 0.95 & 0.01 & 0.01 & 1.4 & 1.1 & 0.0 & 0.0 & 0.36\\
TDE, On-axis & 0.01 & 0.05 & 126.1 & 43.6& 0.19 & 0.18 & 0.94 & 0.14 & 0.10 & 5.0 & 1.9 & 0.03 & 0.0 & 0.91\\
TDE, Off-Axis (spherical) &  0.0 & 0.005 & 437.6 & 178.4 & 0.11 & 0.78 & 0.74 & 0.03 & 0.01 & 15.2 & 8.3 & 0.59 & 0.81 & 0.78\\
NSM: magnetar & 0.0& 0.006 & 451.6 & 210.2 & 0.10 & 0.79 & 0.55 & 0.03 & 0.01 & 16.0 & 7.7 & 0.71 & 0.76 & 0.59\\
FBOTs & 0.1 & 0.01 & 11.3 & 6.0 & 0.19 & 0.24 & 0.99 & 0.14 & 0.01 & 1.0 & 1.0 & 0.0 & 0.0 & 0.0\\
\enddata
\tablecomments{\\
$^{\rm a}$ Number of detected events.\\
$^{\rm b}$ Average redshift of detected events.\\
$^{\rm c}$ Average number of $3\sigma$ detections per event.\\
$^{\rm d}$ Average number of $10\sigma$ detections per event.\\
$^{\rm e}$ Fraction of events with a measured peak.\\
$^{\rm f}$ Fraction of events with measured rise times.\\
$^{\rm g}$ Fraction of events with measured fall times.}
\label{tab:act}
\end{deluxetable*}

\begin{deluxetable*}{lccccccccccccccc}
\tablecolumns{8}
\caption{SPT-3G}
\tablehead{
\colhead{Transient} &
\colhead{N} &
\colhead{$\bar{z}$} &
\colhead{$\bar{n}_{3\sigma}$} & 
\colhead{$\bar{n}_{10\sigma}$} &  
\colhead{$f_{\rm peak}$} &
\colhead{$f_{\rm rise}$} &
\colhead{$f_{\rm fall}$} &
\colhead{N}&
\colhead{$\bar{z}$} &
\colhead{$\bar{n}_{3\sigma}$} & 
\colhead{$\bar{n}_{10\sigma}$} &  
\colhead{$f_{\rm peak}$} &
\colhead{$f_{\rm rise}$} &
\colhead{$f_{\rm fall}$} 
}  
\startdata
& & \multicolumn{7}{c}{0.5 Day Cadence} & \multicolumn{7}{c}{7 Day Stack} \\
\multicolumn8c{--------------------------------------------------------------------------------------------------------} &  \multicolumn8c{-------------------------------------------------------------} \\
LGRBs (on-axis) & 0.53 & 0.17 & 18.9 & 9.5 & 0.99 & 0.96 & 0.99 & 1.9 & 0.26 & 3.3 & 1.7 & 0.28 & 0.0 & 0.99\\
LGRBs ($\theta_{\rm obs} = 0.4$) &  0.01 & 0.03 & 93.5 & 42.6 & 0.99 & 0.99 & 0.99 & 0.06 & 0.05 & 13.7 & 6.0 & 0.95 & 0.99 & 0.99\\
LGRBs ($\theta_{\rm obs} = 0.8$) & 0.0 & 0.01 & 361.2 & 173.2 & 0.97 & 0.98 & 0.97 & 0.0 & 0.01 & 52.4 & 25.4 & 0.98 & 0.98 & 0.98\\
LGRBs RS, $\Gamma = 200$ & 3.8& 1.2& 1.0 &1.0 & \nodata & \nodata & \nodata & \nodata & \nodata & \nodata & \nodata & \nodata & \nodata & \nodata\\
LGRBs RS, $\Gamma = 50$ &10.0 & 0.9 & 1.0 & 1.0& \nodata & \nodata & \nodata & \nodata & \nodata & \nodata & \nodata & \nodata & \nodata & \nodata\\
LGRBs high energy, (on-axis) &5.4& 0.48 & 22.1 & 10.5 & 0.99 & 0.99 & 0.99 & 18.9 & 0.74 & 4.1 & 1.9 & 0.58 & 0.60 & 0.99\\
LGRBs high energy, $\theta_{\rm obs} = 0.4$ & 0.32 & 0.12 & 79.8 & 39.7 & 0.99 & 0.99 & 0.99 & 1.3 & 0.19 & 12.4 & 6.3 & 0.92 & 0.81 & 0.99\\
LGRBs high energy, $\theta_{\rm obs} = 0.8$ & 0.01 & 0.03 & 271.7 & 130.5 & 0.98 & 0.98 & 0.99 & 0.05 & 0.04 & 40.8 & 19.4 & 0.97 & 0.97 & 0.99\\
LGRBs high energy RS, $\Gamma = 200$ & 8.5 & 1.8& 1.0 &1.0 & \nodata & \nodata & \nodata & \nodata & \nodata & \nodata & \nodata & \nodata & \nodata & \nodata\\
LGRBs high energy RS, $\Gamma = 50$ & 39.1 & 1.6& 1.0 &1.0 & \nodata & \nodata & \nodata & \nodata & \nodata & \nodata & \nodata & \nodata & \nodata & \nodata\\
SGRB, On-axis & 0.02 & 0.03 & 59.3 & 30.3 & 0.99 & 0.99 & 0.99 & 0.06 & 0.05 & 8.9 & 4.5 & 0.90 & 0.76 & 0.99\\
SGRB ($\theta_{\rm obs} = 0.4$) & 0.0 & 0.01 & 37.4 & 19.6 & 0.96 & 0.98 & 0.98 & 0.0 & 0.01 & 35.8 & 18.9 & 0.96 & 0.95 & 0.95\\
TDE, On-axis & 0.01 & 0.08 & 1044.9 & 313.1 & 0.95 & 0.95 & 0.98 & 0.04 & 0.13 & 169.6 & 53.5 & 0.56 & 0.21 & 0.95\\
TDE, Off-Axis (spherical) & 0.0 & 0.01 & 4293.7 & 1500.7 & 0.71 & 0.87 & 0.82 & 0.01 & 0.01 & 600.7 & 215.0 & 0.89 & 0.85 & 0.80\\
NSM: magnetar & 0.04 & 0.01 & 5096.5  &2053.1 & 0.59 & 0.82 & 0.76 & 0.01 & 0.02 & 708.1 & 1282.8 & 0.90 & 0.83 & 0.74\\
FBOTs & 0.06 & 0.01 & 85.4 & 41.5 & 0.99 & 0.99 & 0.99 & 0.26 & 0.02 & 12.7 & 6.3 & 0.91 & 0.61 & 0.99\\
\multicolumn8c{--------------------------------------------------------------------------------------------------------} &  \multicolumn8c{-------------------------------------------------------------} \\
& & \multicolumn8c{ } & \multicolumn6c{30 Day Stack} \\
\multicolumn8c{--------------------------------------------------------------------------------------------------------} &  \multicolumn8c{-------------------------------------------------------------} \\
LGRBs (on-axis)& \nodata &  \nodata & \nodata &\nodata & \nodata& \nodata & \nodata & 2.0 & 0.27 & 1.0 & 1.0 & 0.0 & 0.0 & 0.03 \\
LGRBs ($\theta_{\rm obs} = 0.4$) & \nodata &  \nodata & \nodata &\nodata & \nodata& \nodata & \nodata & 0.11 & 0.07 & 3.9 & 2.2 & 0.16 & 0.0 & 0.99\\
LGRBs ($\theta_{\rm obs} = 0.8$) & \nodata &  \nodata & \nodata &\nodata & \nodata& \nodata & \nodata & 0.0 & 0.01 & 13.5 & 6.6 & 0.95 & 0.66 & 0.98\\
LGRBs high energy, (on-axis) & \nodata &  \nodata & \nodata &\nodata & \nodata& \nodata & \nodata & 28.9 & 0.84 & 1.0 & 1.0 & 0.0 & 0.0 & 0.05\\
LGRBs high energy, $\theta_{\rm obs} = 0.4$ & \nodata &  \nodata & \nodata &\nodata & \nodata& \nodata & \nodata & 3.2 & 0.25 & 3.4 & 1.9 & 0.44 & 0.0 & 0.99\\
LGRBs high energy, $\theta_{\rm obs} = 0.8$  & \nodata &  \nodata & \nodata &\nodata & \nodata& \nodata & \nodata & 0.13 & 0.06 & 9.5 & 4.7 & 0.91 & 0.32 & 0.98\\
SGRB, On-axis & \nodata &  \nodata & \nodata &\nodata & \nodata& \nodata & \nodata & 0.12 & 0.06 & 2.4 & 1.6 & 0.15 & 0.0 & 0.29\\
SGRB ($\theta_{\rm obs} = 0.4$) & \nodata &  \nodata & \nodata &\nodata & \nodata& \nodata & \nodata & 0.01 & 0.02 & 8.7 & 4.7 & 0.94 & 0.27 & 0.98\\
TDE, On-axis& \nodata &  \nodata & \nodata &\nodata & \nodata& \nodata & \nodata & 0.11 & 0.19 & 43.6 & 13.9 & 0.72& 0.0 & 0.95 \\
TDE, Off-Axis (spherical) & \nodata &  \nodata & \nodata &\nodata & \nodata& \nodata & \nodata & 0.02 & 0.02 & 146.0 & 53.6 & 0.89 & 0.84 & 0.82\\
NSM: magnetar  & \nodata &  \nodata & \nodata &\nodata & \nodata& \nodata & \nodata & 0.02 & 0.02 & 167.5 & 67.9& 0.91 & 0.84 & 0.72 \\
FBOTs & \nodata &  \nodata & \nodata &\nodata & \nodata& \nodata & \nodata & 0.66 & 0.03 & 3.4 & 1.6 & 0.63 & 0.99 & 0.99\\
\enddata
\tablecomments{Columns are as in Table~\ref{tab:act}.}
\label{tab:spt}
\end{deluxetable*}

\begin{deluxetable*}{lccccccccccccccc}
\tablecolumns{8}
\caption{Simons Observatory}
\tablehead{
\colhead{Transient} &
\colhead{N} &
\colhead{$\bar{z}$} &
\colhead{$\bar{n}_{3\sigma}$} & 
\colhead{$\bar{n}_{10\sigma}$} &  
\colhead{$f_{\rm peak}$} &
\colhead{$f_{\rm rise}$} &
\colhead{$f_{\rm fall}$} &
\colhead{N}&
\colhead{$\bar{z}$} &
\colhead{$\bar{n}_{3\sigma}$} & 
\colhead{$\bar{n}_{10\sigma}$} &  
\colhead{$f_{\rm peak}$} &
\colhead{$f_{\rm rise}$} &
\colhead{$f_{\rm fall}$} 
}  
\startdata
& & \multicolumn{7}{c}{Daily Cadence} & \multicolumn{7}{c}{7 Day Stack} \\
\multicolumn8c{--------------------------------------------------------------------------------------------------------} &  \multicolumn8c{-------------------------------------------------------------} \\
LGRBs (on-axis) & 1.3 & 0.14 & 10.5 & 5.4 & 0.99 & 0.99 & 0.99 & 1.8 & 0.17 & 2.1 & 1.2 & 0.0 & 0.0 & 1.0\\
LGRBs ($\theta_{\rm obs} = 0.4$) &0.03 & 0.03 & 59.9 & 28.4 & 0.98 & 0.99 & 0.98 & 0.14 & 0.04 &  8.7 & 4.1 &  0.64 & 0.50 & 0.99 \\
LGRBs ($\theta_{\rm obs} = 0.8$) & 0.0& 0.004 & 241.0 & 111.8 & 0.94 & 0.97 & 0.95 &  0.002 & 0.007 & 34.0 & 15.9 & 0.95 & 0.95 & 0.95\\
LGRBs RS, $\Gamma = 200$ & 7.6 & 0.7 & 1.0 &1.0 & \nodata & \nodata & \nodata & \nodata & \nodata & \nodata & \nodata & \nodata & \nodata & \nodata\\
LGRBs RS, $\Gamma = 50$  & 12.8& 0.7& 1.0 & 1.0& \nodata & \nodata & \nodata & \nodata & \nodata & \nodata & \nodata & \nodata & \nodata & \nodata\\
LGRBs high energy, (on-axis) & 12.6 & 0.37 & 9.9 & 5.1 & 0.99 & 0.99 & 0.99 & 34.3 &  0.5 &  1.6 & 1.1 & 0.0 & 0.0 & 0.55\\
LGRBs high energy, $\theta_{\rm obs} = 0.4$ &0.8 & 0.09 & 46.9 & 23.9 & 0.98 & 0.99 & 0.99 & 3.0 & 0.15 & 7.1 & 3.5 & 0.58 & 0.99 & 0.98\\
LGRBs high energy, $\theta_{\rm obs} = 0.8$ & 0.03 & 0.02 & 169.0 & 79.7 & 0.94 & 0.96 & 0.96 & 0.11 & 0.03 & 24.2 & 11.9 & 0.96 & 0.97 & 0.97\\
LGRBs high energy RS, $\Gamma = 200$  & 13.6 & 1.5 &1.0 &1.0 & \nodata & \nodata & \nodata & \nodata & \nodata & \nodata & \nodata & \nodata & \nodata & \nodata\\
LGRBs high energy RS, $\Gamma = 50$  & 63.3& 1.4 &1.0 & 1.0& \nodata & \nodata & \nodata & \nodata & \nodata & \nodata & \nodata & \nodata & \nodata & \nodata\\
SGRB, On-axis & 0.04 & 0.02 & 38.0 & 20.0 & 0.99 & 0.99 & 0.99 & 0.1 & 0.04 & 5.7 & 3.1 & 0.48  &0.0 & 0.94 \\
SGRB ($\theta_{\rm obs} = 0.4$) & 0.0 & 0.006 & 162.8 & 85.1 & 0.96 & 0.96 & 0.96 & 0.01 & 0.01 & 23.0 & 12.1 & 0.96 & 0.95 & 0.95\\
TDE, On-axis  & 0.03& 0.07 & 564.1 &195.1 & 0.91 & 0.70 & 0.92 & 0.1 & 0.1 & 90.4 & 33.7 & 0.34 & 0.0 & 0.93\\
TDE, Off-Axis (spherical) & 0.0 &0.01 & 1754.7& 790.8 &0.40& 0.68 &0.67& 0.02 & 0.01 & 247.1 & 116.4 & 0.74 & 0.70 & 0.65\\
NSM: magnetar & 0.0 & 0.008 & 1828.3 & 948.3 & 0.29 & 0.71 & 0.49 & 0.02 & 0.01 & 259.6 & 134.3 & 0.79  &0.71 & 0.49\\
FBOTs & 0.1 & 0.01 & 55.1 &  27.8 & 0.99 & 0.99 & 0.98 & 0.59 & 0.02 & 8.1 & 4.4 & 0.45 & 0.11 & 0.99\\
\multicolumn8c{--------------------------------------------------------------------------------------------------------} &  \multicolumn8c{-------------------------------------------------------------} \\
& & \multicolumn8c{ } & \multicolumn6c{30 Day Stack} \\
\multicolumn8c{--------------------------------------------------------------------------------------------------------} &  \multicolumn8c{-------------------------------------------------------------} \\
LGRBs (on-axis) & \nodata &  \nodata & \nodata &\nodata & \nodata& \nodata & \nodata &1.9 & 0.16 & 1.0 & 1.0 & 0.0 & 0.0 & 0.0\\
LGRBs ($\theta_{\rm obs} = 0.4$)& \nodata &  \nodata & \nodata &\nodata & \nodata& \nodata & \nodata &0.25 & 0.05 & 2.4 & 1.4 & 0.0 & 0.0 & 0.3\\
LGRBs ($\theta_{\rm obs} = 0.8$) & \nodata &  \nodata & \nodata &\nodata & \nodata& \nodata & \nodata & 0.0 & 0.01 & 8.1 & 3.7 & 0.89 & 0.53 & 0.96\\
LGRBs high energy, (on-axis) & \nodata &  \nodata & \nodata &\nodata & \nodata& \nodata & \nodata &16.6 & 0.41 & 1.0 & 1.0 & 0.0 & 0.0 & 0.0 \\
LGRBs high energy, $\theta_{\rm obs} = 0.4$ & \nodata &  \nodata & \nodata &\nodata & \nodata& \nodata & \nodata & 5.9 & 0.18 & 2.1 & 1.2 & 0.0 & 0.0 & 0.98\\
LGRBs high energy, $\theta_{\rm obs} = 0.8$ & \nodata &  \nodata & \nodata &\nodata & \nodata& \nodata & \nodata &0.32 & 0.04 & 5.9 & 2.9 & 0.87 & 0.02 & 0.96\\
SGRB, On-axis & \nodata &  \nodata & \nodata &\nodata & \nodata& \nodata & \nodata & 0.27 & 0.04 & 1.4 & 1.1 & 0.0 & 0.0 & 3.9\\
SGRB ($\theta_{\rm obs} = 0.4$) & \nodata &  \nodata & \nodata &\nodata & \nodata& \nodata & \nodata & 0.03 & 0.01 & 5.9 & 3.0 & 0.78 & 0.96 & 0.96\\
TDE, On-axis & \nodata &  \nodata & \nodata &\nodata & \nodata& \nodata & \nodata & 0.3 & 0.15 & 20.3 &7.5 & 0.54& 0.0 &0.93  & \\
TDE, Off-Axis (spherical)  & \nodata &  \nodata & \nodata &\nodata & \nodata& \nodata & \nodata & 0.05 &0.02  & 57.6   & 26.9&0.78 &0.71  & 0.64\\
NSM: magnetar  & \nodata &  \nodata & \nodata &\nodata & \nodata& \nodata & \nodata  &  0.05& 0.02 &  61.1 &32.5 & 0.82 &0.69  &0.49 \\
FBOTs & \nodata &  \nodata & \nodata &\nodata & \nodata& \nodata & \nodata  & 0.9 & 0.02 & 2.3  & 1.8&  0.0 & 0.0 & 0.31 & \\
\enddata
\tablecomments{Columns are as in Table~\ref{tab:act}.}
\label{tab:simons}
\end{deluxetable*}

\begin{deluxetable*}{lccccccccccccccc}
\tablecolumns{8}
\caption{CMB-S4}
\tablehead{
\colhead{Transient} &
\colhead{N} &
\colhead{$\bar{z}$} &
\colhead{$\bar{n}_{3\sigma}$} & 
\colhead{$\bar{n}_{10\sigma}$} &  
\colhead{$f_{\rm peak}$} &
\colhead{$f_{\rm rise}$} &
\colhead{$f_{\rm fall}$} &
\colhead{N}&
\colhead{$\bar{z}$} &
\colhead{$\bar{n}_{3\sigma}$} & 
\colhead{$\bar{n}_{10\sigma}$} &  
\colhead{$f_{\rm peak}$} &
\colhead{$f_{\rm rise}$} &
\colhead{$f_{\rm fall}$} 
}  
\startdata
& & \multicolumn{7}{c}{Daily Cadence} & \multicolumn{7}{c}{7 Day Stack} \\
\multicolumn8c{--------------------------------------------------------------------------------------------------------} &  \multicolumn8c{-------------------------------------------------------------} \\
LGRBs (on-axis) & 4.5  & 0.17 & 10.4 & 5.4 & 0.99 & 0.99 & 0.99 & 12.4 & 0.25 & 1.7 & 1.1 & 0.08 & 0.0 & 0.66\\
LGRBs ($\theta_{\rm obs} = 0.4$) & 0.12 & 0.03 & 58.6 & 26.6 & 0.99 & 0.99 & 0.99 & 0.46 & 0.05 & 8.6 & 4.0 & 0.73 & 0.50 &  0.99 \\
LGRBs ($\theta_{\rm obs} = 0.8$) & 0.0&  0.006 & 243.6 & 112.6 & 0.95 & 0.97 & 0.96 & 0.0 & 0.01 & 34.5 & 16.0 & 0.99 & 0.97 & 0.96 \\
LGRBs RS, $\Gamma = 200$ & 16.9 & 1.3 & 1.0 & 1.0 & \nodata & \nodata & \nodata & \nodata & \nodata & \nodata & \nodata & \nodata & \nodata & \nodata\\
LGRBs RS, $\Gamma = 50$  & 40.9& 0.9 & 1.0& 1.0& \nodata & \nodata & \nodata & \nodata & \nodata & \nodata & \nodata & \nodata & \nodata & \nodata\\
LGRBs high energy, (on-axis) & 43.7 & 0.47 & 9.8 & 5.0 & 0.99 & 0.99 & 0.99 & 114.3 & 0.65 & 1.6 & 1.1 & 0.04 & 0.0 & 0.5\\
LGRBs high energy, $\theta_{\rm obs} = 0.4$ & 2.4 & 0.12 & 46.7 & 23.7 & 0.99 & 0.99 & 0.99 & 10.4 & 0.19 & 7.3 & 3.7 & 0.68 & 0.99 & 0.99 \\
LGRBs high energy, $\theta_{\rm obs} = 0.8$ & 0.1 & 0.03 & 169.3 & 81.9 & 0.96 & 0.97 & 0.97 & 0.4 & 0.04 & 24.8 & 12.0 & 0.97 & 0.97 & 0.97 \\
LGRBs high energy RS, $\Gamma = 200$  & 35.2 & 1.8& 1.0&1.0 &  \nodata & \nodata & \nodata & \nodata & \nodata & \nodata & \nodata & \nodata & \nodata & \nodata\\
LGRBs high energy RS, $\Gamma = 50$  & 162.5 & 1.7 & 1.0 & 1.0 & \nodata & \nodata & \nodata & \nodata & \nodata & \nodata & \nodata & \nodata & \nodata & \nodata\\
SGRB, On-axis & 0.1 & 0.03 & 37.7 & 19.5 & 0.99 & 0.99 & 0.99 & 0.49 & 0.05 & 5.7 & 3.1 & 0.63 & 0.0 & 0.95 \\
SGRB ($\theta_{\rm obs} = 0.4$) & 0.0 & 0.01 & 166.3 & 87.2 & 0.96 & 0.96 & 0.98 & 0.04 & 0.01 & 23.5 & 12.5 & 0.97 & 0.96 & 0.97  \\
TDE, On-axis & 0.09 & 0.09 & 617.7 & 198.4 & 0.94 & 0.70 & 0.94 & 0.34 & 0.13 & 93.0 & 32.6 & 0.50 & 0.0 & 0.94\\
TDE, Off-Axis (spherical) & 0.01 & 0.01 & 2164.7 & 885.7 & 0.55 & 0.78 & 0.72 & 0.05 & 0.01 & 314.7 & 136.0 & 0.85 & 0.78 & 0.73\\
NSM: magnetar & 0.01 & 0.01 & 2302.5 &1071.3 & 0.45 & 0.75 & 0.57 & 0.06 & 0.02 & 324.7 & 141.0 & 0.86 & 0.78 & 0.58 \\
FBOTs & 0.5& 0.01 & 54.9 & 27.3 & 0.99 & 0.99 & 0.99 & 2.1 & 0.02 & 8.0 & 4.4 & 0.59 & 0.12 & 0.99\\
\multicolumn8c{--------------------------------------------------------------------------------------------------------} &  \multicolumn8c{-------------------------------------------------------------} \\
& & \multicolumn8c{ } & \multicolumn6c{30 Day Stack} \\
\multicolumn8c{--------------------------------------------------------------------------------------------------------} &  \multicolumn8c{-------------------------------------------------------------} \\
LGRBs (on-axis) & \nodata &  \nodata & \nodata &\nodata & \nodata& \nodata & \nodata & 6.3 & 0.2 & 1.0 & 1.0 & 0.0 & 0.0 &0.0 \\
LGRBs ($\theta_{\rm obs} = 0.4$) & \nodata &  \nodata & \nodata &\nodata & \nodata& \nodata & \nodata & 0.89 & 0.07 & 2.4 & 1.4 & 0.12 & 0.0 & 0.31\\
LGRBs ($\theta_{\rm obs} = 0.8$)  & \nodata &  \nodata & \nodata &\nodata & \nodata& \nodata & \nodata & 0.03 & 0.01 & 8.3 & 3.9 & 0.93 & 0.57 & 0.97\\
LGRBs high energy, (on-axis)  & \nodata &  \nodata & \nodata &\nodata & \nodata& \nodata & \nodata & 57.7 & 0.51 & 1.0 & 1.0 & 0.0 & 0.0 & 0.0\\
LGRBs high energy, $\theta_{\rm obs} = 0.4$  & \nodata &  \nodata & \nodata &\nodata & \nodata& \nodata & \nodata &20.5 & 0.23 & 2.1 & 1.2 & 0.16 &0.0  &0.98\\
LGRBs high energy, $\theta_{\rm obs} = 0.8$  & \nodata &  \nodata & \nodata &\nodata & \nodata& \nodata & \nodata & 1.16 & 0.06& 6.0 & 2.9 & 0.90 & 0.03 &0.97 \\
SGRB, On-axis  & \nodata &  \nodata & \nodata &\nodata & \nodata& \nodata & \nodata & 0.92 & 0.06 & 1.5& 1.1 & 0.12 & 0.0 & 0.4\\
SGRB ($\theta_{\rm obs} = 0.4$) & \nodata &  \nodata & \nodata &\nodata & \nodata& \nodata & \nodata & 0.11 & 0.02 & 6.0 & 3.0 & 0.87 & 0.97 & 0.97\\
TDE, On-axis  & \nodata &  \nodata & \nodata &\nodata & \nodata& \nodata & \nodata & 1.0 & 0.19 & 22.1 &7.9 &0.66 &0.0 &0.93\\
TDE, Off-Axis (spherical)  & \nodata &  \nodata & \nodata &\nodata & \nodata& \nodata & \nodata & 0.16& 0.02& 73.0& 30.8 & 0.85 &0.77 & 0.72\\
NSM: magnetar  & \nodata &  \nodata & \nodata &\nodata & \nodata& \nodata & \nodata & 0.2 &0.02 & 78.5 & 36.9 & 0.88 & 0.76& 0.60&\\
FBOTs  & \nodata &  \nodata & \nodata &\nodata & \nodata& \nodata & \nodata & 3.4 & 0.02 & 2.3& 1.8 & 0.02 & 0.0 &0.31\\
\enddata
\tablecomments{Columns are as in Table~\ref{tab:act}.}
\label{tab:cmbs4}
\end{deluxetable*}

\begin{deluxetable*}{lccccccccccccccc}
\tablecolumns{8}
\caption{CMB-HD}
\tablehead{
\colhead{Transient} &
\colhead{N} &
\colhead{$\bar{z}$} &
\colhead{$\bar{n}_{3\sigma}$} & 
\colhead{$\bar{n}_{10\sigma}$} &  
\colhead{$f_{\rm peak}$} &
\colhead{$f_{\rm rise}$} &
\colhead{$f_{\rm fall}$} &
\colhead{N}&
\colhead{$\bar{z}$} &
\colhead{$\bar{n}_{3\sigma}$} & 
\colhead{$\bar{n}_{10\sigma}$} &  
\colhead{$f_{\rm peak}$} &
\colhead{$f_{\rm rise}$} &
\colhead{$f_{\rm fall}$} 
}  
\startdata
& & \multicolumn{7}{c}{Daily Cadence} & \multicolumn{7}{c}{7 Day Stack} \\
\multicolumn8c{--------------------------------------------------------------------------------------------------------} &  \multicolumn8c{-------------------------------------------------------------} \\
LGRBs (on-axis) & 23.5 & 0.30 & 10.5 & 5.5 & 0.99 & 0.99 & 0.99 & 63.1 & 0.41 & 1.8 & 1.1 & 0.58 & 0.0 &  0.68\\
LGRBs ($\theta_{\rm obs} = 0.4$) & 0.67 & 0.06 & 58.2 & 26.4 & 0.99 & 0.99 & 0.99 & 2.6 & 0.09 & 8.6 & 4.0 & 0.91 & 0.48 & 0.99\\
LGRBs ($\theta_{\rm obs} = 0.8$) & 0.01 & 0.01 & 250.8 & 122.8 & 0.93 & 0.96 & 0.99 & 0.05 & 0.02 & 36.4 & 16.8 & 0.98 & 0.97 & 0.98\\
LGRBs RS, $\Gamma = 200$ & 53.1 & 1.7& 1.0 & 1.0 & \nodata & \nodata & \nodata & \nodata & \nodata & \nodata & \nodata & \nodata & \nodata & \nodata\\
LGRBs RS, $\Gamma = 50$  & 197.8 & 1.4 & 1.0 & 1.0 & \nodata & \nodata & \nodata & \nodata & \nodata & \nodata & \nodata & \nodata & \nodata & \nodata\\
LGRBs high energy, (on-axis) & 182.4 & 0.74 & 10.4 & 5.9 & 0.99 & 0.99 & 0.99 & 515.5 & 1.1 & 1.7  & 1.1 & 0.47 & 0.0 & 0.6\\
LGRBs high energy, $\theta_{\rm obs} = 0.4$ & 15.0 & 0.20 & 47.1 & 24.1 & 0.99 & 0.99 & 0.98 & 55.7 & 0.31 & 7.4 & 3.7 & 0.89 & 0.99 & 0.99\\
LGRBs high energy, $\theta_{\rm obs} = 0.8$ & 0.52 & 0.04 & 171.8 & 82.2 & 0.96 & 0.97 & 0.97 & 2.2 & 0.07 & 24.3 & 11.6 & 0.99 & 0.97 & 0.97\\
LGRBs high energy RS, $\Gamma = 200$  & 71.2 & 1.9 & 1.0& 1.0 & \nodata & \nodata & \nodata & \nodata & \nodata & \nodata & \nodata & \nodata & \nodata & \nodata\\
LGRBs high energy RS, $\Gamma = 50$  & 412.6 & 1.9 &1.0 &1.0 & \nodata & \nodata & \nodata & \nodata & \nodata & \nodata & \nodata & \nodata & \nodata & \nodata\\
SGRB, On-axis & 0.7 & 0.05 &  37.6 & 19.3 & 0.99 & 0.99 & 0.99 & 2.7 & 0.08 & 5.8 & 3.2 & 0.88 & 0.0 & 0.95\\
SGRB ($\theta_{\rm obs} = 0.4$) & 0.05 & 0.01 & 162.5 & 84.1 & 0.97 & 0.98 & 0.97 & 0.22 & 0.02 & 23.6 & 12.3 & 0.98 & 0.97 & 0.97\\
TDE, On-axis & 0.4 & 0.15 & 631.5 & 209.4 & 0.93 & 0.70 & 0.96 & 1.9 & 0.22 & 91.9 & 32.4 & 0.83 & 0.0 & 0.94\\
TDE, Off-Axis (spherical) & 0.07 & 0.02  & 2282.1 & 927.6 & 0.56& 0.78 & 0.75 & 0.3 & 0.02 & 325.1 & 134.3 & 0.89 & 0.78 & 0.74\\
NSM: magnetar & 0.07 & 0.02 & 2451.5 & 1091.2 & 0.46 &  0.79 & 0.62 &  0.31 & 0.03 & 345.3 & 156.2 & 0.92 & 0.78 & 0.61\\
FBOTs & 2.0 & 0.02 & 55.4 & 27.7 & 0.99 & 0.99 & 0.99 & 11.5 & 0.03 & 8.0 & 4.3 & 0.89 & 0.11 & 0.99\\
\multicolumn8c{--------------------------------------------------------------------------------------------------------} &  \multicolumn8c{-------------------------------------------------------------} \\
& & \multicolumn8c{ } & \multicolumn6c{30 Day Stack} \\
\multicolumn8c{--------------------------------------------------------------------------------------------------------} &  \multicolumn8c{-------------------------------------------------------------} \\
LGRBs (on-axis)& \nodata &  \nodata & \nodata &\nodata & \nodata& \nodata & \nodata & 33.3 & 0.34 & 1.0 & 1.0 & 0.02 & 0.0 & 0.01 \\
LGRBs ($\theta_{\rm obs} = 0.4$) & \nodata &  \nodata & \nodata &\nodata & \nodata& \nodata & \nodata &4.8 & 0.11 & 2.4 & 1.4 & 0.44 & 0.0 & 0.31\\
LGRBs ($\theta_{\rm obs} = 0.8$) & \nodata &  \nodata & \nodata &\nodata & \nodata& \nodata & \nodata &0.15 & 0.02 & 8.2 & 3.8 & 0.97 & 0.55 & 0.96\\
LGRBs high energy, (on-axis) & \nodata &  \nodata & \nodata &\nodata & \nodata& \nodata & \nodata &282.6 & 0.89 & 1.0 & 1.0 & 0.02 & 0.0 & 0.0\\
LGRBs high energy, $\theta_{\rm obs} = 0.4$ & \nodata &  \nodata & \nodata &\nodata & \nodata& \nodata & \nodata & 117.2 & 0.40 & 2.1 & 1.2 & 0.45 & 0.0 & 0.98\\
LGRBs high energy, $\theta_{\rm obs} = 0.8$  & \nodata &  \nodata & \nodata &\nodata & \nodata& \nodata & \nodata & 6.5 & 0.10 & 6.0 & 2.9 & 0.95 & 0.03 & 0.97 \\
SGRB, On-axis & \nodata &  \nodata & \nodata &\nodata & \nodata& \nodata & \nodata & 5.1 & 0.10 & 1.4 & 1.1 & 0.45 & 0.0 & 0.41\\
SGRB ($\theta_{\rm obs} = 0.4$)  & \nodata &  \nodata & \nodata &\nodata & \nodata& \nodata & \nodata &0.60 & 0.03 & 6.0 & 3.1 & 0.93 & 0.98 & 0.97\\
TDE, On-axis & \nodata &  \nodata & \nodata &\nodata & \nodata& \nodata & \nodata &  5.6 & 0.3 & 22.4 &7.9 & 0.79 &0.0 & 0.94\\
TDE, Off-Axis (spherical) & \nodata &  \nodata & \nodata &\nodata & \nodata& \nodata & \nodata &  0.84 & 0.04 & 75.4 & 31.4 & 0.90 &0.78 &0.73\\
NSM: magnetar & \nodata &  \nodata & \nodata &\nodata & \nodata& \nodata & \nodata & 0.9 & 0.04 & 81.8 & 37.7 & 0.93 & 0.78 & 0.62\\
FBOTs  & \nodata &  \nodata & \nodata &\nodata & \nodata& \nodata & \nodata & 18.2 & 0.04& 2.3 & 1.8 & 0.38 &0.0 &0.30\\
\enddata
\tablecomments{Columns are as in Table~\ref{tab:act}.}
\label{tab:cmbhd}
\end{deluxetable*}

\bibliographystyle{aasjournal}
\bibliography{references}

\begin{thebibliography}{}
\expandafter\ifx\csname natexlab\endcsname\relax\def\natexlab#1{#1}\fi
\providecommand{\url}[1]{\href{#1}{#1}}
\providecommand{\dodoi}[1]{doi:~\href{http://doi.org/#1}{\nolinkurl{#1}}}

\bibitem[{{Abazajian} {et~al.}(2019){Abazajian}, {Addison}, {Adshead}, {Ahmed},
  {Allen}, {Alonso}, {Alvarez}, {Anderson}, {Arnold}, {Baccigalupi}, {Bailey},
  {Barkats}, {Barron}, {Barry}, {Bartlett}, {Basu Thakur}, {Battaglia},
  {Baxter}, {Bean}, {Bebek}, {Bender}, {Benson}, {Berger}, {Bhimani},
  {Bischoff}, {Bleem}, {Bocquet}, {Boddy}, {Bonato}, {Bond}, {Borrill},
  {Bouchet}, {Brown}, {Bryan}, {Burkhart}, {Buza}, {Byrum}, {Calabrese},
  {Calafut}, {Caldwell}, {Carlstrom}, {Carron}, {Cecil}, {Challinor}, {Chang},
  {Chinone}, {Cho}, {Cooray}, {Crawford}, {Crites}, {Cukierman}, {Cyr-Racine},
  {de Haan}, {de Zotti}, {Delabrouille}, {Demarteau}, {Devlin}, {Di Valentino},
  {Dobbs}, {Duff}, {Duivenvoorden}, {Dvorkin}, {Edwards}, {Eimer}, {Errard},
  {Essinger-Hileman}, {Fabbian}, {Feng}, {Ferraro}, {Filippini}, {Flauger},
  {Flaugher}, {Fraisse}, {Frolov}, {Galitzki}, {Galli}, {Ganga}, {Gerbino},
  {Gilchriese}, {Gluscevic}, {Green}, {Grin}, {Grohs}, {Gualtieri}, {Guarino},
  {Gudmundsson}, {Habib}, {Haller}, {Halpern}, {Halverson}, {Hanany},
  {Harrington}, {Hasegawa}, {Hasselfield}, {Hazumi}, {Heitmann}, {Henderson},
  {Henning}, {Hill}, {Hlozek}, {Holder}, {Holzapfel}, {Hubmayr},
  {Huffenberger}, {Huffer}, {Hui}, {Irwin}, {Johnson}, {Johnstone}, {Jones},
  {Karkare}, {Katayama}, {Kerby}, {Kernovsky}, {Keskitalo}, {Kisner}, {Knox},
  {Kosowsky}, {Kovac}, {Kovetz}, {Kuhlmann}, {Kuo}, {Kurita}, {Kusaka},
  {Lahteenmaki}, {Lawrence}, {Lee}, {Lewis}, {Li}, {Linder}, {Loverde},
  {Lowitz}, {Madhavacheril}, {Mantz}, {Matsuda}, {Mauskopf}, {McMahon},
  {McQuinn}, {Meerburg}, {Melin}, {Meyers}, {Millea}, {Mohr}, {Moncelsi},
  {Mroczkowski}, {Mukherjee}, {M{\"u}nchmeyer}, {Nagai}, {Nagy}, {Namikawa},
  {Nati}, {Natoli}, {Negrello}, {Newburgh}, {Niemack}, {Nishino}, {Nordby},
  {Novosad}, {O'Connor}, {Obied}, {Padin}, {Pandey}, {Partridge}, {Pierpaoli},
  {Pogosian}, {Pryke}, {Puglisi}, {Racine}, {Raghunathan}, {Rahlin},
  {Rajagopalan}, {Raveri}, {Reichanadter}, {Reichardt}, {Remazeilles}, {Rocha},
  {Roe}, {Roy}, {Ruhl}, {Salatino}, {Saliwanchik}, {Schaan}, {Schillaci},
  {Schmittfull}, {Scott}, {Sehgal}, {Shandera}, {Sheehy}, {Sherwin},
  {Shirokoff}, {Simon}, {Slosar}, {Somerville}, {Spergel}, {Staggs}, {Stark},
  {Stompor}, {Story}, {Stoughton}, {Suzuki}, {Tajima}, {Teply}, {Thompson},
  {Timbie}, {Tomasi}, {Treu}, {Tristram}, {Tucker}, {Umilt{\`a}}, {van
  Engelen}, {Vieira}, {Vieregg}, {Vogelsberger}, {Wang}, {Watson}, {White},
  {Whitehorn}, {Wollack}, {Kimmy Wu}, {Xu}, {Yasini}, {Yeck}, {Yoon}, {Young},
  \& {Zonca}}]{Abazajian2019}
{Abazajian}, K., {Addison}, G., {Adshead}, P., {et~al.} 2019, arXiv e-prints,
  \hypersetup{urlcolor=magenta}\href{https://arxiv.org/abs/1907.04473}{arXiv}{:}\hypersetup{urlcolor=blue}\href{https://ui.adsabs.harvard.edu/abs/2019arXiv190704473A}{1907.04473}

\bibitem[{{Abbott} {et~al.}(2017){Abbott}, {Abbott}, {Abbott}, {Acernese},
  {Ackley}, {Adams}, {Adams}, {Addesso}, {Adhikari}, {Adya}, {Affeldt},
  {Afrough}, {Agarwal}, {Agathos}, {Agatsuma}, {Aggarwal}, {Aguiar}, {Aiello},
  {Ain}, {Ajith}, {Allen}, {Allen}, {Allocca}, {Altin}, {Amato}, {Ananyeva},
  {Anderson}, {Anderson}, {Angelova}, {Antier}, {Appert}, {Arai}, {Araya},
  {Areeda}, {Arnaud}, {Arun}, {Ascenzi}, {Ashton}, {Ast}, {Aston}, {Astone},
  {Atallah}, {Aufmuth}, {Aulbert}, {AultONeal}, {Austin}, {Avila-Alvarez},
  {Babak}, {Bacon}, {Bader}, {Bae}, {Bailes}, {Baker}, {Baldaccini},
  {Ballardin}, {Ballmer}, {Banagiri}, {Barayoga}, {Barclay}, {Barish},
  {Barker}, {Barkett}, {Barone}, {Barr}, {Barsotti}, {Barsuglia}, {Barta},
  {Barthelmy}, {Bartlett}, {Bartos}, {Bassiri}, {Basti}, {Batch}, {Bawaj},
  {Bayley}, {Bazzan}, {B{\'e}csy}, {Beer}, {Bejger}, {Belahcene}, {Bell},
  {Berger}, {Bergmann}, {Bernuzzi}, {Bero}, {Berry}, {Bersanetti}, {Bertolini},
  {Betzwieser}, {Bhagwat}, {Bhandare}, {Bilenko}, {Billingsley}, {Billman},
  {Birch}, {Birney}, {Birnholtz}, {Biscans}, {Biscoveanu}, {Bisht}, {Bitossi},
  {Biwer}, {Bizouard}, {Blackburn}, {Blackman}, {Blair}, {Blair}, {Blair},
  {Bloemen}, {Bock}, {Bode}, {Boer}, {Bogaert}, {Bohe}, {Bondu}, {Bonilla},
  {Bonnand}, {Boom}, {Bork}, {Boschi}, {Bose}, {Bossie}, {Bouffanais}, {Bozzi},
  {Bradaschia}, {Brady}, {Branchesi}, {Brau}, {Briant}, {Brillet}, {Brinkmann},
  {Brisson}, {Brockill}, {Broida}, {Brooks}, {Brown}, {Brown}, {Brunett},
  {Buchanan}, {Buikema}, {Bulik}, {Bulten}, {Buonanno}, {Buskulic}, {Buy},
  {Byer}, {Cabero}, {Cadonati}, {Cagnoli}, {Cahillane}, {Calder{\'o}n
  Bustillo}, {Callister}, {Calloni}, {Camp}, {Canepa}, {Canizares}, {Cannon},
  {Cao}, {Cao}, {Capano}, {Capocasa}, {Carbognani}, {Caride}, {Carney},
  {Carullo}, {Casanueva Diaz}, {Casentini}, {Caudill}, {Cavagli{\`a}},
  {Cavalier}, {Cavalieri}, {Cella}, {Cepeda}, {Cerd{\'a}-Dur{\'a}n},
  {Cerretani}, {Cesarini}, {Chamberlin}, {Chan}, {Chao}, {Charlton}, {Chase},
  {Chassande-Mottin}, {Chatterjee}, {Chatziioannou}, {Cheeseboro}, {Chen},
  {Chen}, {Chen}, {Cheng}, {Chia}, {Chincarini}, {Chiummo}, {Chmiel}, {Cho},
  {Cho}, {Chow}, {Christensen}, {Chu}, {Chua}, {Chua}, {Chung}, {Chung},
  {Ciani}, {Ciolfi}, {Cirelli}, {Cirone}, {Clara}, {Clark}, {Clearwater},
  {Cleva}, {Cocchieri}, {Coccia}, {Cohadon}, {Cohen}, {Colla}, {Collette},
  {Cominsky}, {Constancio}, {Conti}, {Cooper}, {Corban}, {Corbitt},
  {Cordero-Carri{\'o}n}, {Corley}, {Cornish}, {Corsi}, {Cortese}, {Costa},
  {Coughlin}, {Coughlin}, {Coulon}, {Countryman}, {Couvares}, {Covas}, {Cowan},
  {Coward}, {Cowart}, {Coyne}, {Coyne}, {Creighton}, {Creighton}, {Cripe},
  {Crowder}, {Cullen}, {Cumming}, {Cunningham}, {Cuoco}, {Dal Canton},
  {D{\'a}lya}, {Danilishin}, {D'Antonio}, {Danzmann}, {Dasgupta}, {Da Silva
  Costa}, {Dattilo}, {Dave}, {Davier}, {Davis}, {Daw}, {Day}, {De}, {DeBra},
  {Degallaix}, {De Laurentis}, {Del{\'e}glise}, {Del Pozzo}, {Demos}, {Denker},
  {Dent}, {De Pietri}, {Dergachev}, {De Rosa}, {DeRosa}, {De Rossi}, {DeSalvo},
  {de Varona}, {Devenson}, {Dhurandhar}, {D{\'\i}az}, {Dietrich}, {Di Fiore},
  {Di Giovanni}, {Di Girolamo}, {Di Lieto}, {Di Pace}, {Di Palma}, {Di Renzo},
  {Doctor}, {Dolique}, {Donovan}, {Dooley}, {Doravari}, {Dorrington},
  {Douglas}, {Dovale {\'A}lvarez}, {Downes}, {Drago}, {Dreissigacker},
  {Driggers}, {Du}, {Ducrot}, {Dudi}, {Dupej}, {Dwyer}, {Edo}, {Edwards},
  {Effler}, {Eggenstein}, {Ehrens}, {Eichholz}, {Eikenberry}, {Eisenstein},
  {Essick}, {Estevez}, {Etienne}, {Etzel}, {Evans}, {Evans}, {Factourovich},
  {Fafone}, {Fair}, {Fairhurst}, {Fan}, {Farinon}, {Farr}, {Farr},
  {Fauchon-Jones}, {Favata}, {Fays}, {Fee}, {Fehrmann}, {Feicht}, {Fejer},
  {Fernandez-Galiana}, {Ferrante}, {Ferreira}, {Ferrini}, {Fidecaro},
  {Finstad}, {Fiori}, {Fiorucci}, {Fishbach}, {Fisher}, {Fitz-Axen},
  {Flaminio}, {Fletcher}, {Fong}, {Font}, {Forsyth}, {Forsyth}, {Fournier},
  {Frasca}, {Frasconi}, {Frei}, {Freise}, {Frey}, {Frey}, {Fries}, {Fritschel},
  {Frolov}, {Fulda}, {Fyffe}, {Gabbard}, {Gadre}, {Gaebel}, {Gair},
  {Gammaitoni}, {Ganija}, {Gaonkar}, {Garcia-Quiros}, {Garufi}, {Gateley},
  {Gaudio}, {Gaur}, {Gayathri}, {Gehrels}, {Gemme}, {Genin}, {Gennai},
  {George}, {George}, {Gergely}, {Germain}, {Ghonge}, {Ghosh}, {Ghosh},
  {Ghosh}, {Giaime}, {Giardina}, {Giazotto}, {Gill}, {Glover}, {Goetz},
  {Goetz}, {Gomes}, {Goncharov}, {Gonz{\'a}lez}, {Gonzalez Castro},
  {Gopakumar}, {Gorodetsky}, {Gossan}, {Gosselin}, {Gouaty}, {Grado}, {Graef},
  {Granata}, {Grant}, {Gras}, {Gray}, {Greco}, {Green}, {Gretarsson}, {Groot},
  {Grote}, {Grunewald}, {Gruning}, {Guidi}, {Guo}, {Gupta}, {Gupta}, {Gushwa},
  {Gustafson}, {Gustafson}, {Halim}, {Hall}, {Hall}, {Hamilton}, {Hammond},
  {Haney}, {Hanke}, {Hanks}, {Hanna}, {Hannam}, {Hannuksela}, {Hanson},
  {Hardwick}, {Harms}, {Harry}, {Harry}, {Hart}, {Haster}, {Haughian}, {Healy},
  {Heidmann}, {Heintze}, {Heitmann}, {Hello}, {Hemming}, {Hendry}, {Heng},
  {Hennig}, {Heptonstall}, {Heurs}, {Hild}, {Hinderer}, {Ho}, {Hoak}, {Hofman},
  {Holt}, {Holz}, {Hopkins}, {Horst}, {Hough}, {Houston}, {Howell}, {Hreibi},
  {Hu}, {Huerta}, {Huet}, {Hughey}, {Husa}, {Huttner}, {Huynh-Dinh}, {Indik},
  {Inta}, {Intini}, {Isa}, {Isac}, {Isi}, {Iyer}, {Izumi}, {Jacqmin}, {Jani},
  {Jaranowski}, {Jawahar}, {Jim{\'e}nez-Forteza}, {Johnson},
  {Johnson-McDaniel}, {Jones}, {Jones}, {Jonker}, {Ju}, {Junker}, {Kalaghatgi},
  {Kalogera}, {Kamai}, {Kandhasamy}, {Kang}, {Kanner}, {Kapadia}, {Karki},
  {Karvinen}, {Kasprzack}, {Kastaun}, {Katolik}, {Katsavounidis}, {Katzman},
  {Kaufer}, {Kawabe}, {K{\'e}f{\'e}lian}, {Keitel}, {Kemball}, {Kennedy},
  {Kent}, {Key}, {Khalili}, {Khan}, {Khan}, {Khan}, {Khazanov}, {Kijbunchoo},
  {Kim}, {Kim}, {Kim}, {Kim}, {Kim}, {Kim}, {Kimbrell}, {King}, {King},
  {Kinley-Hanlon}, {Kirchhoff}, {Kissel}, {Kleybolte}, {Klimenko}, {Knowles},
  {Koch}, {Koehlenbeck}, {Koley}, {Kondrashov}, {Kontos}, {Korobko}, {Korth},
  {Kowalska}, {Kozak}, {Kr{\"a}mer}, {Kringel}, {Krishnan}, {Kr{\'o}lak},
  {Kuehn}, {Kumar}, {Kumar}, {Kumar}, {Kuo}, {Kutynia}, {Kwang}, {Lackey},
  {Lai}, {Landry}, {Lang}, {Lange}, {Lantz}, {Lanza}, {Larson},
  {Lartaux-Vollard}, {Lasky}, {Laxen}, {Lazzarini}, {Lazzaro}, {Leaci},
  {Leavey}, {Lee}, {Lee}, {Lee}, {Lee}, {Lee}, {Lehmann}, {Lenon}, {Leon},
  {Leonardi}, {Leroy}, {Letendre}, {Levin}, {Li}, {Linker}, {Littenberg},
  {Liu}, {Liu}, {Lo}, {Lockerbie}, {London}, {Lord}, {Lorenzini}, {Loriette},
  {Lormand}, {Losurdo}, {Lough}, {Lousto}, {Lovelace}, {L{\"u}ck}, {Lumaca},
  {Lundgren}, {Lynch}, {Ma}, {Macas}, {Macfoy}, {Machenschalk}, {MacInnis},
  {Macleod}, {Maga{\~n}a Hernandez}, {Maga{\~n}a-Sandoval}, {Maga{\~n}a
  Zertuche}, {Magee}, {Majorana}, {Maksimovic}, {Man}, {Mandic}, {Mangano},
  {Mansell}, {Manske}, {Mantovani}, {Marchesoni}, {Marion}, {M{\'a}rka},
  {M{\'a}rka}, {Markakis}, {Markosyan}, {Markowitz}, {Maros}, {Marquina},
  {Marsh}, {Martelli}, {Martellini}, {Martin}, {Martin}, {Martynov}, {Marx},
  {Mason}, {Massera}, {Masserot}, {Massinger}, {Masso-Reid}, {Mastrogiovanni},
  {Matas}, {Matichard}, {Matone}, {Mavalvala}, {Mazumder}, {McCarthy},
  {McClelland}, {McCormick}, {McCuller}, {McGuire}, {McIntyre}, {McIver},
  {McManus}, {McNeill}, {McRae}, {McWilliams}, {Meacher}, {Meadors}, {Mehmet},
  {Meidam}, {Mejuto-Villa}, {Melatos}, {Mendell}, {Mercer}, {Merilh},
  {Merzougui}, {Meshkov}, {Messenger}, {Messick}, {Metzdorff}, {Meyers},
  {Miao}, {Michel}, {Middleton}, {Mikhailov}, {Milano}, {Miller}, {Miller},
  {Miller}, {Millhouse}, {Milovich-Goff}, {Minazzoli}, {Minenkov}, {Ming},
  {Mishra}, {Mitra}, {Mitrofanov}, {Mitselmakher}, {Mittleman}, {Moffa},
  {Moggi}, {Mogushi}, {Mohan}, {Mohapatra}, {Molina}, {Montani}, {Moore},
  {Moraru}, {Moreno}, {Morisaki}, {Morriss}, {Mours}, {Mow-Lowry}, {Mueller},
  {Muir}, {Mukherjee}, {Mukherjee}, {Mukherjee}, {Mukund}, {Mullavey}, {Munch},
  {Mu{\~n}iz}, {Muratore}, {Murray}, {Nagar}, {Napier}, {Nardecchia},
  {Naticchioni}, {Nayak}, {Neilson}, {Nelemans}, {Nelson}, {Nery}, {Neunzert},
  {Nevin}, {Newport}, {Newton}, {Ng}, {Nguyen}, {Nguyen}, {Nichols}, {Nielsen},
  {Nissanke}, {Nitz}, {Noack}, {Nocera}, {Nolting}, {North}, {Nuttall},
  {Oberling}, {O'Dea}, {Ogin}, {Oh}, {Oh}, {Ohme}, {Okada}, {Oliver},
  {Oppermann}, {Oram}, {O'Reilly}, {Ormiston}, {Ortega}, {O'Shaughnessy},
  {Ossokine}, {Ottaway}, {Overmier}, {Owen}, {Pace}, {Page}, {Page}, {Pai},
  {Pai}, {Palamos}, {Palashov}, {Palomba}, {Pal-Singh}, {Pan}, {Pan}, {Pang},
  {Pang}, {Pankow}, {Pannarale}, {Pant}, {Paoletti}, {Paoli}, {Papa}, {Parida},
  {Parker}, {Pascucci}, {Pasqualetti}, {Passaquieti}, {Passuello}, {Patil},
  {Patricelli}, {Pearlstone}, {Pedraza}, {Pedurand}, {Pekowsky}, {Pele},
  {Penn}, {Perez}, {Perreca}, {Perri}, {Pfeiffer}, {Phelps}, {Piccinni},
  {Pichot}, {Piergiovanni}, {Pierro}, {Pillant}, {Pinard}, {Pinto}, {Pirello},
  {Pitkin}, {Poe}, {Poggiani}, {Popolizio}, {Porter}, {Post}, {Powell},
  {Prasad}, {Pratt}, {Pratten}, {Predoi}, {Prestegard}, {Prijatelj},
  {Principe}, {Privitera}, {Prix}, {Prodi}, {Prokhorov}, {Puncken}, {Punturo},
  {Puppo}, {P{\"u}rrer}, {Qi}, {Quetschke}, {Quintero}, {Quitzow-James},
  {Raab}, {Rabeling}, {Radkins}, {Raffai}, {Raja}, {Rajan}, {Rajbhandari},
  {Rakhmanov}, {Ramirez}, {Ramos-Buades}, {Rapagnani}, {Raymond}, {Razzano},
  {Read}, {Regimbau}, {Rei}, {Reid}, {Reitze}, {Ren}, {Reyes}, {Ricci},
  {Ricker}, {Rieger}, {Riles}, {Rizzo}, {Robertson}, {Robie}, {Robinet},
  {Rocchi}, {Rolland}, {Rollins}, {Roma}, {Romano}, {Romano}, {Romel}, {Romie},
  {Rosi{\'n}ska}, {Ross}, {Rowan}, {R{\"u}diger}, {Ruggi}, {Rutins}, {Ryan},
  {Sachdev}, {Sadecki}, {Sadeghian}, {Sakellariadou}, {Salconi}, {Saleem},
  {Salemi}, {Samajdar}, {Sammut}, {Sampson}, {Sanchez}, {Sanchez},
  {Sanchis-Gual}, {Sandberg}, {Sanders}, {Sassolas}, {Sathyaprakash},
  {Saulson}, {Sauter}, {Savage}, {Sawadsky}, {Schale}, {Scheel}, {Scheuer},
  {Schmidt}, {Schmidt}, {Schnabel}, {Schofield}, {Sch{\"o}nbeck}, {Schreiber},
  {Schuette}, {Schulte}, {Schutz}, {Schwalbe}, {Scott}, {Scott}, {Seidel},
  {Sellers}, {Sengupta}, {Sentenac}, {Sequino}, {Sergeev}, {Shaddock},
  {Shaffer}, {Shah}, {Shahriar}, {Shaner}, {Shao}, {Shapiro}, {Shawhan},
  {Sheperd}, {Shoemaker}, {Shoemaker}, {Siellez}, {Siemens}, {Sieniawska},
  {Sigg}, {Silva}, {Singer}, {Singh}, {Singhal}, {Sintes}, {Slagmolen},
  {Smith}, {Smith}, {Smith}, {Somala}, {Son}, {Sonnenberg}, {Sorazu},
  {Sorrentino}, {Souradeep}, {Spencer}, {Srivastava}, {Staats}, {Staley},
  {Steinke}, {Steinlechner}, {Steinlechner}, {Steinmeyer}, {Stevenson},
  {Stone}, {Stops}, {Strain}, {Stratta}, {Strigin}, {Strunk}, {Sturani},
  {Stuver}, {Summerscales}, {Sun}, {Sunil}, {Suresh}, {Sutton}, {Swinkels},
  {Szczepa{\'n}czyk}, {Tacca}, {Tait}, {Talbot}, {Talukder}, {Tanner},
  {T{\'a}pai}, {Taracchini}, {Tasson}, {Taylor}, {Taylor}, {Tewari}, {Theeg},
  {Thies}, {Thomas}, {Thomas}, {Thomas}, {Thorne}, {Thorne}, {Thrane},
  {Tiwari}, {Tiwari}, {Tokmakov}, {Toland}, {Tonelli}, {Tornasi},
  {Torres-Forn{\'e}}, {Torrie}, {T{\"o}yr{\"a}}, {Travasso}, {Traylor},
  {Trinastic}, {Tringali}, {Trozzo}, {Tsang}, {Tse}, {Tso}, {Tsukada}, {Tsuna},
  {Tuyenbayev}, {Ueno}, {Ugolini}, {Unnikrishnan}, {Urban}, {Usman},
  {Vahlbruch}, {Vajente}, {Valdes}, {Vallisneri}, {van Bakel}, {van Beuzekom},
  {van den Brand}, {Van Den Broeck}, {Vander-Hyde}, {van der Schaaf}, {van
  Heijningen}, {van Veggel}, {Vardaro}, {Varma}, {Vass}, {Vas{\'u}th},
  {Vecchio}, {Vedovato}, {Veitch}, {Veitch}, {Venkateswara}, {Venugopalan},
  {Verkindt}, {Vetrano}, {Vicer{\'e}}, {Viets}, {Vinciguerra}, {Vine}, {Vinet},
  {Vitale}, {Vo}, {Vocca}, {Vorvick}, {Vyatchanin}, {Wade}, {Wade}, {Wade},
  {Walet}, {Walker}, {Wallace}, {Walsh}, {Wang}, {Wang}, {Wang}, {Wang},
  {Wang}, {Ward}, {Warner}, {Was}, {Watchi}, {Weaver}, {Wei}, {Weinert},
  {Weinstein}, {Weiss}, {Wen}, {Wessel}, {We{\ss}els}, {Westerweck},
  {Westphal}, {Wette}, {Whelan}, {Whitcomb}, {Whiting}, {Whittle}, {Wilken},
  {Williams}, {Williams}, {Williamson}, {Willis}, {Willke}, {Wimmer},
  {Winkler}, {Wipf}, {Wittel}, {Woan}, {Woehler}, {Wofford}, {Wong}, {Worden},
  {Wright}, {Wu}, {Wysocki}, {Xiao}, {Yamamoto}, {Yancey}, {Yang}, {Yap},
  {Yazback}, {Yu}, {Yu}, {Yvert}, {Zadro{\.Z}ny}, {Zanolin}, {Zelenova},
  {Zendri}, {Zevin}, {Zhang}, {Zhang}, {Zhang}, {Zhang}, {Zhao}, {Zhou},
  {Zhou}, {Zhu}, {Zhu}, {Zimmerman}, {Zucker}, {Zweizig}, {LIGO Scientific
  Collaboration}, \& {Virgo Collaboration}}]{LIGO2017}
{Abbott}, B.~P., {Abbott}, R., {Abbott}, T.~D., {et~al.} 2017,
  \hypersetup{urlcolor=magenta}\href{https://dx.doi.org/10.1103/PhysRevLett.119.161101}{\prl},
  \hypersetup{urlcolor=blue}\href{https://ui.adsabs.harvard.edu/abs/2017PhRvL.119p1101A}{119,
  161101}

\bibitem[{{Abbott} {et~al.}(2019){Abbott}, {Abbott}, {Abbott}, {Abraham},
  {Acernese}, {Ackley}, {Adams}, {Adhikari}, {Adya}, {Affeldt}, {Agathos},
  {Agatsuma}, {Aggarwal}, {Aguiar}, {Aiello}, {Ain}, {Ajith}, {Allen},
  {Allocca}, {Aloy}, {Altin}, {Amato}, {Ananyeva}, {Anderson}, {Anderson},
  {Angelova}, {Antier}, {Appert}, {Arai}, {Araya}, {Areeda}, {Ar{\`e}ne},
  {Arnaud}, {Arun}, {Ascenzi}, {Ashton}, {Aston}, {Astone}, {Aubin}, {Aufmuth},
  {AultONeal}, {Austin}, {Avendano}, {Avila-Alvarez}, {Babak}, {Bacon},
  {Badaracco}, {Bader}, {Bae}, {Baker}, {Baldaccini}, {Ballardin}, {Ballmer},
  {Banagiri}, {Barayoga}, {Barclay}, {Barish}, {Barker}, {Barkett}, {Barnum},
  {Barone}, {Barr}, {Barsotti}, {Barsuglia}, {Barta}, {Bartlett}, {Bartos},
  {Bassiri}, {Basti}, {Bawaj}, {Bayley}, {Bazzan}, {B{\'e}csy}, {Bejger},
  {Belahcene}, {Bell}, {Beniwal}, {Berger}, {Bergmann}, {Bernuzzi}, {Bero},
  {Berry}, {Bersanetti}, {Bertolini}, {Betzwieser}, {Bhandare}, {Bidler},
  {Bilenko}, {Bilgili}, {Billingsley}, {Birch}, {Birney}, {Birnholtz},
  {Biscans}, {Biscoveanu}, {Bisht}, {Bitossi}, {Bizouard}, {Blackburn},
  {Blair}, {Blair}, {Blair}, {Bloemen}, {Bode}, {Boer}, {Boetzel}, {Bogaert},
  {Bondu}, {Bonilla}, {Bonnand}, {Booker}, {Boom}, {Booth}, {Bork}, {Boschi},
  {Bose}, {Bossie}, {Bossilkov}, {Bosveld}, {Bouffanais}, {Bozzi},
  {Bradaschia}, {Brady}, {Bramley}, {Branchesi}, {Brau}, {Briant}, {Briggs},
  {Brighenti}, {Brillet}, {Brinkmann}, {Brisson}, {Brockill}, {Brooks},
  {Brown}, {Brunett}, {Buikema}, {Bulik}, {Bulten}, {Buonanno}, {Buscicchio},
  {Buskulic}, {Buy}, {Byer}, {Cabero}, {Cadonati}, {Cagnoli}, {Cahillane},
  {Calder{\'o}n Bustillo}, {Callister}, {Calloni}, {Camp}, {Campbell},
  {Canepa}, {Cannon}, {Cao}, {Cao}, {Capocasa}, {Carbognani}, {Caride},
  {Carney}, {Carullo}, {Casanueva Diaz}, {Casentini}, {Caudill},
  {Cavagli{\`a}}, {Cavalier}, {Cavalieri}, {Cella}, {Cerd{\'a}-Dur{\'a}n},
  {Cerretani}, {Cesarini}, {Chaibi}, {Chakravarti}, {Chamberlin}, {Chan},
  {Chao}, {Charlton}, {Chase}, {Chassande-Mottin}, {Chatterjee}, {Chaturvedi},
  {Chatziioannou}, {Cheeseboro}, {Chen}, {Chen}, {Chen}, {Cheng}, {Cheong},
  {Chia}, {Chincarini}, {Chiummo}, {Cho}, {Cho}, {Cho}, {Christensen}, {Chu},
  {Chua}, {Chung}, {Chung}, {Ciani}, {Ciobanu}, {Ciolfi}, {Cipriano}, {Cirone},
  {Clara}, {Clark}, {Clearwater}, {Cleva}, {Cocchieri}, {Coccia}, {Cohadon},
  {Cohen}, {Colgan}, {Colleoni}, {Collette}, {Collins}, {Cominsky},
  {Constancio}, {Conti}, {Cooper}, {Corban}, {Corbitt}, {Cordero-Carri{\'o}n},
  {Corley}, {Cornish}, {Corsi}, {Cortese}, {Costa}, {Cotesta}, {Coughlin},
  {Coughlin}, {Coulon}, {Countryman}, {Couvares}, {Covas}, {Cowan}, {Coward},
  {Cowart}, {Coyne}, {Coyne}, {Creighton}, {Creighton}, {Cripe}, {Croquette},
  {Crowder}, {Cullen}, {Cumming}, {Cunningham}, {Cuoco}, {Dal Canton},
  {D{\'a}lya}, {Danilishin}, {D'Antonio}, {Danzmann}, {Dasgupta}, {Da Silva
  Costa}, {Datrier}, {Dattilo}, {Dave}, {Davier}, {Davis}, {Daw}, {DeBra},
  {Deenadayalan}, {Degallaix}, {De Laurentis}, {Del{\'e}glise}, {Del Pozzo},
  {DeMarchi}, {Demos}, {Dent}, {De Pietri}, {Derby}, {De Rosa}, {De Rossi},
  {DeSalvo}, {de Varona}, {Dhurandhar}, {D{\'\i}az}, {Dietrich}, {Di Fiore},
  {Di Giovanni}, {Di Girolamo}, {Di Lieto}, {Ding}, {Di Pace}, {Di Palma}, {Di
  Renzo}, {Dmitriev}, {Doctor}, {Donovan}, {Dooley}, {Doravari}, {Dorrington},
  {Downes}, {Drago}, {Driggers}, {Du}, {Ducoin}, {Dupej}, {Dwyer}, {Easter},
  {Edo}, {Edwards}, {Effler}, {Ehrens}, {Eichholz}, {Eikenberry}, {Eisenmann},
  {Eisenstein}, {Essick}, {Estelles}, {Estevez}, {Etienne}, {Etzel}, {Evans},
  {Evans}, {Fafone}, {Fair}, {Fairhurst}, {Fan}, {Farinon}, {Farr}, {Farr},
  {Fauchon-Jones}, {Favata}, {Fays}, {Fazio}, {Fee}, {Feicht}, {Fejer}, {Feng},
  {Fernandez-Galiana}, {Ferrante}, {Ferreira}, {Ferreira}, {Ferrini},
  {Fidecaro}, {Fiori}, {Fiorucci}, {Fishbach}, {Fisher}, {Fishner},
  {Fitz-Axen}, {Flaminio}, {Fletcher}, {Flynn}, {Fong}, {Font}, {Forsyth},
  {Fournier}, {Frasca}, {Frasconi}, {Frei}, {Freise}, {Frey}, {Frey},
  {Fritschel}, {Frolov}, {Fulda}, {Fyffe}, {Gabbard}, {Gadre}, {Gaebel},
  {Gair}, {Gammaitoni}, {Ganija}, {Gaonkar}, {Garcia},
  {Garc{\'\i}a-Quir{\'o}s}, {Garufi}, {Gateley}, {Gaudio}, {Gaur}, {Gayathri},
  {Gemme}, {Genin}, {Gennai}, {George}, {George}, {Gergely}, {Germain},
  {Ghonge}, {Ghosh}, {Ghosh}, {Ghosh}, {Giacomazzo}, {Giaime}, {Giardina},
  {Giazotto}, {Gill}, {Giordano}, {Glover}, {Godwin}, {Goetz}, {Goetz},
  {Goncharov}, {Gonz{\'a}lez}, {Gonzalez Castro}, {Gopakumar}, {Gorodetsky},
  {Gossan}, {Gosselin}, {Gouaty}, {Grado}, {Graef}, {Granata}, {Grant}, {Gras},
  {Grassia}, {Gray}, {Gray}, {Greco}, {Green}, {Green}, {Gretarsson}, {Groot},
  {Grote}, {Grunewald}, {Gruning}, {Guidi}, {Gulati}, {Guo}, {Gupta}, {Gupta},
  {Gustafson}, {Gustafson}, {Haegel}, {Halim}, {Hall}, {Hall}, {Hamilton},
  {Hammond}, {Haney}, {Hanke}, {Hanks}, {Hanna}, {Hannam}, {Hannuksela},
  {Hanson}, {Hardwick}, {Haris}, {Harms}, {Harry}, {Harry}, {Haster},
  {Haughian}, {Hayes}, {Healy}, {Heidmann}, {Heintze}, {Heitmann}, {Hello},
  {Hemming}, {Hendry}, {Heng}, {Hennig}, {Heptonstall}, {Hernandez Vivanco},
  {Heurs}, {Hild}, {Hinderer}, {Hoak}, {Hochheim}, {Hofman}, {Holgado},
  {Holland}, {Holt}, {Holz}, {Hopkins}, {Horst}, {Hough}, {Howell}, {Hoy},
  {Hreibi}, {Huerta}, {Huet}, {Hughey}, {Hulko}, {Husa}, {Huttner},
  {Huynh-Dinh}, {Idzkowski}, {Iess}, {Ingram}, {Inta}, {Intini}, {Irwin},
  {Isa}, {Isac}, {Isi}, {Iyer}, {Izumi}, {Jacqmin}, {Jadhav}, {Jani},
  {Janthalur}, {Jaranowski}, {Jenkins}, {Jiang}, {Johnson}, {Jones}, {Jones},
  {Jones}, {Jonker}, {Ju}, {Junker}, {Kalaghatgi}, {Kalogera}, {Kamai},
  {Kandhasamy}, {Kang}, {Kanner}, {Kapadia}, {Karki}, {Karvinen}, {Kashyap},
  {Kasprzack}, {Katsanevas}, {Katsavounidis}, {Katzman}, {Kaufer}, {Kawabe},
  {Keerthana}, {K{\'e}f{\'e}lian}, {Keitel}, {Kennedy}, {Key}, {Khalili},
  {Khan}, {Khan}, {Khan}, {Khan}, {Khazanov}, {Khursheed}, {Kijbunchoo}, {Kim},
  {Kim}, {Kim}, {Kim}, {Kim}, {Kim}, {Kimball}, {King}, {King},
  {Kinley-Hanlon}, {Kirchhoff}, {Kissel}, {Kleybolte}, {Klika}, {Klimenko},
  {Knowles}, {Koch}, {Koehlenbeck}, {Koekoek}, {Koley}, {Kondrashov}, {Kontos},
  {Koper}, {Korobko}, {Korth}, {Kowalska}, {Kozak}, {Kringel}, {Krishnendu},
  {Kr{\'o}lak}, {Kuehn}, {Kumar}, {Kumar}, {Kumar}, {Kumar}, {Kuo}, {Kutynia},
  {Kwang}, {Lackey}, {Lai}, {Lam}, {Landry}, {Lane}, {Lang}, {Lange}, {Lantz},
  {Lanza}, {Lartaux-Vollard}, {Lasky}, {Laxen}, {Lazzarini}, {Lazzaro},
  {Leaci}, {Leavey}, {Lecoeuche}, {Lee}, {Lee}, {Lee}, {Lee}, {Lee}, {Lee},
  {Lehmann}, {Lenon}, {Leroy}, {Letendre}, {Levin}, {Li}, {Li}, {Li}, {Li},
  {Lin}, {Linde}, {Linker}, {Littenberg}, {Liu}, {Liu}, {Lo}, {Lockerbie},
  {London}, {Longo}, {Lorenzini}, {Loriette}, {Lormand}, {Losurdo}, {Lough},
  {Lousto}, {Lovelace}, {Lower}, {L{\"u}ck}, {Lumaca}, {Lundgren}, {Lynch},
  {Ma}, {Macas}, {Macfoy}, {MacInnis}, {Macleod}, {Macquet},
  {Maga{\~n}a-Sandoval}, {Maga{\~n}a Zertuche}, {Magee}, {Majorana},
  {Maksimovic}, {Malik}, {Man}, {Mandic}, {Mangano}, {Mansell}, {Manske},
  {Mantovani}, {Mapelli}, {Marchesoni}, {Marion}, {M{\'a}rka}, {M{\'a}rka},
  {Markakis}, {Markosyan}, {Markowitz}, {Maros}, {Marquina}, {Marsat},
  {Martelli}, {Martin}, {Martin}, {Martynov}, {Mason}, {Massera}, {Masserot},
  {Massinger}, {Masso-Reid}, {Mastrogiovanni}, {Matas}, {Matichard}, {Matone},
  {Mavalvala}, {Mazumder}, {McCann}, {McCarthy}, {McClelland}, {McCormick},
  {McCuller}, {McGuire}, {McIver}, {McManus}, {McRae}, {McWilliams}, {Meacher},
  {Meadors}, {Mehmet}, {Mehta}, {Meidam}, {Melatos}, {Mendell}, {Mercer},
  {Mereni}, {Merilh}, {Merzougui}, {Meshkov}, {Messenger}, {Messick},
  {Metzdorff}, {Meyers}, {Miao}, {Michel}, {Middleton}, {Mikhailov}, {Milano},
  {Miller}, {Miller}, {Millhouse}, {Mills}, {Milovich-Goff}, {Minazzoli},
  {Minenkov}, {Mishkin}, {Mishra}, {Mistry}, {Mitra}, {Mitrofanov},
  {Mitselmakher}, {Mittleman}, {Mo}, {Moffa}, {Mogushi}, {Mohapatra},
  {Montani}, {Moore}, {Moraru}, {Moreno}, {Morisaki}, {Mours}, {Mow-Lowry},
  {Mukherjee}, {Mukherjee}, {Mukherjee}, {Mukund}, {Mullavey}, {Munch},
  {Mu{\~n}iz}, {Muratore}, {Murray}, {Nagar}, {Nardecchia}, {Naticchioni},
  {Nayak}, {Neilson}, {Nelemans}, {Nelson}, {Nery}, {Neunzert}, {Ng}, {Ng},
  {Nguyen}, {Nichols}, {Nissanke}, {Nocera}, {North}, {Nuttall},
  {Obergaulinger}, {Oberling}, {O'Brien}, {O'Dea}, {Ogin}, {Oh}, {Oh}, {Ohme},
  {Ohta}, {Okada}, {Oliver}, {Oppermann}, {Oram}, {O'Reilly}, {Ormiston},
  {Ortega}, {O'Shaughnessy}, {Ossokine}, {Ottaway}, {Overmier}, {Owen}, {Pace},
  {Pagano}, {Page}, {Pai}, {Pai}, {Palamos}, {Palashov}, {Palomba},
  {Pal-Singh}, {Pan}, {Pang}, {Pang}, {Pankow}, {Pannarale}, {Pant},
  {Paoletti}, {Paoli}, {Parida}, {Parker}, {Pascucci}, {Pasqualetti},
  {Passaquieti}, {Passuello}, {Patil}, {Patricelli}, {Pearlstone}, {Pedersen},
  {Pedraza}, {Pedurand}, {Pele}, {Penn}, {Perez}, {Perreca}, {Pfeiffer},
  {Phelps}, {Phukon}, {Piccinni}, {Pichot}, {Piergiovanni}, {Pillant},
  {Pinard}, {Pirello}, {Pitkin}, {Poggiani}, {Pong}, {Ponrathnam}, {Popolizio},
  {Porter}, {Powell}, {Prajapati}, {Prasad}, {Prasai}, {Prasanna}, {Pratten},
  {Prestegard}, {Privitera}, {Prodi}, {Prokhorov}, {Puncken}, {Punturo},
  {Puppo}, {P{\"u}rrer}, {Qi}, {Quetschke}, {Quinonez}, {Quintero},
  {Quitzow-James}, {Raab}, {Radkins}, {Radulescu}, {Raffai}, {Raja}, {Rajan},
  {Rajbhandari}, {Rakhmanov}, {Ramirez}, {Ramos-Buades}, {Rana}, {Rao},
  {Rapagnani}, {Raymond}, {Razzano}, {Read}, {Regimbau}, {Rei}, {Reid},
  {Reitze}, {Ren}, {Ricci}, {Richardson}, {Richardson}, {Ricker}, {Riles},
  {Rizzo}, {Robertson}, {Robie}, {Robinet}, {Rocchi}, {Rolland}, {Rollins},
  {Roma}, {Romanelli}, {Romano}, {Romel}, {Romie}, {Rose}, {Rosi{\'n}ska},
  {Rosofsky}, {Ross}, {Rowan}, {R{\"u}diger}, {Ruggi}, {Rutins}, {Ryan},
  {Sachdev}, {Sadecki}, {Sakellariadou}, {Salconi}, {Saleem}, {Samajdar},
  {Sammut}, {Sanchez}, {Sanchez}, {Sanchis-Gual}, {Sandberg}, {Sanders},
  {Santiago}, {Sarin}, {Sassolas}, {Sathyaprakash}, {Saulson}, {Sauter},
  {Savage}, {Schale}, {Scheel}, {Scheuer}, {Schmidt}, {Schnabel}, {Schofield},
  {Sch{\"o}nbeck}, {Schreiber}, {Schulte}, {Schutz}, {Schwalbe}, {Scott},
  {Scott}, {Seidel}, {Sellers}, {Sengupta}, {Sennett}, {Sentenac}, {Sequino},
  {Sergeev}, {Setyawati}, {Shaddock}, {Shaffer}, {Shahriar}, {Shaner}, {Shao},
  {Sharma}, {Shawhan}, {Shen}, {Shink}, {Shoemaker}, {Shoemaker},
  {ShyamSundar}, {Siellez}, {Sieniawska}, {Sigg}, {Silva}, {Singer}, {Singh},
  {Singhal}, {Sintes}, {Sitmukhambetov}, {Skliris}, {Slagmolen},
  {Slaven-Blair}, {Smith}, {Smith}, {Somala}, {Son}, {Sorazu}, {Sorrentino},
  {Souradeep}, {Sowell}, {Spencer}, {Spera}, {Srivastava}, {Srivastava},
  {Staats}, {Stachie}, {Standke}, {Steer}, {Steinke}, {Steinlechner},
  {Steinlechner}, {Steinmeyer}, {Stevenson}, {Stocks}, {Stone}, {Stops},
  {Strain}, {Stratta}, {Strigin}, {Strunk}, {Sturani}, {Stuver}, {Sudhir},
  {Summerscales}, {Sun}, {Sunil}, {Suresh}, {Sutton}, {Swinkels},
  {Szczepa{\'n}czyk}, {Tacca}, {Tait}, {Talbot}, {Talukder}, {Tanner},
  {T{\'a}pai}, {Taracchini}, {Tasson}, {Taylor}, {Thies}, {Thomas}, {Thomas},
  {Thondapu}, {Thorne}, {Thrane}, {Tiwari}, {Tiwari}, {Tiwari}, {Toland},
  {Tonelli}, {Tornasi}, {Torres-Forn{\'e}}, {Torrie}, {T{\"o}yr{\"a}},
  {Travasso}, {Traylor}, {Tringali}, {Trovato}, {Trozzo}, {Trudeau}, {Tsang},
  {Tse}, {Tso}, {Tsukada}, {Tsuna}, {Tuyenbayev}, {Ueno}, {Ugolini},
  {Unnikrishnan}, {Urban}, {Usman}, {Vahlbruch}, {Vajente}, {Valdes}, {van
  Bakel}, {van Beuzekom}, {van den Brand}, {Van Den Broeck}, {Vander-Hyde},
  {van der Schaaf}, {van Heijningen}, {van Veggel}, {Vardaro}, {Varma}, {Vass},
  {Vas{\'u}th}, {Vecchio}, {Vedovato}, {Veitch}, {Veitch}, {Venkateswara},
  {Venugopalan}, {Verkindt}, {Vetrano}, {Vicer{\'e}}, {Viets}, {Vine}, {Vinet},
  {Vitale}, {Vo}, {Vocca}, {Vorvick}, {Vyatchanin}, {Wade}, {Wade}, {Wade},
  {Walet}, {Walker}, {Wallace}, {Walsh}, {Wang}, {Wang}, {Wang}, {Wang},
  {Wang}, {Ward}, {Warden}, {Warner}, {Was}, {Watchi}, {Weaver}, {Wei},
  {Weinert}, {Weinstein}, {Weiss}, {Wellmann}, {Wen}, {Wessel}, {We{\ss}els},
  {Westhouse}, {Wette}, {Whelan}, {Whiting}, {Whittle}, {Wilken}, {Williams},
  {Williamson}, {Willis}, {Willke}, {Wimmer}, {Winkler}, {Wipf}, {Wittel},
  {Woan}, {Woehler}, {Wofford}, {Worden}, {Wright}, {Wu}, {Wysocki}, {Xiao},
  {Yamamoto}, {Yancey}, {Yang}, {Yap}, {Yazback}, {Yeeles}, {Yu}, {Yu}, {Yuen},
  {Yvert}, {Zadro{\.z}ny}, {Zanolin}, {Zelenova}, {Zendri}, {Zevin}, {Zhang},
  {Zhang}, {Zhang}, {Zhao}, {Zhou}, {Zhou}, {Zhu}, {Zimmerman}, {Zlochower},
  {Zucker}, {Zweizig}, {LIGO Scientific Collaboration}, \& {Virgo
  Collaboration}}]{LIGO2019}
{Abbott}, B.~P., {Abbott}, R., {Abbott}, T.~D., {et~al.} 2019,
  \hypersetup{urlcolor=magenta}\href{https://dx.doi.org/10.3847/2041-8213/ab3800}{\apjl},
  \hypersetup{urlcolor=blue}\href{https://ui.adsabs.harvard.edu/abs/2019ApJ...882L..24A}{882,
  L24}

\bibitem[{{Abbott} {et~al.}(2021){Abbott}, {Abbott}, {Abraham}, {Acernese},
  {Ackley}, {Adams}, {Adams}, {Adhikari}, {Adya}, {Affeldt}, {Agarwal},
  {Agathos}, {Agatsuma}, {Aggarwal}, {Aguiar}, {Aiello}, {Ain}, {Ajith},
  {Akutsu}, {Aleman}, {Allen}, {Allocca}, {Altin}, {Amato}, {Anand},
  {Ananyeva}, {Anderson}, {Anderson}, {Ando}, {Angelova}, {Ansoldi}, {Antelis},
  {Antier}, {Appert}, {Arai}, {Arai}, {Arai}, {Araki}, {Araya}, {Araya},
  {Areeda}, {Ar{\`e}ne}, {Aritomi}, {Arnaud}, {Aronson}, {Arun}, {Asada},
  {Asali}, {Ashton}, {Aso}, {Aston}, {Astone}, {Aubin}, {Aufmuth}, {Aultoneal},
  {Austin}, {Babak}, {Badaracco}, {Bader}, {Bae}, {Bae}, {Baer}, {Bagnasco},
  {Bai}, {Baiotti}, {Baird}, {Bajpai}, {Ball}, {Ballardin}, {Ballmer}, {Bals},
  {Balsamo}, {Baltus}, {Banagiri}, {Bankar}, {Bankar}, {Barayoga}, {Barbieri},
  {Barish}, {Barker}, {Barneo}, {Barone}, {Barr}, {Barsotti}, {Barsuglia},
  {Barta}, {Bartlett}, {Barton}, {Bartos}, {Bassiri}, {Basti}, {Bawaj},
  {Bayley}, {Baylor}, {Bazzan}, {B{\'e}csy}, {Bedakihale}, {Bejger},
  {Belahcene}, {Benedetto}, {Beniwal}, {Benjamin}, {Benkel}, {Bennett},
  {Bentley}, {Benyaala}, {Bergamin}, {Berger}, {Bernuzzi}, {Berry},
  {Bersanetti}, {Bertolini}, {Betzwieser}, {Bhandare}, {Bhandari},
  {Bhattacharjee}, {Bhaumik}, {Bidler}, {Bilenko}, {Billingsley}, {Birney},
  {Birnholtz}, {Biscans}, {Bischi}, {Biscoveanu}, {Bisht}, {Biswas}, {Bitossi},
  {Bizouard}, {Blackburn}, {Blackman}, {Blair}, {Blair}, {Blair}, {Bobba},
  {Bode}, {Boer}, {Bogaert}, {Boldrini}, {Bondu}, {Bonilla}, {Bonnand},
  {Booker}, {Boom}, {Bork}, {Boschi}, {Bose}, {Bose}, {Bossilkov}, {Boudart},
  {Bouffanais}, {Bozzi}, {Bradaschia}, {Brady}, {Bramley}, {Branch},
  {Branchesi}, {Brau}, {Breschi}, {Briant}, {Briggs}, {Brillet}, {Brinkmann},
  {Brockill}, {Brooks}, {Brooks}, {Brown}, {Brunett}, {Bruno}, {Bruntz},
  {Bryant}, {Buikema}, {Bulik}, {Bulten}, {Buonanno}, {Buscicchio}, {Buskulic},
  {Byer}, {Cadonati}, {Caesar}, {Cagnoli}, {Cahillane}, {Cain}, {Calder{\'o}n
  Bustillo}, {Callaghan}, {Callister}, {Calloni}, {Camp}, {Canepa},
  {Cannavacciuolo}, {Cannon}, {Cao}, {Cao}, {Cao}, {Capocasa}, {Capote},
  {Carapella}, {Carbognani}, {Carlin}, {Carney}, {Carpinelli}, {Carullo},
  {Carver}, {Casanueva Diaz}, {Casentini}, {Castaldi}, {Caudill},
  {Cavagli{\`a}}, {Cavalier}, {Cavalieri}, {Cella}, {Cerd{\'a}-Dur{\'a}n},
  {Cesarini}, {Chaibi}, {Chakravarti}, {Champion}, {Chan}, {Chan}, {Chan},
  {Chan}, {Chandra}, {Chanial}, {Chao}, {Charlton}, {Chase},
  {Chassande-Mottin}, {Chatterjee}, {Chaturvedi}, {Chatziioannou}, {Chen},
  {Chen}, {Chen}, {Chen}, {Chen}, {Chen}, {Chen}, {Chen}, {Chen}, {Cheng},
  {Cheong}, {Cheung}, {Chia}, {Chiadini}, {Chiang}, {Chierici}, {Chincarini},
  {Chiofalo}, {Chiummo}, {Cho}, {Cho}, {Choate}, {Choudhary}, {Choudhary},
  {Christensen}, {Chu}, {Chu}, {Chu}, {Chua}, {Chung}, {Ciani}, {Ciecielag},
  {Cie{\'s}lar}, {Cifaldi}, {Ciobanu}, {Ciolfi}, {Cipriano}, {Cirone}, {Clara},
  {Clark}, {Clark}, {Clarke}, {Clearwater}, {Clesse}, {Cleva}, {Coccia},
  {Cohadon}, {Cohen}, {Cohen}, {Colleoni}, {Collette}, {Colpi}, {Compton},
  {Constancio}, {Conti}, {Cooper}, {Corban}, {Corbitt}, {Cordero-Carri{\'o}n},
  {Corezzi}, {Corley}, {Cornish}, {Corre}, {Corsi}, {Cortese}, {Costa},
  {Cotesta}, {Coughlin}, {Coughlin}, {Coulon}, {Countryman}, {Cousins},
  {Couvares}, {Covas}, {Coward}, {Cowart}, {Coyne}, {Coyne}, {Creighton},
  {Creighton}, {Criswell}, {Croquette}, {Crowder}, {Cudell}, {Cullen},
  {Cumming}, {Cummings}, {Cuoco}, {Cury{\l}o}, {Dal Canton}, {D{\'a}lya},
  {Dana}, {Daneshgaranbajastani}, {D'Angelo}, {Danilishin}, {D'Antonio},
  {Danzmann}, {Darsow-Fromm}, {Dasgupta}, {Datrier}, {Dattilo}, {Dave},
  {Davier}, {Davies}, {Davis}, {Daw}, {Dean}, {Debra}, {Deenadayalan},
  {Degallaix}, {de Laurentis}, {Del{\'e}glise}, {Del Favero}, {de Lillo}, {de
  Lillo}, {Del Pozzo}, {Demarchi}, {de Matteis}, {D'Emilio}, {Demos}, {Dent},
  {Depasse}, {de Pietri}, {De Rosa}, {de Rossi}, {Desalvo}, {de Simone},
  {Dhurandhar}, {D{\'\i}az}, {Diaz-Ortiz}, {Didio}, {Dietrich}, {di Fiore}, {di
  Fronzo}, {di Giorgio}, {di Giovanni}, {di Girolamo}, {di Lieto}, {Ding}, {di
  Pace}, {di Palma}, {di Renzo}, {Divakarla}, {Dmitriev}, {Doctor},
  {D'Onofrio}, {Donovan}, {Dooley}, {Doravari}, {Dorrington}, {Drago},
  {Driggers}, {Drori}, {Du}, {Ducoin}, {Dupej}, {Durante}, {D'Urso}, {Duverne},
  {Dwyer}, {Easter}, {Ebersold}, {Eddolls}, {Edelman}, {Edo}, {Edy}, {Effler},
  {Eguchi}, {Eichholz}, {Eikenberry}, {Eisenmann}, {Eisenstein}, {Ejlli},
  {Enomoto}, {Errico}, {Essick}, {Estell{\'e}s}, {Estevez}, {Etienne}, {Etzel},
  {Evans}, {Evans}, {Ewing}, {Fafone}, {Fair}, {Fairhurst}, {Fan}, {Farah},
  {Farinon}, {Farr}, {Farr}, {Farrow}, {Fauchon-Jones}, {Favata}, {Fays},
  {Fazio}, {Feicht}, {Fejer}, {Feng}, {Fenyvesi}, {Ferguson},
  {Fernandez-Galiana}, {Ferrante}, {Ferreira}, {Fidecaro}, {Figura}, {Fiori},
  {Fishbach}, {Fisher}, {Fittipaldi}, {Fiumara}, {Flaminio}, {Floden}, {Flynn},
  {Fong}, {Font}, {Fornal}, {Forsyth}, {Franke}, {Frasca}, {Frasconi},
  {Frederick}, {Frei}, {Freise}, {Frey}, {Fritschel}, {Frolov}, {Fronz{\'e}},
  {Fujii}, {Fujikawa}, {Fukunaga}, {Fukushima}, {Fulda}, {Fyffe}, {Gabbard},
  {Gadre}, {Gaebel}, {Gair}, {Gais}, {Galaudage}, {Gamba}, {Ganapathy},
  {Ganguly}, {Gao}, {Gaonkar}, {Garaventa}, {Garc{\'\i}a-N{\'u}{\~n}ez},
  {Garc{\'\i}a-Quir{\'o}s}, {Garufi}, {Gateley}, {Gaudio}, {Gayathri}, {Ge},
  {Gemme}, {Gennai}, {George}, {Gergely}, {Gewecke}, {Ghonge}, {Ghosh},
  {Ghosh}, {Ghosh}, {Ghosh}, {Ghosh}, {Giacomazzo}, {Giacoppo}, {Giaime},
  {Giardina}, {Gibson}, {Gier}, {Giesler}, {Giri}, {Gissi}, {Glanzer},
  {Gleckl}, {Godwin}, {Goetz}, {Goetz}, {Gohlke}, {Goncharov}, {Gonz{\'a}lez},
  {Gopakumar}, {Gosselin}, {Gouaty}, {Grace}, {Grado}, {Granata}, {Granata},
  {Grant}, {Gras}, {Grassia}, {Gray}, {Gray}, {Greco}, {Green}, {Green},
  {Gretarsson}, {Gretarsson}, {Griffith}, {Griffiths}, {Griggs}, {Grignani},
  {Grimaldi}, {Grimes}, {Grimm}, {Grote}, {Grunewald}, {Gruning}, {Guerrero},
  {Guidi}, {Guimaraes}, {Guix{\'e}}, {Gulati}, {Guo}, {Guo}, {Gupta}, {Gupta},
  {Gupta}, {Gustafson}, {Gustafson}, {Guzman}, {Ha}, {Haegel}, {Hagiwara},
  {Haino}, {Halim}, {Hall}, {Hamilton}, {Hammond}, {Han}, {Haney}, {Hanks},
  {Hanna}, {Hannam}, {Hannuksela}, {Hansen}, {Hansen}, {Hanson}, {Harder},
  {Hardwick}, {Haris}, {Harms}, {Harry}, {Harry}, {Hartwig}, {Hasegawa},
  {Haskell}, {Hasskew}, {Haster}, {Hattori}, {Haughian}, {Hayakawa}, {Hayama},
  {Hayes}, {Healy}, {Heidmann}, {Heintze}, {Heinze}, {Heinzel}, {Heitmann},
  {Hellman}, {Hello}, {Helmling-Cornell}, {Hemming}, {Hendry}, {Heng},
  {Hennes}, {Hennig}, {Hennig}, {Hernandez Vivanco}, {Heurs}, {Hild}, {Hill},
  {Himemoto}, {Hinderer}, {Hines}, {Hiranuma}, {Hirata}, {Hirose}, {Ho},
  {Hochheim}, {Hofman}, {Hohmann}, {Holgado}, {Holland}, {Hollows}, {Holmes},
  {Holt}, {Holz}, {Hong}, {Hopkins}, {Hough}, {Howell}, {Hoy}, {Hoyland},
  {Hreibi}, {Hsieh}, {Hsu}, {Huang}, {Huang}, {Huang}, {Huang}, {Huang},
  {Huang}, {H{\"u}bner}, {Huddart}, {Huerta}, {Hughey}, {Hui}, {Hui}, {Husa},
  {Huttner}, {Huxford}, {Huynh-Dinh}, {Ide}, {Idzkowski}, {Iess}, {Ikenoue},
  {Imam}, {Inayoshi}, {Inchauspe}, {Ingram}, {Inoue}, {Intini}, {Ioka}, {Isi},
  {Isleif}, {Ito}, {Itoh}, {Iyer}, {Izumi}, {Jaberianhamedan}, {Jacqmin},
  {Jadhav}, {Jadhav}, {James}, {Jan}, {Jani}, {Janssens}, {Janthalur},
  {Jaranowski}, {Jariwala}, {Jaume}, {Jenkins}, {Jeon}, {Jeunon}, {Jia},
  {Jiang}, {Jin}, {Johns}, {Jones}, {Jones}, {Jones}, {Jones}, {Jones},
  {Jonker}, {Ju}, {Jung}, {Jung}, {Junker}, {Kaihotsu}, {Kajita}, {Kakizaki},
  {Kalaghatgi}, {Kalogera}, {Kamai}, {Kamiizumi}, {Kanda}, {Kandhasamy},
  {Kang}, {Kanner}, {Kao}, {Kapadia}, {Kapasi}, {Karat}, {Karathanasis},
  {Karki}, {Kashyap}, {Kasprzack}, {Kastaun}, {Katsanevas}, {Katsavounidis},
  {Katzman}, {Kaur}, {Kawabe}, {Kawaguchi}, {Kawai}, {Kawasaki},
  {K{\'e}f{\'e}lian}, {Keitel}, {Key}, {Khadka}, {Khalili}, {Khan}, {Khan},
  {Khazanov}, {Khetan}, {Khursheed}, {Kijbunchoo}, {Kim}, {Kim}, {Kim}, {Kim},
  {Kim}, {Kim}, {Kimball}, {Kimura}, {King}, {Kinley-Hanlon}, {Kirchhoff},
  {Kissel}, {Kita}, {Kitazawa}, {Kleybolte}, {Klimenko}, {Knee}, {Knowles},
  {Knyazev}, {Koch}, {Koekoek}, {Kojima}, {Kokeyama}, {Koley}, {Kolitsidou},
  {Kolstein}, {Komori}, {Kondrashov}, {Kong}, {Kontos}, {Koper}, {Korobko},
  {Kotake}, {Kovalam}, {Kozak}, {Kozakai}, {Kozu}, {Kringel}, {Krishnendu},
  {Kr{\'o}lak}, {Kuehn}, {Kuei}, {Kumar}, {Kumar}, {Kumar}, {Kumar}, {Kume},
  {Kuns}, {Kuo}, {Kuo}, {Kuromiya}, {Kuroyanagi}, {Kusayanagi}, {Kwak},
  {Kwang}, {Laghi}, {Lalande}, {Lam}, {Lamberts}, {Landry}, {Landry}, {Lane},
  {Lang}, {Lange}, {Lantz}, {La Rosa}, {Lartaux-Vollard}, {Lasky}, {Laxen},
  {Lazzarini}, {Lazzaro}, {Leaci}, {Leavey}, {Lecoeuche}, {Lee}, {Lee}, {Lee},
  {Lee}, {Lee}, {Lee}, {Lehmann}, {Lema{\^\i}tre}, {Leon}, {Leonardi}, {Leroy},
  {Letendre}, {Levin}, {Leviton}, {Li}, {Li}, {Li}, {Li}, {Li}, {Li}, {Lin},
  {Lin}, {Lin}, {Lin}, {Lin}, {Linde}, {Linker}, {Linley}, {Littenberg}, {Liu},
  {Liu}, {Liu}, {Liu}, {Llorens-Monteagudo}, {Lo}, {Lockwood}, {Lollie},
  {London}, {Longo}, {Lopez}, {Lorenzini}, {Loriette}, {Lormand}, {Losurdo},
  {Lough}, {Lousto}, {Lovelace}, {L{\"u}ck}, {Lumaca}, {Lundgren}, {Luo},
  {Macas}, {Macinnis}, {MacLeod}, {MacMillan}, {Macquet}, {Maga{\~n}a
  Hernandez}, {Maga{\~n}a-Sandoval}, {Magazz{\`u}}, {Magee}, {Maggiore},
  {Majorana}, {Makarem}, {Maksimovic}, {Maliakal}, {Malik}, {Man}, {Mandic},
  {Mangano}, {Mango}, {Mansell}, {Manske}, {Mantovani}, {Mapelli},
  {Marchesoni}, {Marchio}, {Marion}, {Mark}, {M{\'a}rka}, {M{\'a}rka},
  {Markakis}, {Markosyan}, {Markowitz}, {Maros}, {Marquina}, {Marsat},
  {Martelli}, {Martin}, {Martin}, {Martinez}, {Martinez}, {Martinovic},
  {Martynov}, {Marx}, {Masalehdan}, {Mason}, {Massera}, {Masserot},
  {Massinger}, {Masso-Reid}, {Mastrogiovanni}, {Matas}, {Mateu-Lucena},
  {Matichard}, {Matiushechkina}, {Mavalvala}, {McCann}, {McCarthy},
  {McClelland}, {McClincy}, {McCormick}, {McCuller}, {McGhee}, {McGuire},
  {McIsaac}, {McIver}, {McManus}, {McRae}, {McWilliams}, {Meacher}, {Mehmet},
  {Mehta}, {Melatos}, {Melchor}, {Mendell}, {Menendez-Vazquez}, {Menoni},
  {Mercer}, {Mereni}, {Merfeld}, {Merilh}, {Merritt}, {Merzougui}, {Meshkov},
  {Messenger}, {Messick}, {Meyers}, {Meylahn}, {Mhaske}, {Miani}, {Miao},
  {Michaloliakos}, {Michel}, {Michimura}, {Middleton}, {Milano}, {Miller},
  {Millhouse}, {Mills}, {Milotti}, {Milovich-Goff}, {Minazzoli}, {Minenkov},
  {Mio}, {Mir}, {Mishkin}, {Mishra}, {Mishra}, {Mistry}, {Mitra}, {Mitrofanov},
  {Mitselmakher}, {Mittleman}, {Miyakawa}, {Miyamoto}, {Miyazaki}, {Miyo},
  {Miyoki}, {Mo}, {Mogushi}, {Mohapatra}, {Mohite}, {Molina}, {Molina-Ruiz},
  {Mondin}, {Montani}, {Moore}, {Moraru}, {Morawski}, {More}, {Moreno},
  {Moreno}, {Mori}, {Morisaki}, {Moriwaki}, {Mours}, {Mow-Lowry}, {Mozzon},
  {Muciaccia}, {Mukherjee}, {Mukherjee}, {Mukherjee}, {Mukherjee}, {Mukund},
  {Mullavey}, {Munch}, {Mu{\~n}iz}, {Murray}, {Musenich}, {Nadji}, {Nagano},
  {Nagano}, {Nagar}, {Nakamura}, {Nakano}, {Nakano}, {Nakashima}, {Nakayama},
  {Nardecchia}, {Narikawa}, {Naticchioni}, {Nayak}, {Nayak}, {Negishi}, {Neil},
  {Neilson}, {Nelemans}, {Nelson}, {Nery}, {Neunzert}, {Ng}, {Ng}, {Nguyen},
  {Nguyen}, {Nguyen}, {Nguyen Quynh}, {Ni}, {Nichols}, {Nishizawa}, {Nissanke},
  {Nocera}, {Noh}, {Norman}, {North}, {Nozaki}, {Nuttall}, {Oberling},
  {O'Brien}, {Obuchi}, {O'Dell}, {Ogaki}, {Oganesyan}, {Oh}, {Oh}, {Oh},
  {Ohashi}, {Ohishi}, {Ohkawa}, {Ohme}, {Ohta}, {Okada}, {Okutani}, {Okutomi},
  {Olivetto}, {Oohara}, {Ooi}, {Oram}, {O'Reilly}, {Ormiston}, {Ormsby},
  {Ortega}, {O'Shaughnessy}, {O'Shea}, {Oshino}, {Ossokine}, {Osthelder},
  {Otabe}, {Ottaway}, {Overmier}, {Pace}, {Pagano}, {Page}, {Pagliaroli},
  {Pai}, {Pai}, {Palamos}, {Palashov}, {Palomba}, {Pan}, {Panda}, {Pang},
  {Pang}, {Pankow}, {Pannarale}, {Pant}, {Paoletti}, {Paoli}, {Paolone},
  {Parisi}, {Park}, {Parker}, {Pascucci}, {Pasqualetti}, {Passaquieti},
  {Passuello}, {Patel}, {Patricelli}, {Payne}, {Pechsiri}, {Pedraza},
  {Pegoraro}, {Pele}, {Pe{\~n}a Arellano}, {Penn}, {Perego}, {Pereira},
  {Pereira}, {Perez}, {P{\'e}rigois}, {Perreca}, {Perri{\`e}s}, {Petermann},
  {Petterson}, {Pfeiffer}, {Pham}, {Phukon}, {Piccinni}, {Pichot},
  {Piendibene}, {Piergiovanni}, {Pierini}, {Pierro}, {Pillant}, {Pilo},
  {Pinard}, {Pinto}, {Piotrzkowski}, {Piotrzkowski}, {Pirello}, {Pitkin},
  {Placidi}, {Plastino}, {Pluchar}, {Poggiani}, {Polini}, {Pong}, {Ponrathnam},
  {Popolizio}, {Porter}, {Powell}, {Pracchia}, {Pradier}, {Prajapati},
  {Prasai}, {Prasanna}, {Pratten}, {Prestegard}, {Principe}, {Prodi},
  {Prokhorov}, {Prosposito}, {Prudenzi}, {Puecher}, {Punturo}, {Puosi},
  {Puppo}, {P{\"u}rrer}, {Qi}, {Quetschke}, {Quinonez}, {Quitzow-James},
  {Raab}, {Raaijmakers}, {Radkins}, {Radulesco}, {Raffai}, {Rail}, {Raja},
  {Rajan}, {Ramirez}, {Ramirez}, {Ramos-Buades}, {Rana}, {Rapagnani}, {Rapol},
  {Ratto}, {Ray}, {Raymond}, {Raza}, {Razzano}, {Read}, {Rees}, {Regimbau},
  {Rei}, {Reid}, {Reitze}, {Relton}, {Rettegno}, {Ricci}, {Richardson},
  {Richardson}, {Richardson}, {Ricker}, {Riemenschneider}, {Riles}, {Rizzo},
  {Robertson}, {Robie}, {Robinet}, {Rocchi}, {Rocha}, {Rodriguez},
  {Rodriguez-Soto}, {Rolland}, {Rollins}, {Roma}, {Romanelli}, {Romano},
  {Romel}, {Romero}, {Romero-Shaw}, {Romie}, {Rose}, {Rosi{\'n}ska},
  {Rosofsky}, {Ross}, {Rowan}, {Rowlinson}, {Roy}, {Roy}, {Rozza}, {Ruggi},
  {Ryan}, {Sachdev}, {Sadecki}, {Sadiq}, {Sago}, {Saito}, {Saito}, {Sakai},
  {Sakai}, {Sakellariadou}, {Sakuno}, {Salafia}, {Salconi}, {Saleem}, {Salemi},
  {Samajdar}, {Sanchez}, {Sanchez}, {Sanchez}, {Sanchis-Gual}, {Sanders},
  {Sanuy}, {Saravanan}, {Sarin}, {Sassolas}, {Satari}, {Sathyaprakash}, {Sato},
  {Sato}, {Sauter}, {Savage}, {Savant}, {Sawada}, {Sawant}, {Sawant}, {Sayah},
  {Schaetzl}, {Scheel}, {Scheuer}, {Schindler-Tyka}, {Schmidt}, {Schnabel},
  {Schneewind}, {Schofield}, {Sch{\"o}nbeck}, {Schulte}, {Schutz}, {Schwartz},
  {Scott}, {Scott}, {Seglar-Arroyo}, {Seidel}, {Sekiguchi}, {Sekiguchi},
  {Sellers}, {Sengupta}, {Sennett}, {Sentenac}, {Seo}, {Sequino}, {Sergeev},
  {Setyawati}, {Shaffer}, {Shahriar}, {Shams}, {Shao}, {Sharifi}, {Sharma},
  {Sharma}, {Shawhan}, {Shcheblanov}, {Shen}, {Shibagaki}, {Shikauchi},
  {Shimizu}, {Shimoda}, {Shimode}, {Shink}, {Shinkai}, {Shishido}, {Shoda},
  {Shoemaker}, {Shoemaker}, {Shukla}, {Shyamsundar}, {Sieniawska}, {Sigg},
  {Singer}, {Singh}, {Singh}, {Singha}, {Sintes}, {Sipala}, {Skliris},
  {Slagmolen}, {Slaven-Blair}, {Smetana}, {Smith}, {Smith}, {Somala}, {Somiya},
  {Son}, {Soni}, {Soni}, {Sorazu}, {Sordini}, {Sorrentino}, {Sorrentino},
  {Sotani}, {Soulard}, {Souradeep}, {Sowell}, {Spagnuolo}, {Spencer}, {Spera},
  {Srivastava}, {Srivastava}, {Staats}, {Stachie}, {Steer}, {Steinlechner},
  {Steinlechner}, {Stops}, {Stevenson}, {Stover}, {Strain}, {Strang},
  {Stratta}, {Strunk}, {Sturani}, {Stuver}, {S{\"u}dbeck}, {Sudhagar},
  {Sudhir}, {Sugimoto}, {Suh}, {Summerscales}, {Sun}, {Sun}, {Sunil}, {Sur},
  {Suresh}, {Sutton}, {Suzuki}, {Suzuki}, {Swinkels}, {Szczepa{\'n}czyk},
  {Szewczyk}, {Tacca}, {Tagoshi}, {Tait}, {Takahashi}, {Takahashi}, {Takamori},
  {Takano}, {Takeda}, {Takeda}, {Talbot}, {Tanaka}, {Tanaka}, {Tanaka},
  {Tanaka}, {Tanaka}, {Tanasijczuk}, {Tanioka}, {Tanner}, {Tao}, {Tapia},
  {Tapia San Martin}, {Tasson}, {Telada}, {Tenorio}, {Terkowski}, {Test},
  {Thirugnanasambandam}, {Thomas}, {Thomas}, {Thompson}, {Thondapu}, {Thorne},
  {Thrane}, {Tiwari}, {Tiwari}, {Tiwari}, {Toland}, {Tolley}, {Tomaru},
  {Tomigami}, {Tomura}, {Tonelli}, {Torres-Forn{\'e}}, {Torrie}, {Tosta E
  Melo}, {T{\"o}yr{\"a}}, {Trapananti}, {Travasso}, {Traylor}, {Tringali},
  {Tripathee}, {Troiano}, {Trovato}, {Trozzo}, {Trudeau}, {Tsai}, {Tsai},
  {Tsang}, {Tsang}, {Tsao}, {Tse}, {Tso}, {Tsubono}, {Tsuchida}, {Tsukada},
  {Tsuna}, {Tsutsui}, {Tsuzuki}, {Turconi}, {Tuyenbayev}, {Ubhi}, {Uchikata},
  {Uchiyama}, {Udall}, {Ueda}, {Uehara}, {Ueno}, {Ueshima}, {Ugolini},
  {Unnikrishnan}, {Uraguchi}, {Urban}, {Ushiba}, {Usman}, {Utina}, {Vahlbruch},
  {Vajente}, {Vajpeyi}, {Valdes}, {Valentini}, {Valsan}, {van Bakel}, {van
  Beuzekom}, {van den Brand}, {van den Broeck}, {Vander-Hyde}, {van der
  Schaaf}, {van Heijningen}, {Vanosky}, {van Putten}, {Vardaro}, {Vargas},
  {Varma}, {Vas{\'u}th}, {Vecchio}, {Vedovato}, {Veitch}, {Veitch},
  {Venkateswara}, {Venneberg}, {Venugopalan}, {Verkindt}, {Verma}, {Veske},
  {Vetrano}, {Vicer{\'e}}, {Viets}, {Villa-Ortega}, {Vinet}, {Vitale}, {Vo},
  {Vocca}, {von Reis}, {von Wrangel}, {Vorvick}, {Vyatchanin}, {Wade}, {Wade},
  {Wagner}, {Walet}, {Walker}, {Wallace}, {Wallace}, {Walsh}, {Wang}, {Wang},
  {Wang}, {Ward}, {Warner}, {Was}, {Washimi}, {Washington}, {Watchi}, {Weaver},
  {Wei}, {Weinert}, {Weinstein}, {Weiss}, {Weller}, {Wellmann}, {Wen},
  {We{\ss}els}, {Westhouse}, {Wette}, {Whelan}, {White}, {Whiting}, {Whittle},
  {Wilken}, {Williams}, {Williams}, {Williamson}, {Willis}, {Willke}, {Wilson},
  {Winkler}, {Wipf}, {Wlodarczyk}, {Woan}, {Woehler}, {Wofford}, {Wong}, {Wu},
  {Wu}, {Wu}, {Wu}, {Wysocki}, {Xiao}, {Xu}, {Yamada}, {Yamamoto}, {Yamamoto},
  {Yamamoto}, {Yamamoto}, {Yamashita}, {Yamazaki}, {Yang}, {Yang}, {Yang},
  {Yang}, {Yang}, {Yap}, {Yeeles}, {Yelikar}, {Ying}, {Yokogawa}, {Yokoyama},
  {Yokozawa}, {Yoon}, {Yoshioka}, {Yu}, {Yu}, {Yuzurihara}, {Zadro{\.z}ny},
  {Zanolin}, {Zappa}, {Zeidler}, {Zelenova}, {Zendri}, {Zevin}, {Zhan},
  {Zhang}, {Zhang}, {Zhang}, {Zhang}, {Zhang}, {Zhao}, {Zhao}, {Zhao}, {Zhao},
  {Zhou}, {Zhu}, {Zhu}, {Zimmerman}, {Zlochower}, {Zucker}, {Zweizig}, {Ligo
  Scientific Collaboration}, {VIRGO Collaboration}, \& {KAGRA
  Collaboration}}]{Abbott2021}
{Abbott}, R., {Abbott}, T.~D., {Abraham}, S., {et~al.} 2021,
  \hypersetup{urlcolor=magenta}\href{https://dx.doi.org/10.3847/2041-8213/ac082e}{\apjl},
  \hypersetup{urlcolor=blue}\href{https://ui.adsabs.harvard.edu/abs/2021ApJ...915L...5A}{915,
  L5}

\bibitem[{{Ade} {et~al.}(2019){Ade}, {Aguirre}, {Ahmed}, {Aiola}, {Ali},
  {Alonso}, {Alvarez}, {Arnold}, {Ashton}, {Austermann}, {Awan}, {Baccigalupi},
  {Baildon}, {Barron}, {Battaglia}, {Battye}, {Baxter}, {Bazarko}, {Beall},
  {Bean}, {Beck}, {Beckman}, {Beringue}, {Bianchini}, {Boada}, {Boettger},
  {Bond}, {Borrill}, {Brown}, {Bruno}, {Bryan}, {Calabrese}, {Calafut},
  {Calisse}, {Carron}, {Challinor}, {Chesmore}, {Chinone}, {Chluba}, {Cho},
  {Choi}, {Coppi}, {Cothard}, {Coughlin}, {Crichton}, {Crowley}, {Crowley},
  {Cukierman}, {D'Ewart}, {D{\"u}nner}, {de Haan}, {Devlin}, {Dicker},
  {Didier}, {Dobbs}, {Dober}, {Duell}, {Duff}, {Duivenvoorden}, {Dunkley},
  {Dusatko}, {Errard}, {Fabbian}, {Feeney}, {Ferraro}, {Flux{\`a}}, {Freese},
  {Frisch}, {Frolov}, {Fuller}, {Fuzia}, {Galitzki}, {Gallardo}, {Tomas Galvez
  Ghersi}, {Gao}, {Gawiser}, {Gerbino}, {Gluscevic}, {Goeckner-Wald}, {Golec},
  {Gordon}, {Gralla}, {Green}, {Grigorian}, {Groh}, {Groppi}, {Guan},
  {Gudmundsson}, {Han}, {Hargrave}, {Hasegawa}, {Hasselfield}, {Hattori},
  {Haynes}, {Hazumi}, {He}, {Healy}, {Henderson}, {Hervias-Caimapo}, {Hill},
  {Hill}, {Hilton}, {Hilton}, {Hincks}, {Hinshaw}, {Hlo{\v{z}}ek}, {Ho}, {Ho},
  {Howe}, {Huang}, {Hubmayr}, {Huffenberger}, {Hughes}, {Ijjas}, {Ikape},
  {Irwin}, {Jaffe}, {Jain}, {Jeong}, {Kaneko}, {Karpel}, {Katayama}, {Keating},
  {Kernasovskiy}, {Keskitalo}, {Kisner}, {Kiuchi}, {Klein}, {Knowles},
  {Koopman}, {Kosowsky}, {Krachmalnicoff}, {Kuenstner}, {Kuo}, {Kusaka},
  {Lashner}, {Lee}, {Lee}, {Leon}, {Leung}, {Lewis}, {Li}, {Li}, {Limon},
  {Linder}, {Lopez-Caraballo}, {Louis}, {Lowry}, {Lungu}, {Madhavacheril},
  {Mak}, {Maldonado}, {Mani}, {Mates}, {Matsuda}, {Maurin}, {Mauskopf}, {May},
  {McCallum}, {McKenney}, {McMahon}, {Meerburg}, {Meyers}, {Miller},
  {Mirmelstein}, {Moodley}, {Munchmeyer}, {Munson}, {Naess}, {Nati},
  {Navaroli}, {Newburgh}, {Nguyen}, {Niemack}, {Nishino}, {Orlowski-Scherer},
  {Page}, {Partridge}, {Peloton}, {Perrotta}, {Piccirillo}, {Pisano},
  {Poletti}, {Puddu}, {Puglisi}, {Raum}, {Reichardt}, {Remazeilles},
  {Rephaeli}, {Riechers}, {Rojas}, {Roy}, {Sadeh}, {Sakurai}, {Salatino},
  {Sathyanarayana Rao}, {Schaan}, {Schmittfull}, {Sehgal}, {Seibert}, {Seljak},
  {Sherwin}, {Shimon}, {Sierra}, {Sievers}, {Sikhosana}, {Silva-Feaver},
  {Simon}, {Sinclair}, {Siritanasak}, {Smith}, {Smith}, {Spergel}, {Staggs},
  {Stein}, {Stevens}, {Stompor}, {Suzuki}, {Tajima}, {Takakura}, {Teply},
  {Thomas}, {Thorne}, {Thornton}, {Trac}, {Tsai}, {Tucker}, {Ullom},
  {Vagnozzi}, {van Engelen}, {Van Lanen}, {Van Winkle}, {Vavagiakis},
  {Verg{\`e}s}, {Vissers}, {Wagoner}, {Walker}, {Ward}, {Westbrook},
  {Whitehorn}, {Williams}, {Williams}, {Wollack}, {Xu}, {Yu}, {Yu}, {Zago},
  {Zhang}, {Zhu}, \& {Simons Observatory Collaboration}}]{Ade2019}
{Ade}, P., {Aguirre}, J., {Ahmed}, Z., {et~al.} 2019,
  \hypersetup{urlcolor=magenta}\href{https://dx.doi.org/10.1088/1475-7516/2019/02/056}{\jcap},
  \hypersetup{urlcolor=blue}\href{https://ui.adsabs.harvard.edu/abs/2019JCAP...02..056A}{2019,
  056}

\bibitem[{{Alexander} {et~al.}(2022){Alexander}, {Battalia}, {Bhandarkar},
  {Clark}, {Devlin}, {Dicker}, {Haridas}, {Hill}, {Laskar}, {Mason},
  {Miller-Jones}, {Moravec}, {Mroczkowski}, {Naess}, {Pasham}, {Perez
  Sarmiento}, {Romero}, {Sarazin}, {Sievers}, \& {van Velzen}}]{Alexander2022}
{Alexander}, K., {Battalia}, N., {Bhandarkar}, T., {et~al.} 2022, The
  Astronomer's Telegram,
  \hypersetup{urlcolor=blue}\href{https://ui.adsabs.harvard.edu/abs/2022ATel15269....1A}{15269,
  1}

\bibitem[{{Alexander} {et~al.}(2020){Alexander}, {van Velzen}, {Horesh}, \&
  {Zauderer}}]{Alexander2020}
{Alexander}, K.~D., {van Velzen}, S., {Horesh}, A., \& {Zauderer}, B.~A. 2020,
  \hypersetup{urlcolor=magenta}\href{https://dx.doi.org/10.1007/s11214-020-00702-w}{\ssr},
  \hypersetup{urlcolor=blue}\href{https://ui.adsabs.harvard.edu/abs/2020SSRv..216...81A}{216,
  81}

\bibitem[{{Anderson} {et~al.}(2020){Anderson}, {Mooley}, {Hallinan}, {Dong},
  {Phinney}, {Horesh}, {Bourke}, {Cenko}, {Frail}, {Kulkarni}, \&
  {Myers}}]{Anderson2020}
{Anderson}, M.~M., {Mooley}, K.~P., {Hallinan}, G., {et~al.} 2020,
  \hypersetup{urlcolor=magenta}\href{https://dx.doi.org/10.3847/1538-4357/abb94b}{\apj},
  \hypersetup{urlcolor=blue}\href{https://ui.adsabs.harvard.edu/abs/2020ApJ...903..116A}{903,
  116}

\bibitem[{{Andreoni} {et~al.}(2022){Andreoni}, {Coughlin}, {Ahumada},
  {Kasliwal}, {Perley}, {Burns}, {Bulla}, {Cenko}, {Anand}, \&
  {Kool}}]{Andreoni2022}
{Andreoni}, I., {Coughlin}, M., {Ahumada}, T., {et~al.} 2022, GRB Coordinates
  Network,
  \hypersetup{urlcolor=blue}\href{https://ui.adsabs.harvard.edu/abs/2022GCN.31590....1A}{31590,
  1}

\bibitem[{{Arcavi} {et~al.}(2016){Arcavi}, {Wolf}, {Howell}, {Bildsten},
  {Leloudas}, {Hardin}, {Prajs}, {Perley}, {Svirski}, {Gal-Yam}, {Katz},
  {McCully}, {Cenko}, {Lidman}, {Sullivan}, {Valenti}, {Astier}, {Balland},
  {Carlberg}, {Conley}, {Fouchez}, {Guy}, {Pain}, {Palanque-Delabrouille},
  {Perrett}, {Pritchet}, {Regnault}, {Rich}, \&
  {Ruhlmann-Kleider}}]{Arcavi2016}
{Arcavi}, I., {Wolf}, W.~M., {Howell}, D.~A., {et~al.} 2016,
  \hypersetup{urlcolor=magenta}\href{https://dx.doi.org/10.3847/0004-637X/819/1/35}{\apj},
  \hypersetup{urlcolor=blue}\href{https://ui.adsabs.harvard.edu/abs/2016ApJ...819...35A}{819,
  35}

\bibitem[{{Astropy Collaboration} {et~al.}(2018){Astropy Collaboration},
  {Price-Whelan}, {Sip{\H{o}}cz}, {G{\"u}nther}, {Lim}, {Crawford}, {Conseil},
  {Shupe}, {Craig}, {Dencheva}, {Ginsburg}, {VanderPlas}, {Bradley},
  {P{\'e}rez-Su{\'a}rez}, {de Val-Borro}, {Aldcroft}, {Cruz}, {Robitaille},
  {Tollerud}, {Ardelean}, {Babej}, {Bach}, {Bachetti}, {Bakanov}, {Bamford},
  {Barentsen}, {Barmby}, {Baumbach}, {Berry}, {Biscani}, {Boquien}, {Bostroem},
  {Bouma}, {Brammer}, {Bray}, {Breytenbach}, {Buddelmeijer}, {Burke},
  {Calderone}, {Cano Rodr{\'\i}guez}, {Cara}, {Cardoso}, {Cheedella}, {Copin},
  {Corrales}, {Crichton}, {D'Avella}, {Deil}, {Depagne}, {Dietrich}, {Donath},
  {Droettboom}, {Earl}, {Erben}, {Fabbro}, {Ferreira}, {Finethy}, {Fox},
  {Garrison}, {Gibbons}, {Goldstein}, {Gommers}, {Greco}, {Greenfield},
  {Groener}, {Grollier}, {Hagen}, {Hirst}, {Homeier}, {Horton}, {Hosseinzadeh},
  {Hu}, {Hunkeler}, {Ivezi{\'c}}, {Jain}, {Jenness}, {Kanarek}, {Kendrew},
  {Kern}, {Kerzendorf}, {Khvalko}, {King}, {Kirkby}, {Kulkarni}, {Kumar},
  {Lee}, {Lenz}, {Littlefair}, {Ma}, {Macleod}, {Mastropietro}, {McCully},
  {Montagnac}, {Morris}, {Mueller}, {Mumford}, {Muna}, {Murphy}, {Nelson},
  {Nguyen}, {Ninan}, {N{\"o}the}, {Ogaz}, {Oh}, {Parejko}, {Parley}, {Pascual},
  {Patil}, {Patil}, {Plunkett}, {Prochaska}, {Rastogi}, {Reddy Janga},
  {Sabater}, {Sakurikar}, {Seifert}, {Sherbert}, {Sherwood-Taylor}, {Shih},
  {Sick}, {Silbiger}, {Singanamalla}, {Singer}, {Sladen}, {Sooley},
  {Sornarajah}, {Streicher}, {Teuben}, {Thomas}, {Tremblay}, {Turner},
  {Terr{\'o}n}, {van Kerkwijk}, {de la Vega}, {Watkins}, {Weaver}, {Whitmore},
  {Woillez}, {Zabalza}, \& {Astropy Contributors}}]{astropy2018}
{Astropy Collaboration}, {Price-Whelan}, A.~M., {Sip{\H{o}}cz}, B.~M., {et~al.}
  2018,
  \hypersetup{urlcolor=magenta}\href{https://dx.doi.org/10.3847/1538-3881/aabc4f}{\aj},
  \hypersetup{urlcolor=blue}\href{https://ui.adsabs.harvard.edu/abs/2018AJ....156..123A}{156,
  123}

\bibitem[{{Bender} {et~al.}(2018){Bender}, {Ade}, {Ahmed}, {Anderson}, {Avva},
  {Aylor}, {Barry}, {Basu Thakur}, {Benson}, {Bleem}, {Bocquet}, {Byrum},
  {Carlstrom}, {Carter}, {Cecil}, {Chang}, {Cho}, {Cliche}, {Crawford},
  {Cukierman}, {de Haan}, {Denison}, {Ding}, {Dobbs}, {Dodelson}, {Dutcher},
  {Everett}, {Foster}, {Gallicchio}, {Gilbert}, {Groh}, {Guns}, {Halverson},
  {Harke-Hosemann}, {Harrington}, {Henning}, {Hilton}, {Holder}, {Holzapfel},
  {Huang}, {Irwin}, {Jeong}, {Jonas}, {Jones}, {Khaire}, {Knox}, {Kofman},
  {Korman}, {Kubik}, {Kuhlmann}, {Kuo}, {Lee}, {Leitch}, {Lowitz}, {Meyer},
  {Michalik}, {Montgomery}, {Nadolski}, {Natoli}, {Ngyuen}, {Noble}, {Novosad},
  {Padin}, {Pan}, {Pearson}, {Posada}, {Quan}, {Raghunathan}, {Rahlin},
  {Reichardt}, {Ruhl}, {Sayre}, {Shirokoff}, {Smecher}, {Sobrin}, {Stark},
  {Story}, {Suzuki}, {Thompson}, {Tucker}, {Vale}, {Vanderlinde}, {Vieira},
  {Wang}, {Whitehorn}, {Wu}, {Yefremenko}, {Yoon}, \& {Young}}]{Bender2018}
{Bender}, A.~N., {Ade}, P.~A.~R., {Ahmed}, Z., {et~al.} 2018, in Society of
  Photo-Optical Instrumentation Engineers (SPIE) Conference Series, Vol. 10708,
  Millimeter, Submillimeter, and Far-Infrared Detectors and Instrumentation for
  Astronomy IX, ed. J.~{Zmuidzinas} \& J.-R. {Gao}, 1070803

\bibitem[{{Berger}(2014)}]{Berger2014}
{Berger}, E. 2014,
  \hypersetup{urlcolor=magenta}\href{https://dx.doi.org/10.1146/annurev-astro-081913-035926}{\araa},
  \hypersetup{urlcolor=blue}\href{https://ui.adsabs.harvard.edu/abs/2014ARA&A..52...43B}{52,
  43}

\bibitem[{{Berger} {et~al.}(2003){Berger}, {Soderberg}, {Frail}, \&
  {Kulkarni}}]{Berger2003}
{Berger}, E., {Soderberg}, A.~M., {Frail}, D.~A., \& {Kulkarni}, S.~R. 2003,
  \hypersetup{urlcolor=magenta}\href{https://dx.doi.org/10.1086/375158}{\apjl},
  \hypersetup{urlcolor=blue}\href{https://ui.adsabs.harvard.edu/abs/2003ApJ...587L...5B}{587,
  L5}

\bibitem[{{Berger} {et~al.}(2012){Berger}, {Zauderer}, {Pooley}, {Soderberg},
  {Sari}, {Brunthaler}, \& {Bietenholz}}]{Berger2012}
{Berger}, E., {Zauderer}, A., {Pooley}, G.~G., {et~al.} 2012,
  \hypersetup{urlcolor=magenta}\href{https://dx.doi.org/10.1088/0004-637X/748/1/36}{\apj},
  \hypersetup{urlcolor=blue}\href{https://ui.adsabs.harvard.edu/abs/2012ApJ...748...36B}{748,
  36}

\bibitem[{{Berger} {et~al.}(2000){Berger}, {Sari}, {Frail}, {Kulkarni},
  {Bertoldi}, {Peck}, {Menten}, {Shepherd}, {Moriarty-Schieven}, {Pooley},
  {Bloom}, {Diercks}, {Galama}, \& {Hurley}}]{Berger2000}
{Berger}, E., {Sari}, R., {Frail}, D.~A., {et~al.} 2000,
  \hypersetup{urlcolor=magenta}\href{https://dx.doi.org/10.1086/317814}{\apj},
  \hypersetup{urlcolor=blue}\href{https://ui.adsabs.harvard.edu/abs/2000ApJ...545...56B}{545,
  56}

\bibitem[{{Bloom} {et~al.}(2011){Bloom}, {Giannios}, {Metzger}, {Cenko},
  {Perley}, {Butler}, {Tanvir}, {Levan}, {O'Brien}, {Strubbe}, {De Colle},
  {Ramirez-Ruiz}, {Lee}, {Nayakshin}, {Quataert}, {King}, {Cucchiara},
  {Guillochon}, {Bower}, {Fruchter}, {Morgan}, \& {van der Horst}}]{Bloom2011}
{Bloom}, J.~S., {Giannios}, D., {Metzger}, B.~D., {et~al.} 2011,
  \hypersetup{urlcolor=magenta}\href{https://dx.doi.org/10.1126/science.1207150}{Science},
  \hypersetup{urlcolor=blue}\href{https://ui.adsabs.harvard.edu/abs/2011Sci...333..203B}{333,
  203}

\bibitem[{{Bock} {et~al.}(2009){Bock}, {Chandra}, {Frail}, \&
  {Kulkarni}}]{Bock2009}
{Bock}, D.~C.~J., {Chandra}, P., {Frail}, D.~A., \& {Kulkarni}, S.~R. 2009, GRB
  Coordinates Network,
  \hypersetup{urlcolor=blue}\href{https://ui.adsabs.harvard.edu/abs/2009GCN..9005....1B}{9005,
  1}

\bibitem[{{Bower} {et~al.}(2013){Bower}, {Metzger}, {Cenko}, {Silverman}, \&
  {Bloom}}]{Bower2013}
{Bower}, G.~C., {Metzger}, B.~D., {Cenko}, S.~B., {Silverman}, J.~M., \&
  {Bloom}, J.~S. 2013,
  \hypersetup{urlcolor=magenta}\href{https://dx.doi.org/10.1088/0004-637X/763/2/84}{\apj},
  \hypersetup{urlcolor=blue}\href{https://ui.adsabs.harvard.edu/abs/2013ApJ...763...84B}{763,
  84}

\bibitem[{{Bower} {et~al.}(2007){Bower}, {Saul}, {Bloom}, {Bolatto},
  {Filippenko}, {Foley}, \& {Perley}}]{Bower2007}
{Bower}, G.~C., {Saul}, D., {Bloom}, J.~S., {et~al.} 2007,
  \hypersetup{urlcolor=magenta}\href{https://dx.doi.org/10.1086/519831}{\apj},
  \hypersetup{urlcolor=blue}\href{https://ui.adsabs.harvard.edu/abs/2007ApJ...666..346B}{666,
  346}

\bibitem[{{Bremer} {et~al.}(1998){Bremer}, {Krichbaum}, {Galama},
  {Castro-Tirado}, {Frontera}, {van Paradijs}, {Mirabel}, \&
  {Costa}}]{Bremer1998}
{Bremer}, M., {Krichbaum}, T.~P., {Galama}, T.~J., {et~al.} 1998, \aap,
  \hypersetup{urlcolor=blue}\href{https://ui.adsabs.harvard.edu/abs/1998A&A...332L..13B}{332,
  L13}

\bibitem[{{Brown} {et~al.}(2015){Brown}, {Levan}, {Stanway}, {Tanvir}, {Cenko},
  {Berger}, {Chornock}, \& {Cucchiaria}}]{Brown2015}
{Brown}, G.~C., {Levan}, A.~J., {Stanway}, E.~R., {et~al.} 2015,
  \hypersetup{urlcolor=magenta}\href{https://dx.doi.org/10.1093/mnras/stv1520}{\mnras},
  \hypersetup{urlcolor=blue}\href{https://ui.adsabs.harvard.edu/abs/2015MNRAS.452.4297B}{452,
  4297}

\bibitem[{{Brown} {et~al.}(2017){Brown}, {Levan}, {Stanway}, {Kr{\"u}hler},
  {Tanvir}, {Davies}, {Fruchter}, {Cenko}, \& {Metzger}}]{Brown2017}
{Brown}, G.~C., {Levan}, A.~J., {Stanway}, E.~R., {et~al.} 2017,
  \hypersetup{urlcolor=magenta}\href{https://dx.doi.org/10.1093/mnras/stx2193}{\mnras},
  \hypersetup{urlcolor=blue}\href{https://ui.adsabs.harvard.edu/abs/2017MNRAS.472.4469B}{472,
  4469}

\bibitem[{{Burrows} {et~al.}(2011){Burrows}, {Kennea}, {Ghisellini}, {Mangano},
  {Zhang}, {Page}, {Eracleous}, {Romano}, {Sakamoto}, {Falcone}, {Osborne},
  {Campana}, {Beardmore}, {Breeveld}, {Chester}, {Corbet}, {Covino},
  {Cummings}, {D'Avanzo}, {D'Elia}, {Esposito}, {Evans}, {Fugazza}, {Gelbord},
  {Hiroi}, {Holland}, {Huang}, {Im}, {Israel}, {Jeon}, {Jeon}, {Jun}, {Kawai},
  {Kim}, {Krimm}, {Marshall}, {P. M{\'e}sz{\'a}ros}, {Negoro}, {Omodei},
  {Park}, {Perkins}, {Sugizaki}, {Sung}, {Tagliaferri}, {Troja}, {Ueda},
  {Urata}, {Usui}, {Antonelli}, {Barthelmy}, {Cusumano}, {Giommi}, {Melandri},
  {Perri}, {Racusin}, {Sbarufatti}, {Siegel}, \& {Gehrels}}]{Burrows2011}
{Burrows}, D.~N., {Kennea}, J.~A., {Ghisellini}, G., {et~al.} 2011,
  \hypersetup{urlcolor=magenta}\href{https://dx.doi.org/10.1038/nature10374}{\nat},
  \hypersetup{urlcolor=blue}\href{https://ui.adsabs.harvard.edu/abs/2011Natur.476..421B}{476,
  421}

\bibitem[{{Carilli} \& {Yun}(1999)}]{Carilli1999}
{Carilli}, C.~L., \& {Yun}, M.~S. 1999,
  \hypersetup{urlcolor=magenta}\href{https://dx.doi.org/10.1086/311909}{\apjl},
  \hypersetup{urlcolor=blue}\href{https://ui.adsabs.harvard.edu/abs/1999ApJ...513L..13C}{513,
  L13}

\bibitem[{{Castro-Tirado} {et~al.}(2007){Castro-Tirado}, {Bremer}, {McBreen},
  {Gorosabel}, {Guziy}, {Fakthullin}, {Sokolov}, {Gonz{\'a}lez Delgado},
  {Bihain}, {Pandey}, {Jel{\'\i}nek}, {de Ugarte Postigo}, {Misra}, {Sagar},
  {Bama}, {Kamble}, {Anupama}, {Licandro}, {P{\'e}rez-Ram{\'\i}rez},
  {Bhattacharya}, {Aceituno}, \& {Neri}}]{Castro2007}
{Castro-Tirado}, A.~J., {Bremer}, M., {McBreen}, S., {et~al.} 2007,
  \hypersetup{urlcolor=magenta}\href{https://dx.doi.org/10.1051/0004-6361:20066748}{\aap},
  \hypersetup{urlcolor=blue}\href{https://ui.adsabs.harvard.edu/abs/2007A&A...475..101C}{475,
  101}

\bibitem[{{Castro-Tirado}
  {et~al.}(2009{\natexlab{\hspace{0pt}a}}){Castro-Tirado}, {Bremer}, {Winters},
  {Gorosabel}, {Guziy}, {Jelinek}, {Kubanek}, {de Ugarte Postigo}, {Santiago},
  \& {Perez-Ramirez}}]{Castro2009a}
{Castro-Tirado}, A.~J., {Bremer}, M., {Winters}, J.~M., {et~al.}
  2009{\natexlab{\hspace{0pt}a}}, GRB Coordinates Network,
  \hypersetup{urlcolor=blue}\href{https://ui.adsabs.harvard.edu/abs/2009GCN..9273....1C}{9273,
  1}

\bibitem[{{Castro-Tirado}
  {et~al.}(2009{\natexlab{\hspace{0pt}b}}){Castro-Tirado}, {Bremer}, {Winters},
  {Gorosabel}, {Guziy}, {Jelinek}, {Kubanek}, {de Ugarte Postigo}, {Santiago},
  \& {Perez-Ramirez}}]{Castro2009b}
{Castro-Tirado}, A.~J., {Bremer}, M., {Winters}, J.~M., {et~al.}
  2009{\natexlab{\hspace{0pt}b}}, GRB Coordinates Network,
  \hypersetup{urlcolor=blue}\href{https://ui.adsabs.harvard.edu/abs/2009GCN..9273....1C}{9273,
  1}

\bibitem[{{Cendes} {et~al.}(2021{\natexlab{\hspace{0pt}a}}){Cendes},
  {Alexander}, {Berger}, {Eftekhari}, {Williams}, \& {Chornock}}]{Cendes2021a}
{Cendes}, Y., {Alexander}, K.~D., {Berger}, E., {et~al.}
  2021{\natexlab{\hspace{0pt}a}}, arXiv e-prints,
  \hypersetup{urlcolor=magenta}\href{https://arxiv.org/abs/2103.06299}{arXiv}{:}\hypersetup{urlcolor=blue}\href{https://ui.adsabs.harvard.edu/abs/2021arXiv210306299C}{2103.06299}

\bibitem[{{Cendes} {et~al.}(2021{\natexlab{\hspace{0pt}b}}){Cendes},
  {Eftekhari}, {Berger}, \& {Polisensky}}]{Cendes2021b}
{Cendes}, Y., {Eftekhari}, T., {Berger}, E., \& {Polisensky}, E.
  2021{\natexlab{\hspace{0pt}b}},
  \hypersetup{urlcolor=magenta}\href{https://dx.doi.org/10.3847/1538-4357/abd323}{\apj},
  \hypersetup{urlcolor=blue}\href{https://ui.adsabs.harvard.edu/abs/2021ApJ...908..125C}{908,
  125}

\bibitem[{{Cenko} {et~al.}(2012){Cenko}, {Krimm}, {Horesh}, {Rau}, {Frail},
  {Kennea}, {Levan}, {Holland}, {Butler}, {Quimby}, {Bloom}, {Filippenko},
  {Gal-Yam}, {Greiner}, {Kulkarni}, {Ofek}, {Olivares E.}, {Schady},
  {Silverman}, {Tanvir}, \& {Xu}}]{Cenko2012}
{Cenko}, S.~B., {Krimm}, H.~A., {Horesh}, A., {et~al.} 2012,
  \hypersetup{urlcolor=magenta}\href{https://dx.doi.org/10.1088/0004-637X/753/1/77}{\apj},
  \hypersetup{urlcolor=blue}\href{https://ui.adsabs.harvard.edu/abs/2012ApJ...753...77C}{753,
  77}

\bibitem[{{Chakraborti} {et~al.}(2013){Chakraborti}, {Petitpas}, {Zauderer},
  {Soderberg}, {Kamble}, {Margutti}, {Milisavljevic}, {Drout}, \&
  {Sanders}}]{sn2013ak}
{Chakraborti}, S., {Petitpas}, G., {Zauderer}, A., {et~al.} 2013, The
  Astronomer's Telegram,
  \hypersetup{urlcolor=blue}\href{https://ui.adsabs.harvard.edu/abs/2013ATel.4947....1C}{4947,
  1}

\bibitem[{{Chandra} {et~al.}(2008){Chandra}, {Cenko}, {Frail}, {Chevalier},
  {Macquart}, {Kulkarni}, {Bock}, {Bertoldi}, {Kasliwal}, {Fox}, {Price},
  {Berger}, {Soderberg}, {Harrison}, {Gal-Yam}, {Ofek}, {Rau}, {Schmidt},
  {Cameron}, {Cowie}, {Cowie}, {Roth}, {Dopita}, {Peterson}, \&
  {Penprase}}]{Chandra2008}
{Chandra}, P., {Cenko}, S.~B., {Frail}, D.~A., {et~al.} 2008,
  \hypersetup{urlcolor=magenta}\href{https://dx.doi.org/10.1086/589807}{\apj},
  \hypersetup{urlcolor=blue}\href{https://ui.adsabs.harvard.edu/abs/2008ApJ...683..924C}{683,
  924}

\bibitem[{{Chen} {et~al.}(2021){Chen}, {Cowperthwaite}, {Metzger}, \&
  {Berger}}]{Chen2021}
{Chen}, H.-Y., {Cowperthwaite}, P.~S., {Metzger}, B.~D., \& {Berger}, E. 2021,
  \hypersetup{urlcolor=magenta}\href{https://dx.doi.org/10.3847/2041-8213/abdab0}{\apjl},
  \hypersetup{urlcolor=blue}\href{https://ui.adsabs.harvard.edu/abs/2021ApJ...908L...4C}{908,
  L4}

\bibitem[{{Chen} {et~al.}(2020){Chen}, {Urata}, {Huang}, {Takahashi},
  {Petitpas}, \& {Asada}}]{Chen2020}
{Chen}, W.~J., {Urata}, Y., {Huang}, K., {et~al.} 2020,
  \hypersetup{urlcolor=magenta}\href{https://dx.doi.org/10.3847/2041-8213/ab76d4}{\apjl},
  \hypersetup{urlcolor=blue}\href{https://ui.adsabs.harvard.edu/abs/2020ApJ...891L..15C}{891,
  L15}

\bibitem[{{CHIME/FRB Collaboration} {et~al.}(2019){CHIME/FRB Collaboration},
  {Amiri}, {Bandura}, {Bhardwaj}, {Boubel}, {Boyce}, {Boyle}, {Brar},
  {Burhanpurkar}, {Chawla}, {Cliche}, {Cubranic}, {Deng}, {Denman}, {Dobbs},
  {Fandino}, {Fonseca}, {Gaensler}, {Gilbert}, {Giri}, {Good}, {Halpern},
  {Hanna}, {Hill}, {Hinshaw}, {H{\"o}fer}, {Josephy}, {Kaspi}, {Landecker},
  {Lang}, {Masui}, {Mckinven}, {Mena-Parra}, {Merryfield}, {Milutinovic},
  {Moatti}, {Naidu}, {Newburgh}, {Ng}, {Patel}, {Pen}, {Pinsonneault-Marotte},
  {Pleunis}, {Rafiei-Ravandi}, {Ransom}, {Renard}, {Scholz}, {Shaw}, {Siegel},
  {Smith}, {Stairs}, {Tendulkar}, {Tretyakov}, {Vanderlinde}, \&
  {Yadav}}]{CHIME2019}
{CHIME/FRB Collaboration}, {Amiri}, M., {Bandura}, K., {et~al.} 2019,
  \hypersetup{urlcolor=magenta}\href{https://dx.doi.org/10.1038/s41586-018-0867-7}{\nat},
  \hypersetup{urlcolor=blue}\href{https://ui.adsabs.harvard.edu/abs/2019Natur.566..230C}{566,
  230}

\bibitem[{{CHIME/FRB Collaboration} {et~al.}(2021){CHIME/FRB Collaboration},
  {:}, {Amiri}, {Andersen}, {Bandura}, {Berger}, {Bhardwaj}, {Boyce}, {Boyle},
  {Brar}, {Breitman}, {Cassanelli}, {Chawla}, {Chen}, {Cliche}, {Cook},
  {Cubranic}, {Curtin}, {Deng}, {Dobbs}, {Fengqiu}, {Dong}, {Eadie}, {Fandino},
  {Fonseca}, {Gaensler}, {Giri}, {Good}, {Halpern}, {Hill}, {Hinshaw},
  {Josephy}, {Kaczmarek}, {Kader}, {Kania}, {Kaspi}, {Landecker}, {Lang},
  {Leung}, {Li}, {Lin}, {Masui}, {Mckinven}, {Mena-Parra}, {Merryfield},
  {Meyers}, {Michilli}, {Milutinovic}, {Mirhosseini}, {M{\"u}nchmeyer},
  {Naidu}, {Newburgh}, {Ng}, {Patel}, {Pen}, {Petroff}, {Pinsonneault-Marotte},
  {Pleunis}, {Rafiei-Ravandi}, {Rahman}, {Ransom}, {Renard}, {Sanghavi},
  {Scholz}, {Shaw}, {Shin}, {Siegel}, {Sikora}, {Singh}, {Smith}, {Stairs},
  {Tan}, {Tendulkar}, {Vanderlinde}, {Wang}, {Wulf}, \& {Zwaniga}}]{CHIME2021}
{CHIME/FRB Collaboration}, {:}, {Amiri}, M., {et~al.} 2021, arXiv e-prints,
  \hypersetup{urlcolor=magenta}\href{https://arxiv.org/abs/2106.04352}{arXiv}{:}\hypersetup{urlcolor=blue}\href{https://ui.adsabs.harvard.edu/abs/2021arXiv210604352T}{2106.04352}

\bibitem[{{Coppejans} {et~al.}(2020){Coppejans}, {Margutti}, {Terreran},
  {Nayana}, {Coughlin}, {Laskar}, {Alexander}, {Bietenholz}, {Caprioli},
  {Chandra}, {Drout}, {Frederiks}, {Frohmaier}, {Hurley}, {Kochanek},
  {MacLeod}, {Meisner}, {Nugent}, {Ridnaia}, {Sand}, {Svinkin}, {Ward}, {Yang},
  {Baldeschi}, {Chilingarian}, {Dong}, {Esquivia}, {Fong}, {Guidorzi},
  {Lundqvist}, {Milisavljevic}, {Paterson}, {Reichart}, {Shappee}, {Stroh},
  {Valenti}, {Zauderer}, \& {Zhang}}]{Coppejans2020}
{Coppejans}, D.~L., {Margutti}, R., {Terreran}, G., {et~al.} 2020,
  \hypersetup{urlcolor=magenta}\href{https://dx.doi.org/10.3847/2041-8213/ab8cc7}{\apjl},
  \hypersetup{urlcolor=blue}\href{https://ui.adsabs.harvard.edu/abs/2020ApJ...895L..23C}{895,
  L23}

\bibitem[{{Corsi} \& {M{\'e}sz{\'a}ros}(2009)}]{Corsi2009}
{Corsi}, A., \& {M{\'e}sz{\'a}ros}, P. 2009,
  \hypersetup{urlcolor=magenta}\href{https://dx.doi.org/10.1088/0004-637X/702/2/1171}{\apj},
  \hypersetup{urlcolor=blue}\href{https://ui.adsabs.harvard.edu/abs/2009ApJ...702.1171C}{702,
  1171}

\bibitem[{{Corsi} {et~al.}(2014){Corsi}, {Ofek}, {Gal-Yam}, {Frail},
  {Kulkarni}, {Fox}, {Kasliwal}, {Sullivan}, {Horesh}, {Carpenter}, {Maguire},
  {Arcavi}, {Cenko}, {Cao}, {Mooley}, {Pan}, {Sesar}, {Sternberg}, {Xu},
  {Bersier}, {James}, {Bloom}, \& {Nugent}}]{Corsi2014}
{Corsi}, A., {Ofek}, E.~O., {Gal-Yam}, A., {et~al.} 2014,
  \hypersetup{urlcolor=magenta}\href{https://dx.doi.org/10.1088/0004-637X/782/1/42}{\apj},
  \hypersetup{urlcolor=blue}\href{https://ui.adsabs.harvard.edu/abs/2014ApJ...782...42C}{782,
  42}

\bibitem[{{de Ugarte Postigo} {et~al.}(2005){de Ugarte Postigo},
  {Castro-Tirado}, {Gorosabel}, {J{\'o}hannesson}, {Bj{\"o}rnsson},
  {Gudmundsson}, {Bremer}, {Pak}, {Tanvir}, {Castro Cer{\'o}n}, {Guzyi},
  {Jel{\'\i}nek}, {Klose}, {P{\'e}rez-Ram{\'\i}rez}, {Aceituno}, {Campo
  Bagat{\'\i}n}, {Covino}, {Cardiel}, {Fathkullin}, {Henden}, {Huferath},
  {Kurata}, {Malesani}, {Mannucci}, {Ruiz-Lapuente}, {Sokolov}, {Thiele},
  {Wisotzki}, {Antonelli}, {Bartolini}, {Boattini}, {Guarnieri}, {Piccioni},
  {Pizzichini}, {del Principe}, {di Paola}, {Fugazza}, {Ghisellini}, {Hunt},
  {Konstantinova}, {Masetti}, {Palazzi}, {Pian}, {Stefanon}, {Testa}, \&
  {Tristram}}]{deUgarte2005}
{de Ugarte Postigo}, A., {Castro-Tirado}, A.~J., {Gorosabel}, J., {et~al.}
  2005,
  \hypersetup{urlcolor=magenta}\href{https://dx.doi.org/10.1051/0004-6361:20052898}{\aap},
  \hypersetup{urlcolor=blue}\href{https://ui.adsabs.harvard.edu/abs/2005A&A...443..841D}{443,
  841}

\bibitem[{{de Ugarte Postigo} {et~al.}(2012){de Ugarte Postigo}, {Lundgren},
  {Mart{\'\i}n}, {Garcia-Appadoo}, {de Gregorio Monsalvo}, {Peck},
  {Micha{\l}owski}, {Th{\"o}ne}, {Campana}, {Gorosabel}, {Tanvir}, {Wiersema},
  {Castro-Tirado}, {Schulze}, {De Breuck}, {Petitpas}, {Hjorth}, {Jakobsson},
  {Covino}, {Fynbo}, {Winters}, {Bremer}, {Levan}, {Llorente},
  {S{\'a}nchez-Ram{\'\i}rez}, {Tello}, \& {Salvaterra}}]{deUgarte2012}
{de Ugarte Postigo}, A., {Lundgren}, A., {Mart{\'\i}n}, S., {et~al.} 2012,
  \hypersetup{urlcolor=magenta}\href{https://dx.doi.org/10.1051/0004-6361/201117848}{\aap},
  \hypersetup{urlcolor=blue}\href{https://ui.adsabs.harvard.edu/abs/2012A&A...538A..44D}{538,
  A44}

\bibitem[{{de Ugarte Postigo} {et~al.}(2018{\natexlab{\hspace{0pt}a}}){de
  Ugarte Postigo}, {Th{\"o}ne}, {Bensch}, {van der Horst}, {Kann}, {Cano},
  {Izzo}, {Goldoni}, {Mart{\'\i}n}, {Filgas}, {Schady}, {Gorosabel}, {Bikmaev},
  {Bremer}, {Burenin}, {Castro-Tirado}, {Covino}, {Fynbo}, {Garcia-Appadoo},
  {de Gregorio-Monsalvo}, {Jel{\'\i}nek}, {Khamitov}, {Kamble}, {Kouveliotou},
  {Kr{\"u}hler}, {Leloudas}, {Melnikov}, {Nardini}, {Perley}, {Petitpas},
  {Pooley}, {Rau}, {Rol}, {S{\'a}nchez-Ram{\'\i}rez}, {Starling}, {Tanvir},
  {Wiersema}, {Wijers}, \& {Zafar}}]{deUgarte2018}
{de Ugarte Postigo}, A., {Th{\"o}ne}, C.~C., {Bensch}, K., {et~al.}
  2018{\natexlab{\hspace{0pt}a}},
  \hypersetup{urlcolor=magenta}\href{https://dx.doi.org/10.1051/0004-6361/201833636}{\aap},
  \hypersetup{urlcolor=blue}\href{https://ui.adsabs.harvard.edu/abs/2018A&A...620A.190D}{620,
  A190}

\bibitem[{{de Ugarte Postigo} {et~al.}(2018{\natexlab{\hspace{0pt}b}}){de
  Ugarte Postigo}, {Th{\"o}ne}, {Bolmer}, {Schulze}, {Mart{\'\i}n}, {Kann},
  {D'Elia}, {Selsing}, {Martin-Carrillo}, {Perley}, {Kim}, {Izzo},
  {S{\'a}nchez-Ram{\'\i}rez}, {Guidorzi}, {Klotz}, {Wiersema}, {Bauer},
  {Bensch}, {Campana}, {Cano}, {Covino}, {Coward}, {De Cia}, {de
  Gregorio-Monsalvo}, {De Pasquale}, {Fynbo}, {Greiner}, {Gomboc}, {Hanlon},
  {Hansen}, {Hartmann}, {Heintz}, {Jakobsson}, {Kobayashi}, {Malesani},
  {Martone}, {Meintjes}, {Micha{\l}owski}, {Mundell}, {Murphy}, {Oates},
  {Salmon}, {van Soelen}, {Tanvir}, {Turpin}, {Xu}, \& {Zafar}}]{grb161023a}
{de Ugarte Postigo}, A., {Th{\"o}ne}, C.~C., {Bolmer}, J., {et~al.}
  2018{\natexlab{\hspace{0pt}b}},
  \hypersetup{urlcolor=magenta}\href{https://dx.doi.org/10.1051/0004-6361/201833094}{\aap},
  \hypersetup{urlcolor=blue}\href{https://ui.adsabs.harvard.edu/abs/2018A&A...620A.119D}{620,
  A119}

\bibitem[{{Dichiara} {et~al.}(2022){Dichiara}, {Troja}, {Lipunov}, {Ricci},
  {Oates}, {Butler}, {Liuzzo}, {Ryan}, {O'Connor}, {Cenko}, {Cosentino},
  {Lien}, {Gorbovskoy}, {Tyurina}, {Balanutsa}, {Vlasenko}, {Gorbunov},
  {Podesta}, {Podesta}, {Rebolo}, {Serra}, \& {Buckley}}]{Dichiara2022}
{Dichiara}, S., {Troja}, E., {Lipunov}, V., {et~al.} 2022,
  \hypersetup{urlcolor=magenta}\href{https://dx.doi.org/10.1093/mnras/stac454}{\mnras},
  \hypersetup{urlcolor=blue}\href{https://ui.adsabs.harvard.edu/abs/2022MNRAS.512.2337D}{512,
  2337}

\bibitem[{{Dobie} {et~al.}(2021){Dobie}, {Murphy}, {Kaplan}, {Hotokezaka},
  {Bonilla Ataides}, {Mahony}, \& {Sadler}}]{Dobie2021}
{Dobie}, D., {Murphy}, T., {Kaplan}, D.~L., {et~al.} 2021,
  \hypersetup{urlcolor=magenta}\href{https://dx.doi.org/10.1093/mnras/stab1468}{\mnras},
  \hypersetup{urlcolor=blue}\href{https://ui.adsabs.harvard.edu/abs/2021MNRAS.505.2647D}{505,
  2647}

\bibitem[{{Dong} {et~al.}(2021){Dong}, {Hallinan}, {Nakar}, {Ho}, {Hughes},
  {Hotokezaka}, {Myers}, {De}, {Mooley}, {Ravi}, {Horesh}, {Kasliwal}, \&
  {Kulkarni}}]{Dong2021}
{Dong}, D.~Z., {Hallinan}, G., {Nakar}, E., {et~al.} 2021,
  \hypersetup{urlcolor=magenta}\href{https://dx.doi.org/10.1126/science.abg6037}{Science},
  \hypersetup{urlcolor=blue}\href{https://ui.adsabs.harvard.edu/abs/2021Sci...373.1125D}{373,
  1125}

\bibitem[{{Driver} {et~al.}(2016){Driver}, {Andrews}, {Davies}, {Robotham},
  {Wright}, {Windhorst}, {Cohen}, {Emig}, {Jansen}, \& {Dunne}}]{Driver2016}
{Driver}, S.~P., {Andrews}, S.~K., {Davies}, L.~J., {et~al.} 2016,
  \hypersetup{urlcolor=magenta}\href{https://dx.doi.org/10.3847/0004-637X/827/2/108}{\apj},
  \hypersetup{urlcolor=blue}\href{https://ui.adsabs.harvard.edu/abs/2016ApJ...827..108D}{827,
  108}

\bibitem[{{Drout} {et~al.}(2014){Drout}, {Chornock}, {Soderberg}, {Sanders},
  {McKinnon}, {Rest}, {Foley}, {Milisavljevic}, {Margutti}, {Berger},
  {Calkins}, {Fong}, {Gezari}, {Huber}, {Kankare}, {Kirshner}, {Leibler},
  {Lunnan}, {Mattila}, {Marion}, {Narayan}, {Riess}, {Roth}, {Scolnic},
  {Smartt}, {Tonry}, {Burgett}, {Chambers}, {Hodapp}, {Jedicke}, {Kaiser},
  {Magnier}, {Metcalfe}, {Morgan}, {Price}, \& {Waters}}]{Drout2014}
{Drout}, M.~R., {Chornock}, R., {Soderberg}, A.~M., {et~al.} 2014,
  \hypersetup{urlcolor=magenta}\href{https://dx.doi.org/10.1088/0004-637X/794/1/23}{\apj},
  \hypersetup{urlcolor=blue}\href{https://ui.adsabs.harvard.edu/abs/2014ApJ...794...23D}{794,
  23}

\bibitem[{{Eftekhari} \& {Berger}(2017)}]{Eftekhari2017}
{Eftekhari}, T., \& {Berger}, E. 2017,
  \hypersetup{urlcolor=magenta}\href{https://dx.doi.org/10.3847/1538-4357/aa90b9}{\apj},
  \hypersetup{urlcolor=blue}\href{https://ui.adsabs.harvard.edu/abs/2017ApJ...849..162E}{849,
  162}

\bibitem[{{Eftekhari} {et~al.}(2018){Eftekhari}, {Berger}, {Zauderer},
  {Margutti}, \& {Alexander}}]{Eftekhari2018}
{Eftekhari}, T., {Berger}, E., {Zauderer}, B.~A., {Margutti}, R., \&
  {Alexander}, K.~D. 2018,
  \hypersetup{urlcolor=magenta}\href{https://dx.doi.org/10.3847/1538-4357/aaa8e0}{\apj},
  \hypersetup{urlcolor=blue}\href{https://ui.adsabs.harvard.edu/abs/2018ApJ...854...86E}{854,
  86}

\bibitem[{{Foley} {et~al.}(2019){Foley}, {Bloom}, {Cenko}, {Chornock},
  {Dimitriadis}, {Dor{\'e}}, {Filippenko}, {Fox}, {Hirata}, {Jha}, {Jones},
  {Kasliwal}, {Kelly}, {Kilpatrick}, {Kirshner}, {Koekemoer}, {Kruk}, {Mandel},
  {Margutti}, {Miranda}, {Nissanke}, {Rest}, {Rhodes}, {Rodney}, {Rose},
  {Sand}, {Scolnic}, {Siellez}, {Smith}, {Spergel}, {Strolger}, {Suntzeff},
  {Wang}, \& {Wollack}}]{Foley2019}
{Foley}, R., {Bloom}, J.~S., {Cenko}, S.~B., {et~al.} 2019, \baas,
  \hypersetup{urlcolor=blue}\href{https://ui.adsabs.harvard.edu/abs/2019BAAS...51c.305F}{51,
  305}

\bibitem[{{Fong} {et~al.}(2015){Fong}, {Berger}, {Margutti}, \&
  {Zauderer}}]{Fong2015}
{Fong}, W., {Berger}, E., {Margutti}, R., \& {Zauderer}, B.~A. 2015,
  \hypersetup{urlcolor=magenta}\href{https://dx.doi.org/10.1088/0004-637X/815/2/102}{\apj},
  \hypersetup{urlcolor=blue}\href{https://ui.adsabs.harvard.edu/abs/2015ApJ...815..102F}{815,
  102}

\bibitem[{{Fowler} {et~al.}(2007){Fowler}, {Niemack}, {Dicker}, {Aboobaker},
  {Ade}, {Battistelli}, {Devlin}, {Fisher}, {Halpern}, {Hargrave}, {Hincks},
  {Kaul}, {Klein}, {Lau}, {Limon}, {Marriage}, {Mauskopf}, {Page}, {Staggs},
  {Swetz}, {Switzer}, {Thornton}, \& {Tucker}}]{Fowler2007}
{Fowler}, J.~W., {Niemack}, M.~D., {Dicker}, S.~R., {et~al.} 2007,
  \hypersetup{urlcolor=magenta}\href{https://dx.doi.org/10.1364/AO.46.003444}{\ao},
  \hypersetup{urlcolor=blue}\href{https://ui.adsabs.harvard.edu/abs/2007ApOpt..46.3444F}{46,
  3444}

\bibitem[{{Frail} {et~al.}(2012){Frail}, {Kulkarni}, {Ofek}, {Bower}, \&
  {Nakar}}]{Frail2012}
{Frail}, D.~A., {Kulkarni}, S.~R., {Ofek}, E.~O., {Bower}, G.~C., \& {Nakar},
  E. 2012,
  \hypersetup{urlcolor=magenta}\href{https://dx.doi.org/10.1088/0004-637X/747/1/70}{\apj},
  \hypersetup{urlcolor=blue}\href{https://ui.adsabs.harvard.edu/abs/2012ApJ...747...70F}{747,
  70}

\bibitem[{{Frail} {et~al.}(2001){Frail}, {Kulkarni}, {Sari}, {Djorgovski},
  {Bloom}, {Galama}, {Reichart}, {Berger}, {Harrison}, {Price}, {Yost},
  {Diercks}, {Goodrich}, \& {Chaffee}}]{Frail2001}
{Frail}, D.~A., {Kulkarni}, S.~R., {Sari}, R., {et~al.} 2001,
  \hypersetup{urlcolor=magenta}\href{https://dx.doi.org/10.1086/338119}{\apjl},
  \hypersetup{urlcolor=blue}\href{https://ui.adsabs.harvard.edu/abs/2001ApJ...562L..55F}{562,
  L55}

\bibitem[{{Galama} {et~al.}(2000){Galama}, {Bremer}, {Bertoldi}, {Menten},
  {Lisenfeld}, {Shepherd}, {Mason}, {Walter}, {Pooley}, {Frail}, {Sari},
  {Kulkarni}, {Berger}, {Bloom}, {Castro-Tirado}, \& {Granot}}]{Galama2000}
{Galama}, T.~J., {Bremer}, M., {Bertoldi}, F., {et~al.} 2000,
  \hypersetup{urlcolor=magenta}\href{https://dx.doi.org/10.1086/312904}{\apjl},
  \hypersetup{urlcolor=blue}\href{https://ui.adsabs.harvard.edu/abs/2000ApJ...541L..45G}{541,
  L45}

\bibitem[{{Generozov} {et~al.}(2017){Generozov}, {Mimica}, {Metzger}, {Stone},
  {Giannios}, \& {Aloy}}]{Generozov2017}
{Generozov}, A., {Mimica}, P., {Metzger}, B.~D., {et~al.} 2017,
  \hypersetup{urlcolor=magenta}\href{https://dx.doi.org/10.1093/mnras/stw2439}{\mnras},
  \hypersetup{urlcolor=blue}\href{https://ui.adsabs.harvard.edu/abs/2017MNRAS.464.2481G}{464,
  2481}

\bibitem[{{Ghirlanda} {et~al.}(2014){Ghirlanda}, {Burlon}, {Ghisellini},
  {Salvaterra}, {Bernardini}, {Campana}, {Covino}, {D'Avanzo}, {D'Elia},
  {Melandri}, {Murphy}, {Nava}, {Vergani}, \& {Tagliaferri}}]{Ghirlanda2014}
{Ghirlanda}, G., {Burlon}, D., {Ghisellini}, G., {et~al.} 2014,
  \hypersetup{urlcolor=magenta}\href{https://dx.doi.org/10.1017/pasa.2014.14}{\pasa},
  \hypersetup{urlcolor=blue}\href{https://ui.adsabs.harvard.edu/abs/2014PASA...31...22G}{31,
  e022}

\bibitem[{{Giannios} \& {Metzger}(2011)}]{Giannios2011}
{Giannios}, D., \& {Metzger}, B.~D. 2011,
  \hypersetup{urlcolor=magenta}\href{https://dx.doi.org/10.1111/j.1365-2966.2011.19188.x}{\mnras},
  \hypersetup{urlcolor=blue}\href{https://ui.adsabs.harvard.edu/abs/2011MNRAS.416.2102G}{416,
  2102}

\bibitem[{{Gorosabel} {et~al.}(2010){Gorosabel}, {de Ugarte Postigo},
  {Castro-Tirado}, {Agudo}, {Jel{\'\i}nek}, {Leon}, {Augusteijn}, {Fynbo},
  {Hjorth}, {Micha{\l}owski}, {Xu}, {Ferrero}, {Kann}, {Klose}, {Rossi},
  {Madrid}, {Llorente}, {Bremer}, \& {Winters}}]{Gorosabel2010}
{Gorosabel}, J., {de Ugarte Postigo}, A., {Castro-Tirado}, A.~J., {et~al.}
  2010,
  \hypersetup{urlcolor=magenta}\href{https://dx.doi.org/10.1051/0004-6361/200913263}{\aap},
  \hypersetup{urlcolor=blue}\href{https://ui.adsabs.harvard.edu/abs/2010A&A...522A..14G}{522,
  A14}

\bibitem[{{Granot} \& {Sari}(2002)}]{Granot2002}
{Granot}, J., \& {Sari}, R. 2002,
  \hypersetup{urlcolor=magenta}\href{https://dx.doi.org/10.1086/338966}{\apj},
  \hypersetup{urlcolor=blue}\href{https://ui.adsabs.harvard.edu/abs/2002ApJ...568..820G}{568,
  820}

\bibitem[{{Granot} \& {Taylor}(2005)}]{Granot2005}
{Granot}, J., \& {Taylor}, G.~B. 2005,
  \hypersetup{urlcolor=magenta}\href{https://dx.doi.org/10.1086/429536}{\apj},
  \hypersetup{urlcolor=blue}\href{https://ui.adsabs.harvard.edu/abs/2005ApJ...625..263G}{625,
  263}

\bibitem[{{Greiner} {et~al.}(2009){Greiner}, {Kr{\"u}hler}, {McBreen},
  {Ajello}, {Giannios}, {Schwarz}, {Savaglio}, {Yolda{\c{s}}}, {Clemens},
  {Stefanescu}, {Sala}, {Bertoldi}, {Szokoly}, \& {Klose}}]{Greiner2009}
{Greiner}, J., {Kr{\"u}hler}, T., {McBreen}, S., {et~al.} 2009,
  \hypersetup{urlcolor=magenta}\href{https://dx.doi.org/10.1088/0004-637X/693/2/1912}{\apj},
  \hypersetup{urlcolor=blue}\href{https://ui.adsabs.harvard.edu/abs/2009ApJ...693.1912G}{693,
  1912}

\bibitem[{{Greiner} {et~al.}(2013){Greiner}, {Kr{\"u}hler}, {Nardini},
  {Filgas}, {Moin}, {de Breuck}, {Montenegro-Montes}, {Lundgren}, {Klose},
  {fonso}, {Bertoldi}, {Elliott}, {Kann}, {Knust}, {Menten}, {Nicuesa
  Guelbenzu}, {Olivares E.}, {Rau}, {Rossi}, {Schady}, {Schmidl}, {Siringo},
  {Spezzi}, {Sudilovsky}, {Tingay}, {Updike}, {Wang}, {Weiss}, {Wieringa}, \&
  {Wyrowski}}]{Greiner2013}
{Greiner}, J., {Kr{\"u}hler}, T., {Nardini}, M., {et~al.} 2013,
  \hypersetup{urlcolor=magenta}\href{https://dx.doi.org/10.1051/0004-6361/201321284}{\aap},
  \hypersetup{urlcolor=blue}\href{https://ui.adsabs.harvard.edu/abs/2013A&A...560A..70G}{560,
  A70}

\bibitem[{{Guns} {et~al.}(2021){Guns}, {Foster}, {Daley}, {Rahlin},
  {Whitehorn}, {Ade}, {Ahmed}, {Anderes}, {Anderson}, {Archipley}, {Avva},
  {Aylor}, {Balkenhol}, {Barry}, {Basu Thakur}, {Benabed}, {Bender}, {Benson},
  {Bianchini}, {Bleem}, {Bouchet}, {Bryant}, {Byrum}, {Carlstrom}, {Carter},
  {Cecil}, {Chang}, {Chaubal}, {Chen}, {Cho}, {Chou}, {Cliche}, {Crawford},
  {Cukierman}, {de Haan}, {Denison}, {Dibert}, {Ding}, {Dobbs}, {Dutcher},
  {Everett}, {Feng}, {Ferguson}, {Fu}, {Galli}, {Gambrel}, {Gardner},
  {Goeckner-Wald}, {Gualtieri}, {Gupta}, {Guyser}, {Halverson},
  {Harke-Hosemann}, {Harrington}, {Henning}, {Hilton}, {Hivon}, {Holder},
  {Holzapfel}, {Hood}, {Howe}, {Huang}, {Irwin}, {Jeong}, {Jonas}, {Jones},
  {Khaire}, {Knox}, {Kofman}, {Korman}, {Kubik}, {Kuhlmann}, {Kuo}, {Lee},
  {Leitch}, {Lowitz}, {Lu}, {Marrone}, {Meyer}, {Michalik}, {Millea},
  {Montgomery}, {Nadolski}, {Natoli}, {Nguyen}, {Noble}, {Novosad}, {Omori},
  {Padin}, {Pan}, {Paschos}, {Pearson}, {Phadke}, {Posada}, {Prabhu}, {Quan},
  {Reichardt}, {Riebel}, {Riedel}, {Rouble}, {Ruhl}, {Sayre}, {Schiappucci},
  {Shirokoff}, {Smecher}, {Sobrin}, {Stark}, {Stephen}, {Story}, {Suzuki},
  {Thompson}, {Thorne}, {Tucker}, {Umilta}, {Vale}, {Vieira}, {Wang}, {Wu},
  {Yefremenko}, {Yoon}, {Young}, \& {Zhang}}]{Guns2021}
{Guns}, S., {Foster}, A., {Daley}, C., {et~al.} 2021, arXiv e-prints,
  \hypersetup{urlcolor=magenta}\href{https://arxiv.org/abs/2103.06166}{arXiv}{:}\hypersetup{urlcolor=blue}\href{https://ui.adsabs.harvard.edu/abs/2021arXiv210306166G}{2103.06166}

\bibitem[{{Hall} {et~al.}(2020){Hall}, {Kuns}, {Smith}, {Bai}, {Wipf},
  {Biscans}, {Adhikari}, {Arai}, {Ballmer}, {Barsotti}, {Chen}, {Evans},
  {Fritschel}, {Harms}, {Kamai}, {Graef Rollins}, {Shoemaker}, {Slagmolen},
  {Weiss}, \& {Yamamoto}}]{Hall2020}
{Hall}, E.~D., {Kuns}, K., {Smith}, J.~R., {et~al.} 2020, arXiv e-prints,
  \hypersetup{urlcolor=magenta}\href{https://arxiv.org/abs/2012.03608}{arXiv}{:}\hypersetup{urlcolor=blue}\href{https://ui.adsabs.harvard.edu/abs/2020arXiv201203608H}{2012.03608}

\bibitem[{{Ho} {et~al.}(2019){Ho}, {Phinney}, {Ravi}, {Kulkarni}, {Petitpas},
  {Emonts}, {Bhalerao}, {Blundell}, {Cenko}, {Dobie}, {Howie}, {Kamraj},
  {Kasliwal}, {Murphy}, {Perley}, {Sridharan}, \& {Yoon}}]{Ho2019}
{Ho}, A. Y.~Q., {Phinney}, E.~S., {Ravi}, V., {et~al.} 2019,
  \hypersetup{urlcolor=magenta}\href{https://dx.doi.org/10.3847/1538-4357/aaf473}{\apj},
  \hypersetup{urlcolor=blue}\href{https://ui.adsabs.harvard.edu/abs/2019ApJ...871...73H}{871,
  73}

\bibitem[{{Ho} {et~al.}(2020){Ho}, {Perley}, {Beniamini}, {Cenko}, {Kulkarni},
  {Andreoni}, {Singer}, {De}, {Kasliwal}, {Fremling}, {Bellm}, {Dekany},
  {Delacroix}, {Duev}, {Goldstein}, {Golkhou}, {Goobar}, {Graham}, {Hale},
  {Kupfer}, {Laher}, {Masci}, {Miller}, {Neill}, {Riddle}, {Rusholme}, {Shupe},
  {Smith}, {Sollerman}, \& {van Roestel}}]{Ho2020}
{Ho}, A. Y.~Q., {Perley}, D.~A., {Beniamini}, P., {et~al.} 2020,
  \hypersetup{urlcolor=magenta}\href{https://dx.doi.org/10.3847/1538-4357/abc34d}{\apj},
  \hypersetup{urlcolor=blue}\href{https://ui.adsabs.harvard.edu/abs/2020ApJ...905...98H}{905,
  98}

\bibitem[{{Ho} {et~al.}(2021{\natexlab{\hspace{0pt}a}}){Ho}, {Margalit},
  {Bremer}, {Perley}, {Yao}, {Dobie}, {Kaplan}, {O'Brien}, {Petitpas}, \&
  {Zic}}]{Ho2021xnd}
{Ho}, A. Y.~Q., {Margalit}, B., {Bremer}, M., {et~al.}
  2021{\natexlab{\hspace{0pt}a}}, arXiv e-prints,
  \hypersetup{urlcolor=magenta}\href{https://arxiv.org/abs/2110.05490}{arXiv}{:}\hypersetup{urlcolor=blue}\href{https://ui.adsabs.harvard.edu/abs/2021arXiv211005490H}{2110.05490}

\bibitem[{{Ho} {et~al.}(2021{\natexlab{\hspace{0pt}b}}){Ho}, {Perley},
  {Gal-Yam}, {Lunnan}, {Sollerman}, {Schulze}, {Das}, {Dobie}, {Yao},
  {Fremling}, {Adams}, {Anand}, {Andreoni}, {Bellm}, {Bruch}, {Burdge},
  {Castro-Tirado}, {Dahiwale}, {De}, {Dekany}, {Drake}, {Duev}, {Graham},
  {Helou}, {Kaplan}, {Karambelkar}, {Kasliwal}, {Kool}, {Kulkarni}, {Mahabal},
  {Medford}, {Miller}, {Nordin}, {Ofek}, {Petitpas}, {Riddle}, {Sharma},
  {Smith}, {Stewart}, {Taggart}, {Tartaglia}, {Tzanidakis}, \&
  {Winters}}]{Ho2021}
{Ho}, A. Y.~Q., {Perley}, D.~A., {Gal-Yam}, A., {et~al.}
  2021{\natexlab{\hspace{0pt}b}}, arXiv e-prints,
  \hypersetup{urlcolor=magenta}\href{https://arxiv.org/abs/2105.08811}{arXiv}{:}\hypersetup{urlcolor=blue}\href{https://ui.adsabs.harvard.edu/abs/2021arXiv210508811H}{2105.08811}

\bibitem[{{Horesh} {et~al.}(2013{\natexlab{\hspace{0pt}a}}){Horesh}, {Cao},
  {Mooley}, \& {Carpenter}}]{Horesh2013_iptf}
{Horesh}, A., {Cao}, Y., {Mooley}, K., \& {Carpenter}, J.
  2013{\natexlab{\hspace{0pt}a}}, The Astronomer's Telegram,
  \hypersetup{urlcolor=blue}\href{https://ui.adsabs.harvard.edu/abs/2013ATel.5198....1H}{5198,
  1}

\bibitem[{{Horesh} {et~al.}(2016){Horesh}, {Hotokezaka}, {Piran}, {Nakar}, \&
  {Hancock}}]{Horesh+16}
{Horesh}, A., {Hotokezaka}, K., {Piran}, T., {Nakar}, E., \& {Hancock}, P.
  2016,
  \hypersetup{urlcolor=magenta}\href{https://dx.doi.org/10.3847/2041-8205/819/2/L22}{\apjl},
  \hypersetup{urlcolor=blue}\href{https://ui.adsabs.harvard.edu/abs/2016ApJ...819L..22H}{819,
  L22}

\bibitem[{{Horesh} {et~al.}(2013{\natexlab{\hspace{0pt}b}}){Horesh},
  {Stockdale}, {Fox}, {Frail}, {Carpenter}, {Kulkarni}, {Ofek}, {Gal-Yam},
  {Kasliwal}, {Arcavi}, {Quimby}, {Cenko}, {Nugent}, {Bloom}, {Law},
  {Poznanski}, {Gorbikov}, {Polishook}, {Yaron}, {Ryder}, {Weiler}, {Bauer},
  {Van Dyk}, {Immler}, {Panagia}, {Pooley}, \& {Kassim}}]{Horesh2013}
{Horesh}, A., {Stockdale}, C., {Fox}, D.~B., {et~al.}
  2013{\natexlab{\hspace{0pt}b}},
  \hypersetup{urlcolor=magenta}\href{https://dx.doi.org/10.1093/mnras/stt1645}{\mnras},
  \hypersetup{urlcolor=blue}\href{https://ui.adsabs.harvard.edu/abs/2013MNRAS.436.1258H}{436,
  1258}

\bibitem[{{IceCube Collaboration} {et~al.}(2018){IceCube Collaboration},
  {Aartsen}, {Ackermann}, {Adams}, {Aguilar}, {Ahlers}, {Ahrens}, {Samarai},
  {Altmann}, {Andeen}, {Anderson}, {Ansseau}, {Anton}, {Arg{\"u}elles},
  {Arsioli}, {Auffenberg}, {Axani}, {Bagherpour}, {Bai}, {Barron}, {Barwick},
  {Baum}, {Bay}, {Beatty}, {Becker Tjus}, {Becker}, {BenZvi}, {Berley},
  {Bernardini}, {Besson}, {Binder}, {Bindig}, {Blaufuss}, {Blot}, {Bohm},
  {B{\"o}rner}, {Bos}, {B{\"o}ser}, {Botner}, {Bourbeau}, {Bourbeau},
  {Bradascio}, {Braun}, {Brenzke}, {Bretz}, {Bron}, {Brostean-Kaiser},
  {Burgman}, {Busse}, {Carver}, {Cheung}, {Chirkin}, {Christov}, {Clark},
  {Classen}, {Coenders}, {Collin}, {Conrad}, {Coppin}, {Correa}, {Cowen},
  {Cross}, {Dave}, {Day}, {de Andr{\'e}}, {De Clercq}, {DeLaunay}, {Dembinski},
  {DeRidder}, {Desiati}, {de Vries}, {de Wasseige}, {de With}, {DeYoung},
  {D{\'\i}az-V{\'e}lez}, {di Lorenzo}, {Dujmovic}, {Dumm}, {Dunkman}, {Dvorak},
  {Eberhardt}, {Ehrhardt}, {Eichmann}, {Eller}, {Evenson}, {Fahey}, {Fazely},
  {Felde}, {Filimonov}, {Finley}, {Flis}, {Franckowiak}, {Friedman}, {Fritz},
  {Gaisser}, {Gallagher}, {Gerhardt}, {Ghorbani}, {Giommi}, {Glauch},
  {Gl{\"u}senkamp}, {Goldschmidt}, {Gonzalez}, {Grant}, {Griffith}, {Haack},
  {Hallgren}, {Halzen}, {Hanson}, {Hebecker}, {Heereman}, {Helbing},
  {Hellauer}, {Hickford}, {Hignight}, {Hill}, {Hoffman}, {Hoffmann}, {Hoinka},
  {Hokanson-Fasig}, {Hoshina}, {Huang}, {Huber}, {Hultqvist}, {H{\"u}nnefeld},
  {Hussain}, {In}, {Iovine}, {Ishihara}, {Jacobi}, {Japaridze}, {Jeong},
  {Jero}, {Jones}, {Kalaczynski}, {Kang}, {Kappes}, {Kappesser}, {Karg},
  {Karle}, {Katz}, {Kauer}, {Keivani}, {Kelley}, {Kheirandish}, {Kim}, {Kim},
  {Kintscher}, {Kiryluk}, {Kittler}, {Klein}, {Koirala}, {Kolanoski},
  {K{\"o}pke}, {Kopper}, {Kopper}, {Koschinsky}, {Koskinen}, {Kowalski},
  {Krammer}, {Krings}, {Kroll}, {Kr{\"u}ckl}, {Kunwar}, {Kurahashi},
  {Kuwabara}, {Kyriacou}, {Labare}, {Lanfranchi}, {Larson}, {Lauber},
  {Leonard}, {Lesiak-Bzdak}, {Leuermann}, {Liu}, {Lozano Mariscal}, {Lu},
  {L{\"u}nemann}, {Luszczak}, {Madsen}, {Maggi}, {Mahn}, {Mancina}, {Maruyama},
  {Mase}, {Maunu}, {Meagher}, {Medici}, {Meier}, {Menne}, {Merino}, {Meures},
  {Miarecki}, {Micallef}, {Moment{\'e}}, {Montaruli}, {Moore}, {Morse},
  {Moulai}, {Nahnhauer}, {Nakarmi}, {Naumann}, {Neer}, {Niederhausen},
  {Nowicki}, {Nygren}, {Obertacke Pollmann}, {Olivas}, {O'Murchadha},
  {O'Sullivan}, {Padovani}, {Palczewski}, {Pandya}, {Pankova}, {Peiffer},
  {Pepper}, {P{\'e}rez de los Heros}, {Pieloth}, {Pinat}, {Plum}, {Price},
  {Przybylski}, {Raab}, {R{\"a}del}, {Rameez}, {Rawlins}, {Rea}, {Reimann},
  {Relethford}, {Relich}, {Resconi}, {Rhode}, {Richman}, {Robertson}, {Rongen},
  {Rott}, {Ruhe}, {Ryckbosch}, {Rysewyk}, {Safa}, {Sahakyan}, {S{\"a}lzer},
  {Sanchez Herrera}, {Sandrock}, {Sandroos}, {Santander}, {Sarkar}, {Sarkar},
  {Satalecka}, {Schlunder}, {Schmidt}, {Schneider}, {Schoenen},
  {Sch{\"o}neberg}, {Schumacher}, {Sclafani}, {Seckel}, {Seunarine},
  {Soedingrekso}, {Soldin}, {Song}, {Spiczak}, {Spiering}, {Stachurska},
  {Stamatikos}, {Stanev}, {Stasik}, {Stettner}, {Steuer}, {Stezelberger},
  {Stokstad}, {St{\"o}{\ss}l}, {Strotjohann}, {Stuttard}, {Sullivan},
  {Sutherland}, {Taboada}, {Tatar}, {Tenholt}, {Ter-Antonyan}, {Terliuk},
  {Tilav}, {Toale}, {Tobin}, {Toennis}, {Toscano}, {Tosi}, {Tselengidou},
  {Tung}, {Turcati}, {Turley}, {Ty}, {Unger}, {Usner}, {Vandenbroucke}, {Van
  Driessche}, {van Eijk}, {van Eijndhoven}, {Vanheule}, {van Santen}, {Vogel},
  {Vraeghe}, {Walck}, {Wallace}, {Wallraff}, {Wandler}, {Wandkowsky}, {Waza},
  {Weaver}, {Weiss}, {Wendt}, {Werthebach}, {Westerhoff}, {Whelan},
  {Whitehorn}, {Wiebe}, {Wiebusch}, {Wille}, {Williams}, {Wills}, {Wolf},
  {Wood}, {Wood}, {Woschnagg}, {Xu}, {Xu}, {Xu}, {Yanez}, {Yodh}, {Yoshida}, \&
  {Yuan}}]{IceCube2018}
{IceCube Collaboration}, {Aartsen}, M.~G., {Ackermann}, M., {et~al.} 2018,
  \hypersetup{urlcolor=magenta}\href{https://dx.doi.org/10.1126/science.aat2890}{Science},
  \hypersetup{urlcolor=blue}\href{https://ui.adsabs.harvard.edu/abs/2018Sci...361..147I}{361,
  147}

\bibitem[{{Kellermann} \& {Pauliny-Toth}(1969)}]{Kellermann1969}
{Kellermann}, K.~I., \& {Pauliny-Toth}, I.~I.~K. 1969,
  \hypersetup{urlcolor=magenta}\href{https://dx.doi.org/10.1086/180305}{\apjl},
  \hypersetup{urlcolor=blue}\href{https://ui.adsabs.harvard.edu/abs/1969ApJ...155L..71K}{155,
  L71}

\bibitem[{Kobayashi(2000)}]{kob00}
Kobayashi, S. 2000,
  \hypersetup{urlcolor=magenta}\href{https://dx.doi.org/10.1086/317869}{\apj},
  \hypersetup{urlcolor=blue}\href{http://adsabs.harvard.edu/abs/2000ApJ...545..807K}{545,
  807}

\bibitem[{Kobayashi \& Sari(2000)}]{ks00}
Kobayashi, S., \& Sari, R. 2000,
  \hypersetup{urlcolor=magenta}\href{https://dx.doi.org/10.1086/317021}{\apj},
  \hypersetup{urlcolor=blue}\href{http://adsabs.harvard.edu/abs/2000ApJ...542..819K}{542,
  819}

\bibitem[{Kobayashi \& Zhang(2003)}]{kz03}
Kobayashi, S., \& Zhang, B. 2003,
  \hypersetup{urlcolor=magenta}\href{https://dx.doi.org/10.1086/367691}{\apjl},
  \hypersetup{urlcolor=blue}\href{http://adsabs.harvard.edu/abs/2003ApJ...582L..75K}{582,
  L75}

\bibitem[{{Kohno} {et~al.}(2005){Kohno}, {Tosaki}, {Okuda}, {Nakanishi},
  {Kamazaki}, {Muraoka}, {Onodera}, {Sofue}, {Okumura}, {Kuno}, {Nakai},
  {Ohta}, {Ishizuki}, {Kawabe}, \& {Kawai}}]{Kohno2005}
{Kohno}, K., {Tosaki}, T., {Okuda}, T., {et~al.} 2005,
  \hypersetup{urlcolor=magenta}\href{https://dx.doi.org/10.1093/pasj/57.1.147}{\pasj},
  \hypersetup{urlcolor=blue}\href{https://ui.adsabs.harvard.edu/abs/2005PASJ...57..147K}{57,
  147}

\bibitem[{{Kulkarni} {et~al.}(1998){Kulkarni}, {Frail}, {Wieringa}, {Ekers},
  {Sadler}, {Wark}, {Higdon}, {Phinney}, \& {Bloom}}]{Kulkarni1998}
{Kulkarni}, S.~R., {Frail}, D.~A., {Wieringa}, M.~H., {et~al.} 1998,
  \hypersetup{urlcolor=magenta}\href{https://dx.doi.org/10.1038/27139}{\nat},
  \hypersetup{urlcolor=blue}\href{https://ui.adsabs.harvard.edu/abs/1998Natur.395..663K}{395,
  663}

\bibitem[{{Kuno} {et~al.}(2004){Kuno}, {Sato}, {Nakanishi}, {Yamauchi},
  {Nakai}, \& {Kawai}}]{Kuno2004}
{Kuno}, N., {Sato}, N., {Nakanishi}, H., {et~al.} 2004,
  \hypersetup{urlcolor=magenta}\href{https://dx.doi.org/10.1093/pasj/56.2.L1}{\pasj},
  \hypersetup{urlcolor=blue}\href{https://ui.adsabs.harvard.edu/abs/2004PASJ...56L...1K}{56,
  L1}

\bibitem[{{Lacy} {et~al.}(2020){Lacy}, {Baum}, {Chandler}, {Chatterjee},
  {Clarke}, {Deustua}, {English}, {Farnes}, {Gaensler}, {Gugliucci},
  {Hallinan}, {Kent}, {Kimball}, {Law}, {Lazio}, {Marvil}, {Mao}, {Medlin},
  {Mooley}, {Murphy}, {Myers}, {Osten}, {Richards}, {Rosolowsky}, {Rudnick},
  {Schinzel}, {Sivakoff}, {Sjouwerman}, {Taylor}, {White}, {Wrobel},
  {Andernach}, {Beasley}, {Berger}, {Bhatnager}, {Birkinshaw}, {Bower},
  {Brandt}, {Brown}, {Burke-Spolaor}, {Butler}, {Comerford}, {Demorest}, {Fu},
  {Giacintucci}, {Golap}, {G{\"u}th}, {Hales}, {Hiriart}, {Hodge}, {Horesh},
  {Ivezi{\'c}}, {Jarvis}, {Kamble}, {Kassim}, {Liu}, {Loinard}, {Lyons},
  {Masters}, {Mezcua}, {Moellenbrock}, {Mroczkowski}, {Nyland}, {O'Dea},
  {O'Sullivan}, {Peters}, {Radford}, {Rao}, {Robnett}, {Salcido}, {Shen},
  {Sobotka}, {Witz}, {Vaccari}, {van Weeren}, {Vargas}, {Williams}, \&
  {Yoon}}]{Lacy2020}
{Lacy}, M., {Baum}, S.~A., {Chandler}, C.~J., {et~al.} 2020,
  \hypersetup{urlcolor=magenta}\href{https://dx.doi.org/10.1088/1538-3873/ab63eb}{\pasp},
  \hypersetup{urlcolor=blue}\href{https://ui.adsabs.harvard.edu/abs/2020PASP..132c5001L}{132,
  035001}

\bibitem[{{Laskar}(2021)}]{grb210702a_b}
{Laskar}, T. 2021, GRB Coordinates Network,
  \hypersetup{urlcolor=blue}\href{https://ui.adsabs.harvard.edu/abs/2021GCN.30479....1L}{30479,
  1}

\bibitem[{{Laskar}(2022)}]{grb220101a}
{Laskar}, T. 2022, GRB Coordinates Network,
  \hypersetup{urlcolor=blue}\href{https://ui.adsabs.harvard.edu/abs/2022GCN.31372....1L}{31372,
  1}

\bibitem[{{Laskar} \& {a larger Collaboration}(2019)}]{Laskar2019_grb191221b}
{Laskar}, T., \& {a larger Collaboration}. 2019, GRB Coordinates Network,
  \hypersetup{urlcolor=blue}\href{https://ui.adsabs.harvard.edu/abs/2019GCN.26564....1L}{26564,
  1}

\bibitem[{{Laskar} {et~al.}(2021{\natexlab{\hspace{0pt}a}}){Laskar},
  {Alexander}, {Berger}, {Fong}, {Margutti}, {Mundell}, \&
  {Schady}}]{grb210619b}
{Laskar}, T., {Alexander}, K.~D., {Berger}, E., {et~al.}
  2021{\natexlab{\hspace{0pt}a}}, GRB Coordinates Network,
  \hypersetup{urlcolor=blue}\href{https://ui.adsabs.harvard.edu/abs/2021GCN.30386....1L}{30386,
  1}

\bibitem[{{Laskar} {et~al.}(2021{\natexlab{\hspace{0pt}b}}){Laskar},
  {Alexander}, {Margutti}, {Berger}, {Fong}, {Chornock}, {Mundell}, \&
  {Schady}}]{grb210905a}
{Laskar}, T., {Alexander}, K.~D., {Margutti}, R., {et~al.}
  2021{\natexlab{\hspace{0pt}b}}, GRB Coordinates Network,
  \hypersetup{urlcolor=blue}\href{https://ui.adsabs.harvard.edu/abs/2021GCN.30783....1L}{30783,
  1}

\bibitem[{{Laskar} {et~al.}(2018{\natexlab{\hspace{0pt}a}}){Laskar}, {Berger},
  {Chornock}, {Margutti}, {Fong}, \& {Zauderer}}]{Laskar2018_140311}
{Laskar}, T., {Berger}, E., {Chornock}, R., {et~al.}
  2018{\natexlab{\hspace{0pt}a}},
  \hypersetup{urlcolor=magenta}\href{https://dx.doi.org/10.3847/1538-4357/aab8f5}{\apj},
  \hypersetup{urlcolor=blue}\href{https://ui.adsabs.harvard.edu/abs/2018ApJ...858...65L}{858,
  65}

\bibitem[{{Laskar} {et~al.}(2015){Laskar}, {Berger}, {Margutti}, {Perley},
  {Zauderer}, {Sari}, \& {Fong}}]{Laskar2015_eneinjection}
{Laskar}, T., {Berger}, E., {Margutti}, R., {et~al.} 2015,
  \hypersetup{urlcolor=magenta}\href{https://dx.doi.org/10.1088/0004-637X/814/1/1}{\apj},
  \hypersetup{urlcolor=blue}\href{https://ui.adsabs.harvard.edu/abs/2015ApJ...814....1L}{814,
  1}

\bibitem[{{Laskar} {et~al.}(2020){Laskar}, {Hull}, \& {Cortes}}]{Laskar2020}
{Laskar}, T., {Hull}, C. L.~H., \& {Cortes}, P. 2020,
  \hypersetup{urlcolor=magenta}\href{https://dx.doi.org/10.3847/1538-4357/ab88cc}{\apj},
  \hypersetup{urlcolor=blue}\href{https://ui.adsabs.harvard.edu/abs/2020ApJ...895...64L}{895,
  64}

\bibitem[{{Laskar} \& {Perley}(2021)}]{grb210702a_a}
{Laskar}, T., \& {Perley}, D. 2021, GRB Coordinates Network,
  \hypersetup{urlcolor=blue}\href{https://ui.adsabs.harvard.edu/abs/2021GCN.30423....1L}{30423,
  1}

\bibitem[{{Laskar} {et~al.}(2013){Laskar}, {Berger}, {Zauderer}, {Margutti},
  {Soderberg}, {Chakraborti}, {Lunnan}, {Chornock}, {Chandra}, \&
  {Ray}}]{Laskar2013}
{Laskar}, T., {Berger}, E., {Zauderer}, B.~A., {et~al.} 2013,
  \hypersetup{urlcolor=magenta}\href{https://dx.doi.org/10.1088/0004-637X/776/2/119}{\apj},
  \hypersetup{urlcolor=blue}\href{https://ui.adsabs.harvard.edu/abs/2013ApJ...776..119L}{776,
  119}

\bibitem[{{Laskar} {et~al.}(2016){Laskar}, {Alexander}, {Berger}, {Fong},
  {Margutti}, {Shivvers}, {Williams}, {Kopa{\v{c}}}, {Kobayashi}, {Mundell},
  {Gomboc}, {Zheng}, {Menten}, {Graham}, \& {Filippenko}}]{Laskar2016}
{Laskar}, T., {Alexander}, K.~D., {Berger}, E., {et~al.} 2016,
  \hypersetup{urlcolor=magenta}\href{https://dx.doi.org/10.3847/1538-4357/833/1/88}{\apj},
  \hypersetup{urlcolor=blue}\href{https://ui.adsabs.harvard.edu/abs/2016ApJ...833...88L}{833,
  88}

\bibitem[{{Laskar} {et~al.}(2018{\natexlab{\hspace{0pt}b}}){Laskar}, {Berger},
  {Margutti}, {Zauderer}, {Williams}, {Fong}, {Sari}, {Alexander}, \&
  {Kamble}}]{Laskar2018_140304}
{Laskar}, T., {Berger}, E., {Margutti}, R., {et~al.}
  2018{\natexlab{\hspace{0pt}b}},
  \hypersetup{urlcolor=magenta}\href{https://dx.doi.org/10.3847/1538-4357/aabfd8}{\apj},
  \hypersetup{urlcolor=blue}\href{https://ui.adsabs.harvard.edu/abs/2018ApJ...859..134L}{859,
  134}

\bibitem[{{Laskar} {et~al.}(2018{\natexlab{\hspace{0pt}c}}){Laskar},
  {Alexander}, {Berger}, {Guidorzi}, {Margutti}, {Fong}, {Kilpatrick}, {Milne},
  {Drout}, {Mundell}, {Kobayashi}, {Lunnan}, {Barniol Duran}, {Menten}, {Ioka},
  \& {Williams}}]{Laskar2018}
{Laskar}, T., {Alexander}, K.~D., {Berger}, E., {et~al.}
  2018{\natexlab{\hspace{0pt}c}},
  \hypersetup{urlcolor=magenta}\href{https://dx.doi.org/10.3847/1538-4357/aacbcc}{\apj},
  \hypersetup{urlcolor=blue}\href{https://ui.adsabs.harvard.edu/abs/2018ApJ...862...94L}{862,
  94}

\bibitem[{{Laskar} {et~al.}(2019{\natexlab{\hspace{0pt}a}}){Laskar}, {van
  Eerten}, {Schady}, {Mundell}, {Alexander}, {Barniol Duran}, {Berger},
  {Bolmer}, {Chornock}, {Coppejans}, {Fong}, {Gomboc}, {Jordana-Mitjans},
  {Kobayashi}, {Margutti}, {Menten}, {Sari}, {Yamazaki}, {Lipunov},
  {Gorbovskoy}, {Kornilov}, {Tyurina}, {Zimnukhov}, {Podesta}, {Levato},
  {Buckley}, {Tlatov}, {Rebolo}, \& {Serra-Ricart}}]{Laskar2019}
{Laskar}, T., {van Eerten}, H., {Schady}, P., {et~al.}
  2019{\natexlab{\hspace{0pt}a}},
  \hypersetup{urlcolor=magenta}\href{https://dx.doi.org/10.3847/1538-4357/ab40ce}{\apj},
  \hypersetup{urlcolor=blue}\href{https://ui.adsabs.harvard.edu/abs/2019ApJ...884..121L}{884,
  121}

\bibitem[{{Laskar} {et~al.}(2019{\natexlab{\hspace{0pt}b}}){Laskar},
  {Alexander}, {Gill}, {Granot}, {Berger}, {Mundell}, {Barniol Duran},
  {Bolmer}, {Duffell}, {van Eerten}, {Fong}, {Kobayashi}, {Margutti}, \&
  {Schady}}]{Laskar2019_grb190114c}
{Laskar}, T., {Alexander}, K.~D., {Gill}, R., {et~al.}
  2019{\natexlab{\hspace{0pt}b}},
  \hypersetup{urlcolor=magenta}\href{https://dx.doi.org/10.3847/2041-8213/ab2247}{\apjl},
  \hypersetup{urlcolor=blue}\href{https://ui.adsabs.harvard.edu/abs/2019ApJ...878L..26L}{878,
  L26}

\bibitem[{{Laskar} {et~al.}(2021{\natexlab{\hspace{0pt}c}}){Laskar},
  {Alexander}, {Kilpatrick}, {Schroeder}, {Fong}, {Berger}, {Margutti},
  {Mundell}, {Schady}, \& {a larger Collaboration}}]{Laskar2021}
{Laskar}, T., {Alexander}, K.~D., {Kilpatrick}, C., {et~al.}
  2021{\natexlab{\hspace{0pt}c}}, GRB Coordinates Network,
  \hypersetup{urlcolor=blue}\href{https://ui.adsabs.harvard.edu/abs/2021GCN.30217....1L}{30217,
  1}

\bibitem[{{Law} {et~al.}(2019){Law}, {Gaensler}, {Metzger}, {Ofek}, \&
  {Sironi}}]{Law2019}
{Law}, C., {Gaensler}, B., {Metzger}, B., {Ofek}, E., \& {Sironi}, L. 2019, in
  American Astronomical Society Meeting Abstracts, Vol. 233, American
  Astronomical Society Meeting Abstracts \#233, 424.05

\bibitem[{{Lei} {et~al.}(2016){Lei}, {Yuan}, {Zhang}, \& {Wang}}]{Lei2016}
{Lei}, W.-H., {Yuan}, Q., {Zhang}, B., \& {Wang}, D. 2016,
  \hypersetup{urlcolor=magenta}\href{https://dx.doi.org/10.3847/0004-637X/816/1/20}{\apj},
  \hypersetup{urlcolor=blue}\href{https://ui.adsabs.harvard.edu/abs/2016ApJ...816...20L}{816,
  20}

\bibitem[{{Leung} {et~al.}(2021){Leung}, {Murphy}, {Ghirlanda}, {Kaplan},
  {Lenc}, {Dobie}, {Banfield}, {Hale}, {Hotan}, {McConnell}, {Moss},
  {Pritchard}, {Raja}, {Stewart}, \& {Whiting}}]{Leung2021}
{Leung}, J.~K., {Murphy}, T., {Ghirlanda}, G., {et~al.} 2021,
  \hypersetup{urlcolor=magenta}\href{https://dx.doi.org/10.1093/mnras/stab326}{\mnras},
  \hypersetup{urlcolor=blue}\href{https://ui.adsabs.harvard.edu/abs/2021MNRAS.503.1847L}{503,
  1847}

\bibitem[{{Levan} {et~al.}(2011){Levan}, {Tanvir}, {Cenko}, {Perley},
  {Wiersema}, {Bloom}, {Fruchter}, {de Ugarte Postigo}, {O'Brien}, {Butler},
  {van der Horst}, {Leloudas}, {Morgan}, {Misra}, {Bower}, {Farihi},
  {Tunnicliffe}, {Modjaz}, {Silverman}, {Hjorth}, {Th{\"o}ne}, {Cucchiara},
  {Cer{\'o}n}, {Castro-Tirado}, {Arnold}, {Bremer}, {Brodie}, {Carroll},
  {Cooper}, {Curran}, {Cutri}, {Ehle}, {Forbes}, {Fynbo}, {Gorosabel},
  {Graham}, {Hoffman}, {Guziy}, {Jakobsson}, {Kamble}, {Kerr}, {Kasliwal},
  {Kouveliotou}, {Kocevski}, {Law}, {Nugent}, {Ofek}, {Poznanski}, {Quimby},
  {Rol}, {Romanowsky}, {S{\'a}nchez-Ram{\'\i}rez}, {Schulze}, {Singh}, {van
  Spaandonk}, {Starling}, {Strom}, {Tello}, {Vaduvescu}, {Wheatley}, {Wijers},
  {Winters}, \& {Xu}}]{Levan2011}
{Levan}, A.~J., {Tanvir}, N.~R., {Cenko}, S.~B., {et~al.} 2011,
  \hypersetup{urlcolor=magenta}\href{https://dx.doi.org/10.1126/science.1207143}{Science},
  \hypersetup{urlcolor=blue}\href{https://ui.adsabs.harvard.edu/abs/2011Sci...333..199L}{333,
  199}

\bibitem[{{LSST Science Collaboration} {et~al.}(2009){LSST Science
  Collaboration}, {Abell}, {Allison}, {Anderson}, {Andrew}, {Angel}, {Armus},
  {Arnett}, {Asztalos}, {Axelrod}, {Bailey}, {Ballantyne}, {Bankert},
  {Barkhouse}, {Barr}, {Barrientos}, {Barth}, {Bartlett}, {Becker}, {Becla},
  {Beers}, {Bernstein}, {Biswas}, {Blanton}, {Bloom}, {Bochanski}, {Boeshaar},
  {Borne}, {Bradac}, {Brandt}, {Bridge}, {Brown}, {Brunner}, {Bullock},
  {Burgasser}, {Burge}, {Burke}, {Cargile}, {Chandrasekharan}, {Chartas},
  {Chesley}, {Chu}, {Cinabro}, {Claire}, {Claver}, {Clowe}, {Connolly}, {Cook},
  {Cooke}, {Cooray}, {Covey}, {Culliton}, {de Jong}, {de Vries}, {Debattista},
  {Delgado}, {Dell'Antonio}, {Dhital}, {Di Stefano}, {Dickinson}, {Dilday},
  {Djorgovski}, {Dobler}, {Donalek}, {Dubois-Felsmann}, {Durech},
  {Eliasdottir}, {Eracleous}, {Eyer}, {Falco}, {Fan}, {Fassnacht}, {Ferguson},
  {Fernandez}, {Fields}, {Finkbeiner}, {Figueroa}, {Fox}, {Francke}, {Frank},
  {Frieman}, {Fromenteau}, {Furqan}, {Galaz}, {Gal-Yam}, {Garnavich},
  {Gawiser}, {Geary}, {Gee}, {Gibson}, {Gilmore}, {Grace}, {Green}, {Gressler},
  {Grillmair}, {Habib}, {Haggerty}, {Hamuy}, {Harris}, {Hawley}, {Heavens},
  {Hebb}, {Henry}, {Hileman}, {Hilton}, {Hoadley}, {Holberg}, {Holman},
  {Howell}, {Infante}, {Ivezic}, {Jacoby}, {Jain}, {R}, {Jedicke}, {Jee},
  {Garrett Jernigan}, {Jha}, {Johnston}, {Jones}, {Juric}, {Kaasalainen},
  {Styliani}, {Kafka}, {Kahn}, {Kaib}, {Kalirai}, {Kantor}, {Kasliwal},
  {Keeton}, {Kessler}, {Knezevic}, {Kowalski}, {Krabbendam}, {Krughoff},
  {Kulkarni}, {Kuhlman}, {Lacy}, {Lepine}, {Liang}, {Lien}, {Lira}, {Long},
  {Lorenz}, {Lotz}, {Lupton}, {Lutz}, {Macri}, {Mahabal}, {Mandelbaum},
  {Marshall}, {May}, {McGehee}, {Meadows}, {Meert}, {Milani}, {Miller},
  {Miller}, {Mills}, {Minniti}, {Monet}, {Mukadam}, {Nakar}, {Neill}, {Newman},
  {Nikolaev}, {Nordby}, {O'Connor}, {Oguri}, {Oliver}, {Olivier}, {Olsen},
  {Olsen}, {Olszewski}, {Oluseyi}, {Padilla}, {Parker}, {Pepper}, {Peterson},
  {Petry}, {Pinto}, {Pizagno}, {Popescu}, {Prsa}, {Radcka}, {Raddick},
  {Rasmussen}, {Rau}, {Rho}, {Rhoads}, {Richards}, {Ridgway}, {Robertson},
  {Roskar}, {Saha}, {Sarajedini}, {Scannapieco}, {Schalk}, {Schindler},
  {Schmidt}, {Schmidt}, {Schneider}, {Schumacher}, {Scranton}, {Sebag},
  {Seppala}, {Shemmer}, {Simon}, {Sivertz}, {Smith}, {Allyn Smith}, {Smith},
  {Spitz}, {Stanford}, {Stassun}, {Strader}, {Strauss}, {Stubbs}, {Sweeney},
  {Szalay}, {Szkody}, {Takada}, {Thorman}, {Trilling}, {Trimble}, {Tyson}, {Van
  Berg}, {Vanden Berk}, {VanderPlas}, {Verde}, {Vrsnak}, {Walkowicz},
  {Wandelt}, {Wang}, {Wang}, {Warner}, {Wechsler}, {West}, {Wiecha},
  {Williams}, {Willman}, {Wittman}, {Wolff}, {Wood-Vasey}, {Wozniak}, {Young},
  {Zentner}, \& {Zhan}}]{LSST2009}
{LSST Science Collaboration}, {Abell}, P.~A., {Allison}, J., {et~al.} 2009,
  arXiv e-prints,
  \hypersetup{urlcolor=magenta}\href{https://arxiv.org/abs/0912.0201}{arXiv}{:}\hypersetup{urlcolor=blue}\href{https://ui.adsabs.harvard.edu/abs/2009arXiv0912.0201L}{0912.0201}

\bibitem[{{Madau} \& {Dickinson}(2014)}]{Madau2014}
{Madau}, P., \& {Dickinson}, M. 2014,
  \hypersetup{urlcolor=magenta}\href{https://dx.doi.org/10.1146/annurev-astro-081811-125615}{\araa},
  \hypersetup{urlcolor=blue}\href{https://ui.adsabs.harvard.edu/abs/2014ARA&A..52..415M}{52,
  415}

\bibitem[{{Maeda} {et~al.}(2021){Maeda}, {Chandra}, {Matsuoka}, {Ryder},
  {Moriya}, {Kuncarayakti}, {Lee}, {Kundu}, {Patnaude}, {Saito}, \&
  {Folatelli}}]{Maeda2021}
{Maeda}, K., {Chandra}, P., {Matsuoka}, T., {et~al.} 2021,
  \hypersetup{urlcolor=magenta}\href{https://dx.doi.org/10.3847/1538-4357/ac0dbc}{\apj},
  \hypersetup{urlcolor=blue}\href{https://ui.adsabs.harvard.edu/abs/2021ApJ...918...34M}{918,
  34}

\bibitem[{{Margalit} \& {Metzger}(2019)}]{Margalit2019}
{Margalit}, B., \& {Metzger}, B.~D. 2019,
  \hypersetup{urlcolor=magenta}\href{https://dx.doi.org/10.3847/2041-8213/ab2ae2}{\apjl},
  \hypersetup{urlcolor=blue}\href{https://ui.adsabs.harvard.edu/abs/2019ApJ...880L..15M}{880,
  L15}

\bibitem[{{Margutti} {et~al.}(2019){Margutti}, {Metzger}, {Chornock}, {Vurm},
  {Roth}, {Grefenstette}, {Savchenko}, {Cartier}, {Steiner}, {Terreran},
  {Margalit}, {Migliori}, {Milisavljevic}, {Alexander}, {Bietenholz},
  {Blanchard}, {Bozzo}, {Brethauer}, {Chilingarian}, {Coppejans}, {Ducci},
  {Ferrigno}, {Fong}, {G{\"o}tz}, {Guidorzi}, {Hajela}, {Hurley}, {Kuulkers},
  {Laurent}, {Mereghetti}, {Nicholl}, {Patnaude}, {Ubertini}, {Banovetz},
  {Bartel}, {Berger}, {Coughlin}, {Eftekhari}, {Frederiks}, {Kozlova},
  {Laskar}, {Svinkin}, {Drout}, {MacFadyen}, \& {Paterson}}]{Margutti2019}
{Margutti}, R., {Metzger}, B.~D., {Chornock}, R., {et~al.} 2019,
  \hypersetup{urlcolor=magenta}\href{https://dx.doi.org/10.3847/1538-4357/aafa01}{\apj},
  \hypersetup{urlcolor=blue}\href{https://ui.adsabs.harvard.edu/abs/2019ApJ...872...18M}{872,
  18}

\bibitem[{{Martin} {et~al.}(2010){Martin}, {Petitpas}, {de Ugarte Postigo},
  {Gurwell}, {Castro-Tirado}, {Gorosabel}, {Garcia-Apaadoo}, {De Breuck}, \&
  {Lundgren}}]{Martin2010}
{Martin}, S., {Petitpas}, G., {de Ugarte Postigo}, A., {et~al.} 2010, GRB
  Coordinates Network,
  \hypersetup{urlcolor=blue}\href{https://ui.adsabs.harvard.edu/abs/2010GCN.10630....1M}{10630,
  1}

\bibitem[{{Mattila} {et~al.}(2018){Mattila}, {P{\'e}rez-Torres}, {Efstathiou},
  {Mimica}, {Fraser}, {Kankare}, {Alberdi}, {Aloy}, {Heikkil{\"a}}, {Jonker},
  {Lundqvist}, {Mart{\'\i}-Vidal}, {Meikle}, {Romero-Ca{\~n}izales}, {Smartt},
  {Tsygankov}, {Varenius}, {Alonso-Herrero}, {Bondi}, {Fransson},
  {Herrero-Illana}, {Kangas}, {Kotak}, {Ram{\'\i}rez-Olivencia},
  {V{\"a}is{\"a}nen}, {Beswick}, {Clements}, {Greimel}, {Harmanen},
  {Kotilainen}, {Nandra}, {Reynolds}, {Ryder}, {Walton}, {Wiik}, \&
  {{\"O}stlin}}]{Mattila2018}
{Mattila}, S., {P{\'e}rez-Torres}, M., {Efstathiou}, A., {et~al.} 2018,
  \hypersetup{urlcolor=magenta}\href{https://dx.doi.org/10.1126/science.aao4669}{Science},
  \hypersetup{urlcolor=blue}\href{https://ui.adsabs.harvard.edu/abs/2018Sci...361..482M}{361,
  482}

\bibitem[{{Melandri} {et~al.}(2010){Melandri}, {Kobayashi}, {Mundell},
  {Guidorzi}, {de Ugarte Postigo}, {Pooley}, {Yoshida}, {Bersier},
  {Castro-Tirado}, {Jel{\'\i}nek}, {Gomboc}, {Gorosabel}, {Kub{\'a}nek},
  {Bremer}, {Winters}, {Steele}, {de Gregorio-Monsalvo}, {Smith},
  {Garc{\'\i}a-Appadoo}, {Sota}, \& {Lundgren}}]{Mealandri2010}
{Melandri}, A., {Kobayashi}, S., {Mundell}, C.~G., {et~al.} 2010,
  \hypersetup{urlcolor=magenta}\href{https://dx.doi.org/10.1088/0004-637X/723/2/1331}{\apj},
  \hypersetup{urlcolor=blue}\href{https://ui.adsabs.harvard.edu/abs/2010ApJ...723.1331M}{723,
  1331}

\bibitem[{{M{\'e}sz{\'a}ros} \& {Rees}(1997)}]{Meszaros1997}
{M{\'e}sz{\'a}ros}, P., \& {Rees}, M.~J. 1997,
  \hypersetup{urlcolor=magenta}\href{https://dx.doi.org/10.1086/310692}{\apjl},
  \hypersetup{urlcolor=blue}\href{https://ui.adsabs.harvard.edu/abs/1997ApJ...482L..29M}{482,
  L29}

\bibitem[{{Metzger} \& {Bower}(2014)}]{Metzger2014}
{Metzger}, B.~D., \& {Bower}, G.~C. 2014,
  \hypersetup{urlcolor=magenta}\href{https://dx.doi.org/10.1093/mnras/stt2010}{\mnras},
  \hypersetup{urlcolor=blue}\href{https://ui.adsabs.harvard.edu/abs/2014MNRAS.437.1821M}{437,
  1821}

\bibitem[{{Metzger} {et~al.}(2012){Metzger}, {Giannios}, \&
  {Mimica}}]{Metzger2012}
{Metzger}, B.~D., {Giannios}, D., \& {Mimica}, P. 2012,
  \hypersetup{urlcolor=magenta}\href{https://dx.doi.org/10.1111/j.1365-2966.2011.20273.x}{\mnras},
  \hypersetup{urlcolor=blue}\href{https://ui.adsabs.harvard.edu/abs/2012MNRAS.420.3528M}{420,
  3528}

\bibitem[{{Metzger} {et~al.}(2008){Metzger}, {Quataert}, \&
  {Thompson}}]{Metzger2008}
{Metzger}, B.~D., {Quataert}, E., \& {Thompson}, T.~A. 2008,
  \hypersetup{urlcolor=magenta}\href{https://dx.doi.org/10.1111/j.1365-2966.2008.12923.x}{\mnras},
  \hypersetup{urlcolor=blue}\href{https://ui.adsabs.harvard.edu/abs/2008MNRAS.385.1455M}{385,
  1455}

\bibitem[{{Metzger} {et~al.}(2015){Metzger}, {Williams}, \&
  {Berger}}]{Metzger2015}
{Metzger}, B.~D., {Williams}, P.~K.~G., \& {Berger}, E. 2015,
  \hypersetup{urlcolor=magenta}\href{https://dx.doi.org/10.1088/0004-637X/806/2/224}{\apj},
  \hypersetup{urlcolor=blue}\href{https://ui.adsabs.harvard.edu/abs/2015ApJ...806..224M}{806,
  224}

\bibitem[{{Murphy} {et~al.}(2013){Murphy}, {Chatterjee}, {Kaplan}, {Banyer},
  {Bell}, {Bignall}, {Bower}, {Cameron}, {Coward}, {Cordes}, {Croft}, {Curran},
  {Djorgovski}, {Farrell}, {Frail}, {Gaensler}, {Galloway}, {Gendre}, {Green},
  {Hancock}, {Johnston}, {Kamble}, {Law}, {Lazio}, {Lo}, {Macquart}, {Rea},
  {Rebbapragada}, {Reynolds}, {Ryder}, {Schmidt}, {Soria}, {Stairs}, {Tingay},
  {Torkelsson}, {Wagstaff}, {Walker}, {Wayth}, \& {Williams}}]{Murphy2013}
{Murphy}, T., {Chatterjee}, S., {Kaplan}, D.~L., {et~al.} 2013,
  \hypersetup{urlcolor=magenta}\href{https://dx.doi.org/10.1017/pasa.2012.006}{\pasa},
  \hypersetup{urlcolor=blue}\href{https://ui.adsabs.harvard.edu/abs/2013PASA...30....6M}{30,
  e006}

\bibitem[{{Naess} {et~al.}(2021){Naess}, {Battaglia}, {Richard Bond},
  {Calabrese}, {Choi}, {Cothard}, {Devlin}, {Duell}, {Duivenvoorden},
  {Dunkley}, {D{\"u}nner}, {Gallardo}, {Gralla}, {Guan}, {Halpern}, {Colin
  Hill}, {Hilton}, {Huffenberger}, {Koopman}, {Kosowsky}, {Madhavacheril},
  {McMahon}, {Nati}, {Niemack}, {Page}, {Partridge}, {Salatino}, {Sehgal},
  {Spergel}, {Staggs}, {Wollack}, \& {Xu}}]{Naess2021}
{Naess}, S., {Battaglia}, N., {Richard Bond}, J., {et~al.} 2021,
  \hypersetup{urlcolor=magenta}\href{https://dx.doi.org/10.3847/1538-4357/abfe6d}{\apj},
  \hypersetup{urlcolor=blue}\href{https://ui.adsabs.harvard.edu/abs/2021ApJ...915...14N}{915,
  14}

\bibitem[{{Nakar} \& {Piran}(2004)}]{Nakar2004}
{Nakar}, E., \& {Piran}, T. 2004,
  \hypersetup{urlcolor=magenta}\href{https://dx.doi.org/10.1111/j.1365-2966.2004.08099.x}{\mnras},
  \hypersetup{urlcolor=blue}\href{https://ui.adsabs.harvard.edu/abs/2004MNRAS.353..647N}{353,
  647}

\bibitem[{{Nakar} \& {Piran}(2011)}]{Nakar2011}
{Nakar}, E., \& {Piran}, T. 2011,
  \hypersetup{urlcolor=magenta}\href{https://dx.doi.org/10.1038/nature10365}{\nat},
  \hypersetup{urlcolor=blue}\href{https://ui.adsabs.harvard.edu/abs/2011Natur.478...82N}{478,
  82}

\bibitem[{{Ofek} {et~al.}(2010{\natexlab{\hspace{0pt}a}}){Ofek}, {Breslauer},
  {Gal-Yam}, {Frail}, {Kasliwal}, {Kulkarni}, \& {Waxman}}]{Ofek2010a}
{Ofek}, E.~O., {Breslauer}, B., {Gal-Yam}, A., {et~al.}
  2010{\natexlab{\hspace{0pt}a}},
  \hypersetup{urlcolor=magenta}\href{https://dx.doi.org/10.1088/0004-637X/711/1/517}{\apj},
  \hypersetup{urlcolor=blue}\href{https://ui.adsabs.harvard.edu/abs/2010ApJ...711..517O}{711,
  517}

\bibitem[{{Ofek} {et~al.}(2010{\natexlab{\hspace{0pt}b}}){Ofek}, {Rabinak},
  {Neill}, {Arcavi}, {Cenko}, {Waxman}, {Kulkarni}, {Gal-Yam}, {Nugent},
  {Bildsten}, {Bloom}, {Filippenko}, {Forster}, {Howell}, {Jacobsen},
  {Kasliwal}, {Law}, {Martin}, {Poznanski}, {Quimby}, {Shen}, {Sullivan},
  {Dekany}, {Rahmer}, {Hale}, {Smith}, {Zolkower}, {Velur}, {Walters},
  {Henning}, {Bui}, \& {McKenna}}]{Ofek2010}
{Ofek}, E.~O., {Rabinak}, I., {Neill}, J.~D., {et~al.}
  2010{\natexlab{\hspace{0pt}b}},
  \hypersetup{urlcolor=magenta}\href{https://dx.doi.org/10.1088/0004-637X/724/2/1396}{\apj},
  \hypersetup{urlcolor=blue}\href{https://ui.adsabs.harvard.edu/abs/2010ApJ...724.1396O}{724,
  1396}

\bibitem[{{Panaitescu} \& {Kumar}(2002)}]{Panaitescu2002}
{Panaitescu}, A., \& {Kumar}, P. 2002,
  \hypersetup{urlcolor=magenta}\href{https://dx.doi.org/10.1086/340094}{\apj},
  \hypersetup{urlcolor=blue}\href{https://ui.adsabs.harvard.edu/abs/2002ApJ...571..779P}{571,
  779}

\bibitem[{{Pandey} {et~al.}(2009){Pandey}, {Castro-Tirado}, {Jel{\'\i}nek},
  {Kamble}, {Gorosabel}, {de Ugarte Postigo}, {Prins}, {Oreiro}, {Chantry},
  {Trushkin}, {Bremer}, {Winters}, {Pozanenko}, {Krugly}, {Slyusarev},
  {Kornienko}, {Erofeeva}, {Misra}, {Ramprakash}, {Mohan}, {Bhattacharya},
  {Volnova}, {Pl{\'a}}, {Ibrahimov}, {Im}, {Volvach}, \& {Wijers}}]{Pandey2009}
{Pandey}, S.~B., {Castro-Tirado}, A.~J., {Jel{\'\i}nek}, M., {et~al.} 2009,
  \hypersetup{urlcolor=magenta}\href{https://dx.doi.org/10.1051/0004-6361/200811135}{\aap},
  \hypersetup{urlcolor=blue}\href{https://ui.adsabs.harvard.edu/abs/2009A&A...504...45P}{504,
  45}

\bibitem[{{Pasham} {et~al.}(2015){Pasham}, {Cenko}, {Levan}, {Bower}, {Horesh},
  {Brown}, {Dolan}, {Wiersema}, {Filippenko}, {Fruchter}, {Greiner}, {O'Brien},
  {Page}, {Rau}, \& {Tanvir}}]{Pasham2015}
{Pasham}, D.~R., {Cenko}, S.~B., {Levan}, A.~J., {et~al.} 2015,
  \hypersetup{urlcolor=magenta}\href{https://dx.doi.org/10.1088/0004-637X/805/1/68}{\apj},
  \hypersetup{urlcolor=blue}\href{https://ui.adsabs.harvard.edu/abs/2015ApJ...805...68P}{805,
  68}

\bibitem[{{Perley}(2013{\natexlab{\hspace{0pt}a}})}]{Perley2013_grb130418a}
{Perley}, D.~A. 2013{\natexlab{\hspace{0pt}a}}, GRB Coordinates Network,
  \hypersetup{urlcolor=blue}\href{https://ui.adsabs.harvard.edu/abs/2013GCN.14387....1P}{14387,
  1}

\bibitem[{{Perley}(2013{\natexlab{\hspace{0pt}b}})}]{Perley2013_grb131108a}
{Perley}, D.~A. 2013{\natexlab{\hspace{0pt}b}}, GRB Coordinates Network,
  \hypersetup{urlcolor=blue}\href{https://ui.adsabs.harvard.edu/abs/2013GCN.15478....1P}{15478,
  1}

\bibitem[{{Perley}(2014)}]{Perley2014_grb131231A}
{Perley}, D.~A. 2014, GRB Coordinates Network,
  \hypersetup{urlcolor=blue}\href{https://ui.adsabs.harvard.edu/abs/2014GCN.15680....1P}{15680,
  1}

\bibitem[{{Perley} {et~al.}(2022){Perley}, {Ho}, {Petitpas}, \&
  {Keating}}]{Perley2022}
{Perley}, D.~A., {Ho}, A.~Y.~Q., {Petitpas}, G., \& {Keating}, G. 2022, GRB
  Coordinates Network,
  \hypersetup{urlcolor=blue}\href{https://ui.adsabs.harvard.edu/abs/2022.31627....1P}{31627,
  1}

\bibitem[{{Perley} \& {Keating}(2013)}]{Perley2013_grb130215a}
{Perley}, D.~A., \& {Keating}, G. 2013, GRB Coordinates Network,
  \hypersetup{urlcolor=blue}\href{https://ui.adsabs.harvard.edu/abs/2013GCN.14210....1P}{14210,
  1}

\bibitem[{{Perley} {et~al.}(2017){Perley}, {Schulze}, \& {de Ugarte
  Postigo}}]{Perley2017}
{Perley}, D.~A., {Schulze}, S., \& {de Ugarte Postigo}, A. 2017, GRB
  Coordinates Network,
  \hypersetup{urlcolor=blue}\href{https://ui.adsabs.harvard.edu/abs/2017.22252....1P}{22252,
  1}

\bibitem[{{Perley} {et~al.}(2014){Perley}, {Cenko}, {Corsi}, {Tanvir}, {Levan},
  {Kann}, {Sonbas}, {Wiersema}, {Zheng}, {Zhao}, {Bai}, {Bremer},
  {Castro-Tirado}, {Chang}, {Clubb}, {Frail}, {Fruchter},
  {G{\"o}{\u{g}}{\"u}{\textcommabelow s}}, {Greiner}, {G{\"u}ver}, {Horesh},
  {Filippenko}, {Klose}, {Mao}, {Morgan}, {Pozanenko}, {Schmidl}, {Stecklum},
  {Tanga}, {Volnova}, {Volvach}, {Wang}, {Winters}, \& {Xin}}]{Perley2014}
{Perley}, D.~A., {Cenko}, S.~B., {Corsi}, A., {et~al.} 2014,
  \hypersetup{urlcolor=magenta}\href{https://dx.doi.org/10.1088/0004-637X/781/1/37}{\apj},
  \hypersetup{urlcolor=blue}\href{https://ui.adsabs.harvard.edu/abs/2014ApJ...781...37P}{781,
  37}

\bibitem[{{Perley} {et~al.}(2019){Perley}, {Mazzali}, {Yan}, {Cenko}, {Gezari},
  {Taggart}, {Blagorodnova}, {Fremling}, {Mockler}, {Singh}, {Tominaga},
  {Tanaka}, {Watson}, {Ahumada}, {Anupama}, {Ashall}, {Becerra}, {Bersier},
  {Bhalerao}, {Bloom}, {Butler}, {Copperwheat}, {Coughlin}, {De}, {Drake},
  {Duev}, {Frederick}, {Gonz{\'a}lez}, {Goobar}, {Heida}, {Ho}, {Horst},
  {Hung}, {Itoh}, {Jencson}, {Kasliwal}, {Kawai}, {Khanam}, {Kulkarni},
  {Kumar}, {Kumar}, {Kutyrev}, {Lee}, {Maeda}, {Mahabal}, {Murata}, {Neill},
  {Ngeow}, {Penprase}, {Pian}, {Quimby}, {Ramirez-Ruiz}, {Richer},
  {Rom{\'a}n-Z{\'u}{\~n}iga}, {Sahu}, {Srivastav}, {Socia}, {Sollerman},
  {Tachibana}, {Taddia}, {Tinyanont}, {Troja}, {Ward}, {Wee}, \&
  {Yu}}]{Perley2019}
{Perley}, D.~A., {Mazzali}, P.~A., {Yan}, L., {et~al.} 2019,
  \hypersetup{urlcolor=magenta}\href{https://dx.doi.org/10.1093/mnras/sty3420}{\mnras},
  \hypersetup{urlcolor=blue}\href{https://ui.adsabs.harvard.edu/abs/2019MNRAS.484.1031P}{484,
  1031}

\bibitem[{{Perley} {et~al.}(2021){Perley}, {Ho}, {Yao}, {Fremling}, {Anderson},
  {Schulze}, {Kumar}, {Anupama}, {Barway}, {Bellm}, {Bhalerao}, {Chen}, {Duev},
  {Galbany}, {Graham}, {Gromadzki}, {Guti{\'e}rrez}, {Ihanec}, {Inserram},
  {Kasliwal}, {Kool}, {Kulkarni}, {Laher}, {Masci}, {Neill}, {Nicholl},
  {Pursiainen}, {van Roestel}, {Sharma}, {Sollerman}, {Walters}, \&
  {Wiseman}}]{Perley2021}
{Perley}, D.~A., {Ho}, A. Y.~Q., {Yao}, Y., {et~al.} 2021, arXiv e-prints,
  \hypersetup{urlcolor=magenta}\href{https://arxiv.org/abs/2103.01968}{arXiv}{:}\hypersetup{urlcolor=blue}\href{https://ui.adsabs.harvard.edu/abs/2021arXiv210301968P}{2103.01968}

\bibitem[{{Planck Collaboration} {et~al.}(2016){Planck Collaboration}, {Adam},
  {Ade}, {Aghanim}, {Akrami}, {Alves}, {Arg{\"u}eso}, {Arnaud}, {Arroja},
  {Ashdown}, {Aumont}, {Baccigalupi}, {Ballardini}, {Band ay}, {Barreiro},
  {Bartlett}, {Bartolo}, {Basak}, {Battaglia}, {Battaner}, {Battye}, {Benabed},
  {Beno{\^\i}t}, {Benoit-L{\'e}vy}, {Bernard}, {Bersanelli}, {Bertincourt},
  {Bielewicz}, {Bikmaev}, {Bock}, {B{\"o}hringer}, {Bonaldi}, {Bonavera},
  {Bond}, {Borrill}, {Bouchet}, {Boulanger}, {Bucher}, {Burenin}, {Burigana},
  {Butler}, {Calabrese}, {Cardoso}, {Carvalho}, {Casaponsa}, {Castex},
  {Catalano}, {Challinor}, {Chamballu}, {Chary}, {Chiang}, {Chluba}, {Chon},
  {Christensen}, {Church}, {Clemens}, {Clements}, {Colombi}, {Colombo},
  {Combet}, {Comis}, {Contreras}, {Couchot}, {Coulais}, {Crill}, {Cruz},
  {Curto}, {Cuttaia}, {Danese}, {Davies}, {Davis}, {de Bernardis}, {de Rosa},
  {de Zotti}, {Delabrouille}, {Delouis}, {D{\'e}sert}, {Di Valentino},
  {Dickinson}, {Diego}, {Dolag}, {Dole}, {Donzelli}, {Dor{\'e}}, {Douspis},
  {Ducout}, {Dunkley}, {Dupac}, {Efstathiou}, {Eisenhardt}, {Elsner},
  {En{\ss}lin}, {Eriksen}, {Falgarone}, {Fantaye}, {Farhang}, {Feeney},
  {Fergusson}, {Fernandez-Cobos}, {Feroz}, {Finelli}, {Florido}, {Forni},
  {Frailis}, {Fraisse}, {Franceschet}, {Franceschi}, {Frejsel}, {Frolov},
  {Galeotta}, {Galli}, {Ganga}, {Gauthier}, {G{\'e}nova-Santos}, {Gerbino},
  {Ghosh}, {Giard}, {Giraud-H{\'e}raud}, {Giusarma}, {Gjerl{\o}w},
  {Gonz{\'a}lez-Nuevo}, {G{\'o}rski}, {Grainge}, {Gratton}, {Gregorio},
  {Gruppuso}, {Gudmundsson}, {Hamann}, {Handley}, {Hansen}, {Hanson},
  {Harrison}, {Heavens}, {Helou}, {Henrot-Versill{\'e}},
  {Hern{\'a}ndez-Monteagudo}, {Herranz}, {Hildebrandt}, {Hivon}, {Hobson},
  {Holmes}, {Hornstrup}, {Hovest}, {Huang}, {Huffenberger}, {Hurier},
  {Ili{\'c}}, {Jaffe}, {Jaffe}, {Jin}, {Jones}, {Juvela}, {Karakci},
  {Keih{\"a}nen}, {Keskitalo}, {Khamitov}, {Kiiveri}, {Kim}, {Kisner},
  {Kneissl}, {Knoche}, {Knox}, {Krachmalnicoff}, {Kunz}, {Kurki-Suonio},
  {Lacasa}, {Lagache}, {L{\"a}hteenm{\"a}ki}, {Lamarre}, {Langer}, {Lasenby},
  {Lattanzi}, {Lawrence}, {Le Jeune}, {Leahy}, {Lellouch}, {Leonardi},
  {Le{\'o}n-Tavares}, {Lesgourgues}, {Levrier}, {Lewis}, {Liguori}, {Lilje},
  {Lilley}, {Linden-V{\o}rnle}, {Lindholm}, {Liu}, {L{\'o}pez-Caniego},
  {Lubin}, {Ma}, {Mac{\'\i}as-P{\'e}rez}, {Maggio}, {Maino}, {Mak},
  {Mandolesi}, {Mangilli}, {Marchini}, {Marcos-Caballero}, {Marinucci},
  {Maris}, {Marshall}, {Martin}, {Martinelli}, {Mart{\'\i}nez-Gonz{\'a}lez},
  {Masi}, {Matarrese}, {Mazzotta}, {McEwen}, {McGehee}, {Mei}, {Meinhold},
  {Melchiorri}, {Melin}, {Mendes}, {Mennella}, {Migliaccio}, {Mikkelsen},
  {Millea}, {Mitra}, {Miville-Desch{\^e}nes}, {Molinari}, {Moneti}, {Montier},
  {Moreno}, {Morgante}, {Mortlock}, {Moss}, {Mottet}, {M{\"u}nchmeyer},
  {Munshi}, {Murphy}, {Narimani}, {Naselsky}, {Nastasi}, {Nati}, {Natoli},
  {Negrello}, {Netterfield}, {N{\o}rgaard-Nielsen}, {Noviello}, {Novikov},
  {Novikov}, {Olamaie}, {Oppermann}, {Orlando}, {Oxborrow}, {Paci}, {Pagano},
  {Pajot}, {Paladini}, {Pandolfi}, {Paoletti}, {Partridge}, {Pasian},
  {Patanchon}, {Pearson}, {Peel}, {Peiris}, {Pelkonen}, {Perdereau}, {Perotto},
  {Perrott}, {Perrotta}, {Pettorino}, {Piacentini}, {Piat}, {Pierpaoli},
  {Pietrobon}, {Plaszczynski}, {Pogosyan}, {Pointecouteau}, {Polenta}, {Popa},
  {Pratt}, {Pr{\'e}zeau}, {Prunet}, {Puget}, {Rachen}, {Racine}, {Reach},
  {Rebolo}, {Reinecke}, {Remazeilles}, {Renault}, {Renzi}, {Ristorcelli},
  {Rocha}, {Roman}, {Romelli}, {Rosset}, {Rossetti}, {Rotti}, {Roudier},
  {Rouill{\'e} d'Orfeuil}, {Rowan-Robinson}, {Rubi{\~n}o-Mart{\'\i}n},
  {Ruiz-Granados}, {Rumsey}, {Rusholme}, {Said}, {Salvatelli}, {Salvati},
  {Sandri}, {Sanghera}, {Santos}, {Saunders}, {Sauv{\'e}}, {Savelainen},
  {Savini}, {Schaefer}, {Schammel}, {Scott}, {Seiffert}, {Serra}, {Shellard},
  {Shimwell}, {Shiraishi}, {Smith}, {Souradeep}, {Spencer}, {Spinelli},
  {Stanford}, {Stern}, {Stolyarov}, {Stompor}, {Strong}, {Sudiwala}, {Sunyaev},
  {Sutter}, {Sutton}, {Suur-Uski}, {Sygnet}, {Tauber}, {Tavagnacco}, {Terenzi},
  {Texier}, {Toffolatti}, {Tomasi}, {Tornikoski}, {Tramonte}, {Tristram},
  {Troja}, {Trombetti}, {Tucci}, {Tuovinen}, {T{\"u}rler}, {Umana},
  {Valenziano}, {Valiviita}, {Van Tent}, {Vassallo}, {Vibert}, {Vidal}, {Viel},
  {Vielva}, {Villa}, {Wade}, {Walter}, {Wand elt}, {Watson}, {Wehus},
  {Welikala}, {Weller}, {White}, {White}, {Wilkinson}, {Yvon}, {Zacchei},
  {Zibin}, \& {Zonca}}]{Planck2016}
{Planck Collaboration}, {Adam}, R., {Ade}, P.~A.~R., {et~al.} 2016,
  \hypersetup{urlcolor=magenta}\href{https://dx.doi.org/10.1051/0004-6361/201527101}{\aap},
  \hypersetup{urlcolor=blue}\href{https://ui.adsabs.harvard.edu/abs/2016A&A...594A...1P}{594,
  A1}

\bibitem[{{Prentice} {et~al.}(2018){Prentice}, {Maguire}, {Smartt}, {Magee},
  {Schady}, {Sim}, {Chen}, {Clark}, {Colin}, {Fulton}, {McBrien}, {O'Neill},
  {Smith}, {Ashall}, {Chambers}, {Denneau}, {Flewelling}, {Heinze}, {Holoien},
  {Huber}, {Kochanek}, {Mazzali}, {Prieto}, {Rest}, {Shappee}, {Stalder},
  {Stanek}, {Stritzinger}, {Thompson}, \& {Tonry}}]{Prentice2018}
{Prentice}, S.~J., {Maguire}, K., {Smartt}, S.~J., {et~al.} 2018,
  \hypersetup{urlcolor=magenta}\href{https://dx.doi.org/10.3847/2041-8213/aadd90}{\apjl},
  \hypersetup{urlcolor=blue}\href{https://ui.adsabs.harvard.edu/abs/2018ApJ...865L...3P}{865,
  L3}

\bibitem[{{Pursiainen} {et~al.}(2018){Pursiainen}, {Childress}, {Smith},
  {Prajs}, {Sullivan}, {Davis}, {Foley}, {Asorey}, {Calcino}, {Carollo},
  {Curtin}, {D'Andrea}, {Glazebrook}, {Gutierrez}, {Hinton}, {Hoormann},
  {Inserra}, {Kessler}, {King}, {Kuehn}, {Lewis}, {Lidman}, {Macaulay},
  {M{\"o}ller}, {Nichol}, {Sako}, {Sommer}, {Swann}, {Tucker}, {Uddin},
  {Wiseman}, {Zhang}, {Abbott}, {Abdalla}, {Allam}, {Annis}, {Avila}, {Brooks},
  {Buckley-Geer}, {Burke}, {Carnero Rosell}, {Carrasco Kind}, {Carretero},
  {Castander}, {Cunha}, {Davis}, {De Vicente}, {Diehl}, {Doel}, {Eifler},
  {Flaugher}, {Fosalba}, {Frieman}, {Garc{\'\i}a-Bellido}, {Gruen}, {Gruendl},
  {Gutierrez}, {Hartley}, {Hollowood}, {Honscheid}, {James}, {Jeltema},
  {Kuropatkin}, {Li}, {Lima}, {Maia}, {Martini}, {Menanteau}, {Ogando},
  {Plazas}, {Roodman}, {Sanchez}, {Scarpine}, {Schindler}, {Smith},
  {Soares-Santos}, {Sobreira}, {Suchyta}, {Swanson}, {Tarle}, {Tucker},
  {Walker}, \& {DES Collaboration}}]{Pursiainen2018}
{Pursiainen}, M., {Childress}, M., {Smith}, M., {et~al.} 2018,
  \hypersetup{urlcolor=magenta}\href{https://dx.doi.org/10.1093/mnras/sty2309}{\mnras},
  \hypersetup{urlcolor=blue}\href{https://ui.adsabs.harvard.edu/abs/2018MNRAS.481..894P}{481,
  894}

\bibitem[{{Ravi} {et~al.}(2021){Ravi}, {Dykaar}, {Codd}, {Zaccagnini}, {Dong},
  {Drout}, {Gaensler}, {Hallinan}, \& {Law}}]{Ravi2021}
{Ravi}, V., {Dykaar}, H., {Codd}, J., {et~al.} 2021, arXiv e-prints,
  \hypersetup{urlcolor=magenta}\href{https://arxiv.org/abs/2102.05795}{arXiv}{:}\hypersetup{urlcolor=blue}\href{https://ui.adsabs.harvard.edu/abs/2021arXiv210205795R}{2102.05795}

\bibitem[{{Resmi} {et~al.}(2005){Resmi}, {Ishwara-Chandra}, {Castro-Tirado},
  {Bhattacharya}, {Rao}, {Bremer}, {Pandey}, {Sahu}, {Bhatt}, {Sagar},
  {Anupama}, {Subramaniam}, {Lundgren}, {Gorosabel}, {Guziy}, {de Ugarte
  Postigo}, {Castro Cer{\'o}n}, \& {Wiklind}}]{Resmi2005}
{Resmi}, L., {Ishwara-Chandra}, C.~H., {Castro-Tirado}, A.~J., {et~al.} 2005,
  \hypersetup{urlcolor=magenta}\href{https://dx.doi.org/10.1051/0004-6361:20041642}{\aap},
  \hypersetup{urlcolor=blue}\href{https://ui.adsabs.harvard.edu/abs/2005A&A...440..477R}{440,
  477}

\bibitem[{{Salafia} {et~al.}(2021){Salafia}, {Ravasio}, {Yang}, {An},
  {Orienti}, {Ghirlanda}, {Nava}, {Giroletti}, {Mohan}, {Spinelli}, {Zhang},
  {Marcote}, {Cim{\`o}}, {Wu}, \& {Li}}]{sry+21}
{Salafia}, O.~S., {Ravasio}, M.~E., {Yang}, J., {et~al.} 2021, arXiv e-prints,
  \hypersetup{urlcolor=magenta}\href{https://arxiv.org/abs/2106.07169}{arXiv}{:}\hypersetup{urlcolor=blue}\href{https://ui.adsabs.harvard.edu/abs/2021arXiv210607169S}{2106.07169}

\bibitem[{{S{\'a}nchez-Ram{\'\i}rez} {et~al.}(2017){S{\'a}nchez-Ram{\'\i}rez},
  {Hancock}, {J{\'o}hannesson}, {Murphy}, {de Ugarte Postigo}, {Gorosabel},
  {Kann}, {Kr{\"u}hler}, {Oates}, {Japelj}, {Th{\"o}ne}, {Lundgren}, {Perley},
  {Malesani}, {de Gregorio Monsalvo}, {Castro-Tirado}, {D'Elia}, {Fynbo},
  {Garcia-Appadoo}, {Goldoni}, {Greiner}, {Hu}, {Jel{\'\i}nek}, {Jeong},
  {Kamble}, {Klose}, {Kuin}, {Llorente}, {Mart{\'\i}n}, {Nicuesa Guelbenzu},
  {Rossi}, {Schady}, {Sparre}, {Sudilovsky}, {Tello}, {Updike}, {Wiersema}, \&
  {Zhang}}]{Sanchez2017}
{S{\'a}nchez-Ram{\'\i}rez}, R., {Hancock}, P.~J., {J{\'o}hannesson}, G.,
  {et~al.} 2017,
  \hypersetup{urlcolor=magenta}\href{https://dx.doi.org/10.1093/mnras/stw2608}{\mnras},
  \hypersetup{urlcolor=blue}\href{https://ui.adsabs.harvard.edu/abs/2017MNRAS.464.4624S}{464,
  4624}

\bibitem[{{Sari} \& {Piran}(1995)}]{Sari1995}
{Sari}, R., \& {Piran}, T. 1995,
  \hypersetup{urlcolor=magenta}\href{https://dx.doi.org/10.1086/309835}{\apjl},
  \hypersetup{urlcolor=blue}\href{https://ui.adsabs.harvard.edu/abs/1995ApJ...455L.143S}{455,
  L143}

\bibitem[{{Sari} {et~al.}(1998){Sari}, {Piran}, \& {Narayan}}]{Sari1998}
{Sari}, R., {Piran}, T., \& {Narayan}, R. 1998,
  \hypersetup{urlcolor=magenta}\href{https://dx.doi.org/10.1086/311269}{\apjl},
  \hypersetup{urlcolor=blue}\href{https://ui.adsabs.harvard.edu/abs/1998ApJ...497L..17S}{497,
  L17}

\bibitem[{{Schroeder} {et~al.}(2020){Schroeder}, {Margalit}, {Fong}, {Metzger},
  {Williams}, {Paterson}, {Alexander}, {Laskar}, {Goyal}, \&
  {Berger}}]{Schroeder+20}
{Schroeder}, G., {Margalit}, B., {Fong}, W.-f., {et~al.} 2020,
  \hypersetup{urlcolor=magenta}\href{https://dx.doi.org/10.3847/1538-4357/abb407}{\apj},
  \hypersetup{urlcolor=blue}\href{https://ui.adsabs.harvard.edu/abs/2020ApJ...902...82S}{902,
  82}

\bibitem[{{Sehgal} {et~al.}(2019){Sehgal}, {Aiola}, {Akrami}, {Basu},
  {Boylan-Kolchin}, {Bryan}, {Clesse}, {Cyr-Racine}, {Di Mascolo}, {Dicker},
  {Essinger-Hileman}, {Ferraro}, {Fuller}, {Han}, {Hasselfield}, {Holder},
  {Jain}, {Johnson}, {Johnson}, {Klaassen}, {Madhavacheril}, {Mauskopf},
  {Meerburg}, {Meyers}, {Mroczkowski}, {M{\"u}nchmeyer}, {Naess}, {Nagai},
  {Namikawa}, {Newburgh}, {Nguyen}, {Niemack}, {Oppenheimer}, {Pierpaoli},
  {Schaan}, {Slosar}, {Spergel}, {Switzer}, {van Engelen}, \&
  {Wollack}}]{Sehgal2019}
{Sehgal}, N., {Aiola}, S., {Akrami}, Y., {et~al.} 2019, in Bulletin of the
  American Astronomical Society, Vol.~51, 6

\bibitem[{{Shannon} {et~al.}(2018){Shannon}, {Macquart}, {Bannister}, {Ekers},
  {James}, {Os{\l}owski}, {Qiu}, {Sammons}, {Hotan}, {Voronkov}, {Beresford},
  {Brothers}, {Brown}, {Bunton}, {Chippendale}, {Haskins}, {Leach},
  {Marquarding}, {McConnell}, {Pilawa}, {Sadler}, {Troup}, {Tuthill},
  {Whiting}, {Allison}, {Anderson}, {Bell}, {Collier}, {G{\"u}rkan}, {Heald},
  \& {Riseley}}]{Shannon2018}
{Shannon}, R.~M., {Macquart}, J.~P., {Bannister}, K.~W., {et~al.} 2018,
  \hypersetup{urlcolor=magenta}\href{https://dx.doi.org/10.1038/s41586-018-0588-y}{\nat},
  \hypersetup{urlcolor=blue}\href{https://ui.adsabs.harvard.edu/abs/2018Natur.562..386S}{562,
  386}

\bibitem[{{Sheth} {et~al.}(2003){Sheth}, {Frail}, {White}, {Das}, {Bertoldi},
  {Walter}, {Kulkarni}, \& {Berger}}]{Sheth2003}
{Sheth}, K., {Frail}, D.~A., {White}, S., {et~al.} 2003,
  \hypersetup{urlcolor=magenta}\href{https://dx.doi.org/10.1086/378933}{\apjl},
  \hypersetup{urlcolor=blue}\href{https://ui.adsabs.harvard.edu/abs/2003ApJ...595L..33S}{595,
  L33}

\bibitem[{{Sijacki} {et~al.}(2015){Sijacki}, {Vogelsberger}, {Genel},
  {Springel}, {Torrey}, {Snyder}, {Nelson}, \& {Hernquist}}]{Sijacki2015}
{Sijacki}, D., {Vogelsberger}, M., {Genel}, S., {et~al.} 2015,
  \hypersetup{urlcolor=magenta}\href{https://dx.doi.org/10.1093/mnras/stv1340}{\mnras},
  \hypersetup{urlcolor=blue}\href{https://ui.adsabs.harvard.edu/abs/2015MNRAS.452..575S}{452,
  575}

\bibitem[{{Singer} {et~al.}(2013){Singer}, {Cenko}, {Kasliwal}, {Perley},
  {Ofek}, {Brown}, {Nugent}, {Kulkarni}, {Corsi}, {Frail}, {Bellm}, {Mulchaey},
  {Arcavi}, {Barlow}, {Bloom}, {Cao}, {Gehrels}, {Horesh}, {Masci}, {McEnery},
  {Rau}, {Surace}, \& {Yaron}}]{Singer2013}
{Singer}, L.~P., {Cenko}, S.~B., {Kasliwal}, M.~M., {et~al.} 2013,
  \hypersetup{urlcolor=magenta}\href{https://dx.doi.org/10.1088/2041-8205/776/2/L34}{\apjl},
  \hypersetup{urlcolor=blue}\href{https://ui.adsabs.harvard.edu/abs/2013ApJ...776L..34S}{776,
  L34}

\bibitem[{{Smartt} {et~al.}(2018){Smartt}, {Clark}, {Smith}, {McBrien},
  {Maguire}, {O'Neil}, {Fulton}, {Magee}, {Prentice}, {Colin}, {Tonry},
  {Denneau}, {Stalder}, {Heinze}, {Weiland}, {Flewelling}, \&
  {Rest}}]{Smartt2018}
{Smartt}, S.~J., {Clark}, P., {Smith}, K.~W., {et~al.} 2018, The Astronomer's
  Telegram,
  \hypersetup{urlcolor=blue}\href{https://ui.adsabs.harvard.edu/abs/2018ATel11727....1S}{11727,
  1}

\bibitem[{{Smith} {et~al.}(2022){Smith}, {Perley}, \& {Tanvir}}]{Smith2022}
{Smith}, I.~A., {Perley}, D.~A., \& {Tanvir}, N.~R. 2022, GRB Coordinates
  Network,
  \hypersetup{urlcolor=blue}\href{https://ui.adsabs.harvard.edu/abs/2022GCN.31654....1S}{31654,
  1}

\bibitem[{{Smith} \& {Tanvir}(2017)}]{Smith2017}
{Smith}, I.~A., \& {Tanvir}, N.~R. 2017, GRB Coordinates Network,
  \hypersetup{urlcolor=blue}\href{https://ui.adsabs.harvard.edu/abs/2017GCN.22242....1S}{22242,
  1}

\bibitem[{{Smith} {et~al.}(1999){Smith}, {Tilanus}, {van Paradijs}, {Galama},
  {Groot}, {Vreeswijk}, {Kouveliotou}, {Wijers}, \& {Tanvir}}]{Smith1999}
{Smith}, I.~A., {Tilanus}, R.~P.~J., {van Paradijs}, J., {et~al.} 1999, \aap,
  \hypersetup{urlcolor=blue}\href{https://ui.adsabs.harvard.edu/abs/1999A&A...347...92S}{347,
  92}

\bibitem[{{Soderberg} {et~al.}(2008){Soderberg}, {Berger}, {Page}, {Schady},
  {Parrent}, {Pooley}, {Wang}, {Ofek}, {Cucchiara}, {Rau}, {Waxman}, {Simon},
  {Bock}, {Milne}, {Page}, {Barentine}, {Barthelmy}, {Beardmore}, {Bietenholz},
  {Brown}, {Burrows}, {Burrows}, {Byrngelson}, {Cenko}, {Chandra}, {Cummings},
  {Fox}, {Gal-Yam}, {Gehrels}, {Immler}, {Kasliwal}, {Kong}, {Krimm},
  {Kulkarni}, {Maccarone}, {M{\'e}sz{\'a}ros}, {Nakar}, {O'Brien}, {Overzier},
  {de Pasquale}, {Racusin}, {Rea}, \& {York}}]{Soderberg2008}
{Soderberg}, A.~M., {Berger}, E., {Page}, K.~L., {et~al.} 2008,
  \hypersetup{urlcolor=magenta}\href{https://dx.doi.org/10.1038/nature06997}{\nat},
  \hypersetup{urlcolor=blue}\href{https://ui.adsabs.harvard.edu/abs/2008Natur.453..469S}{453,
  469}

\bibitem[{{Stroh} {et~al.}(2021){Stroh}, {Terreran}, {Coppejans}, {Bright},
  {Margutti}, {Bietenholz}, {De Colle}, {DeMarchi}, {Barniol Duran},
  {Milisavljevic}, {Murase}, {Paterson}, \& {Williams}}]{Stroh2021}
{Stroh}, M.~C., {Terreran}, G., {Coppejans}, D.~L., {et~al.} 2021, arXiv
  e-prints,
  \hypersetup{urlcolor=magenta}\href{https://arxiv.org/abs/2106.09737}{arXiv}{:}\hypersetup{urlcolor=blue}\href{https://ui.adsabs.harvard.edu/abs/2021arXiv210609737S}{2106.09737}

\bibitem[{{Sun} {et~al.}(2015){Sun}, {Zhang}, \& {Li}}]{Sun2015}
{Sun}, H., {Zhang}, B., \& {Li}, Z. 2015,
  \hypersetup{urlcolor=magenta}\href{https://dx.doi.org/10.1088/0004-637X/812/1/33}{\apj},
  \hypersetup{urlcolor=blue}\href{https://ui.adsabs.harvard.edu/abs/2015ApJ...812...33S}{812,
  33}

\bibitem[{{Urata} {et~al.}(2012){Urata}, {Huang}, {Takahashi}, \&
  {Petitpas}}]{Urata2012}
{Urata}, Y., {Huang}, K.~Y., {Takahashi}, S., \& {Petitpas}, G. 2012, GRB
  Coordinates Network,
  \hypersetup{urlcolor=blue}\href{https://ui.adsabs.harvard.edu/abs/2012GCN.13136....1U}{13136,
  1}

\bibitem[{{Urata} {et~al.}(2014){Urata}, {Huang}, {Takahashi}, {Im}, {Yamaoka},
  {Tashiro}, {Kim}, {Jang}, \& {Pak}}]{Urata2014}
{Urata}, Y., {Huang}, K., {Takahashi}, S., {et~al.} 2014,
  \hypersetup{urlcolor=magenta}\href{https://dx.doi.org/10.1088/0004-637X/789/2/146}{\apj},
  \hypersetup{urlcolor=blue}\href{https://ui.adsabs.harvard.edu/abs/2014ApJ...789..146U}{789,
  146}

\bibitem[{{van Eerten} {et~al.}(2012){van Eerten}, {van der Horst}, \&
  {MacFadyen}}]{vanEerten2012}
{van Eerten}, H., {van der Horst}, A., \& {MacFadyen}, A. 2012,
  \hypersetup{urlcolor=magenta}\href{https://dx.doi.org/10.1088/0004-637X/749/1/44}{\apj},
  \hypersetup{urlcolor=blue}\href{https://ui.adsabs.harvard.edu/abs/2012ApJ...749...44V}{749,
  44}

\bibitem[{{van Eerten} \& {MacFadyen}(2011)}]{vanEerten2011}
{van Eerten}, H.~J., \& {MacFadyen}, A.~I. 2011,
  \hypersetup{urlcolor=magenta}\href{https://dx.doi.org/10.1088/2041-8205/733/2/L37}{\apjl},
  \hypersetup{urlcolor=blue}\href{https://ui.adsabs.harvard.edu/abs/2011ApJ...733L..37V}{733,
  L37}

\bibitem[{{van Velzen} {et~al.}(2013){van Velzen}, {Frail}, {K{\"o}rding}, \&
  {Falcke}}]{vanVelzen2013}
{van Velzen}, S., {Frail}, D.~A., {K{\"o}rding}, E., \& {Falcke}, H. 2013,
  \hypersetup{urlcolor=magenta}\href{https://dx.doi.org/10.1051/0004-6361/201220426}{\aap},
  \hypersetup{urlcolor=blue}\href{https://ui.adsabs.harvard.edu/abs/2013A&A...552A...5V}{552,
  A5}

\bibitem[{{Wanderman} \& {Piran}(2015)}]{Wanderman2015}
{Wanderman}, D., \& {Piran}, T. 2015,
  \hypersetup{urlcolor=magenta}\href{https://dx.doi.org/10.1093/mnras/stv123}{\mnras},
  \hypersetup{urlcolor=blue}\href{https://ui.adsabs.harvard.edu/abs/2015MNRAS.448.3026W}{448,
  3026}

\bibitem[{{Weiler} {et~al.}(2007){Weiler}, {Williams}, {Panagia}, {Stockdale},
  {Kelley}, {Sramek}, {Van Dyk}, \& {Marcaide}}]{Weiler2007}
{Weiler}, K.~W., {Williams}, C.~L., {Panagia}, N., {et~al.} 2007,
  \hypersetup{urlcolor=magenta}\href{https://dx.doi.org/10.1086/523258}{\apj},
  \hypersetup{urlcolor=blue}\href{https://ui.adsabs.harvard.edu/abs/2007ApJ...671.1959W}{671,
  1959}

\bibitem[{{Whitehorn} {et~al.}(2016){Whitehorn}, {Natoli}, {Ade}, {Austermann},
  {Beall}, {Bender}, {Benson}, {Bleem}, {Carlstrom}, {Chang}, {Chiang}, {Cho},
  {Citron}, {Crawford}, {Crites}, {de Haan}, {Dobbs}, {Everett}, {Gallicchio},
  {George}, {Gilbert}, {Halverson}, {Harrington}, {Henning}, {Hilton},
  {Holder}, {Holzapfel}, {Hoover}, {Hou}, {Hrubes}, {Huang}, {Hubmayr},
  {Irwin}, {Keisler}, {Knox}, {Lee}, {Leitch}, {Li}, {McMahon}, {Meyer},
  {Mocanu}, {Nibarger}, {Novosad}, {Padin}, {Pryke}, {Reichardt}, {Ruhl},
  {Saliwanchik}, {Sayre}, {Schaffer}, {Smecher}, {Stark}, {Story}, {Tucker},
  {Vanderlinde}, {Vieira}, {Wang}, \& {Yefremenko}}]{Whitehorn2016}
{Whitehorn}, N., {Natoli}, T., {Ade}, P.~A.~R., {et~al.} 2016,
  \hypersetup{urlcolor=magenta}\href{https://dx.doi.org/10.3847/0004-637X/830/2/143}{\apj},
  \hypersetup{urlcolor=blue}\href{https://ui.adsabs.harvard.edu/abs/2016ApJ...830..143W}{830,
  143}

\bibitem[{{Whitesides} {et~al.}(2017){Whitesides}, {Lunnan}, {Kasliwal},
  {Perley}, {Corsi}, {Cenko}, {Blagorodnova}, {Cao}, {Cook}, {Doran},
  {Frederiks}, {Fremling}, {Hurley}, {Karamehmetoglu}, {Kulkarni}, {Leloudas},
  {Masci}, {Nugent}, {Ritter}, {Rubin}, {Savchenko}, {Sollerman}, {Svinkin},
  {Taddia}, {Vreeswijk}, \& {Wozniak}}]{Whitesides2017}
{Whitesides}, L., {Lunnan}, R., {Kasliwal}, M.~M., {et~al.} 2017,
  \hypersetup{urlcolor=magenta}\href{https://dx.doi.org/10.3847/1538-4357/aa99de}{\apj},
  \hypersetup{urlcolor=blue}\href{https://ui.adsabs.harvard.edu/abs/2017ApJ...851..107W}{851,
  107}

\bibitem[{{Woosley} \& {Bloom}(2006)}]{Woosley2006}
{Woosley}, S.~E., \& {Bloom}, J.~S. 2006,
  \hypersetup{urlcolor=magenta}\href{https://dx.doi.org/10.1146/annurev.astro.43.072103.150558}{\araa},
  \hypersetup{urlcolor=blue}\href{https://ui.adsabs.harvard.edu/abs/2006ARA&A..44..507W}{44,
  507}

\bibitem[{{Yadlapalli} {et~al.}(2022){Yadlapalli}, {Ravi}, \&
  {Ho}}]{Yadlapalli2022}
{Yadlapalli}, N., {Ravi}, V., \& {Ho}, A. Y.~Q. 2022, arXiv e-prints,
  \hypersetup{urlcolor=magenta}\href{https://arxiv.org/abs/2206.03518}{arXiv}{:}\hypersetup{urlcolor=blue}\href{https://ui.adsabs.harvard.edu/abs/2022arXiv220603518Y}{2206.03518}

\bibitem[{{Yuan} {et~al.}(2016){Yuan}, {Wang}, {Lei}, {Gao}, \&
  {Zhang}}]{Yuan2016}
{Yuan}, Q., {Wang}, Q.~D., {Lei}, W.-H., {Gao}, H., \& {Zhang}, B. 2016,
  \hypersetup{urlcolor=magenta}\href{https://dx.doi.org/10.1093/mnras/stw1543}{\mnras},
  \hypersetup{urlcolor=blue}\href{https://ui.adsabs.harvard.edu/abs/2016MNRAS.461.3375Y}{461,
  3375}

\bibitem[{{Zauderer} {et~al.}(2013{\natexlab{\hspace{0pt}a}}){Zauderer},
  {Soderberg}, {Chakraborti}, {Drout}, {Kamble}, {Milisavljevic}, \&
  {Sanders}}]{sn2013ak_2}
{Zauderer}, A., {Soderberg}, A., {Chakraborti}, S., {et~al.}
  2013{\natexlab{\hspace{0pt}a}}, The Astronomer's Telegram,
  \hypersetup{urlcolor=blue}\href{https://ui.adsabs.harvard.edu/abs/2013ATel.4946....1Z}{4946,
  1}

\bibitem[{{Zauderer} {et~al.}(2013{\natexlab{\hspace{0pt}b}}){Zauderer},
  {Berger}, {Margutti}, {Pooley}, {Sari}, {Soderberg}, {Brunthaler}, \&
  {Bietenholz}}]{Zauderer2013}
{Zauderer}, B.~A., {Berger}, E., {Margutti}, R., {et~al.}
  2013{\natexlab{\hspace{0pt}b}},
  \hypersetup{urlcolor=magenta}\href{https://dx.doi.org/10.1088/0004-637X/767/2/152}{\apj},
  \hypersetup{urlcolor=blue}\href{https://ui.adsabs.harvard.edu/abs/2013ApJ...767..152Z}{767,
  152}

\bibitem[{{Zauderer} {et~al.}(2014){Zauderer}, {Kamble}, {Chakraborti}, \&
  {Soderberg}}]{Zauderer2014}
{Zauderer}, B.~A., {Kamble}, A., {Chakraborti}, S., \& {Soderberg}, A. 2014,
  The Astronomer's Telegram,
  \hypersetup{urlcolor=blue}\href{https://ui.adsabs.harvard.edu/abs/2014ATel.5764....1Z}{5764,
  1}

\bibitem[{{Zauderer} {et~al.}(2011){Zauderer}, {Berger}, {Soderberg}, {Loeb},
  {Narayan}, {Frail}, {Petitpas}, {Brunthaler}, {Chornock}, {Carpenter},
  {Pooley}, {Mooley}, {Kulkarni}, {Margutti}, {Fox}, {Nakar}, {Patel},
  {Volgenau}, {Culverhouse}, {Bietenholz}, {Rupen}, {Max-Moerbeck}, {Readhead},
  {Richards}, {Shepherd}, {Storm}, \& {Hull}}]{Zauderer2011}
{Zauderer}, B.~A., {Berger}, E., {Soderberg}, A.~M., {et~al.} 2011,
  \hypersetup{urlcolor=magenta}\href{https://dx.doi.org/10.1038/nature10366}{\nat},
  \hypersetup{urlcolor=blue}\href{https://ui.adsabs.harvard.edu/abs/2011Natur.476..425Z}{476,
  425}

\bibitem[{{Zhang} \& {Kobayashi}(2005)}]{Zhang2005}
{Zhang}, B., \& {Kobayashi}, S. 2005,
  \hypersetup{urlcolor=magenta}\href{https://dx.doi.org/10.1086/429787}{\apj},
  \hypersetup{urlcolor=blue}\href{https://ui.adsabs.harvard.edu/abs/2005ApJ...628..315Z}{628,
  315}

\end{thebibliography}

\end{document}